\documentclass{aastex63}
\usepackage{amsmath}
\usepackage{amsfonts}
\usepackage{amssymb}
\usepackage{multirow}
\usepackage{comment}
\usepackage{graphicx}
\usepackage{soul}
\usepackage{xcolor}  

\def \apjs {{\rm {ApJS}}}

\def \aap {{\rm {A\&A}}}

\def \apjl{\rm {ApJL}}

\def \deg{$^\circ$}
\def \mj{\,$M_{\rm Jup}$\,}
\def \mstar{\,$M_\star$\,}

\def \msun{\,$M_\odot$\,}
\def \me{\,$M_\oplus$\,}
\def \masyr{\,mas\,yr$^{-1}$}

\def \auyr{\,au\,yr$^{-1}$}

\newcommand{\msyr}{\mbox{\,m s$^{-1}$yr$^{-1}$}}

\date{\today}
\received{xxx}
\revised{xxx}
\accepted{xxx}
\submitjournal{xx}

\shorttitle{3-D selection of 167 sub-stellar companions to nearby stars}
\shortauthors{Feng et al.}

\begin{document}
\title{3-D selection of 167 sub-stellar companions to nearby stars}
\author[0000-0001-6039-0555]{Fabo Feng}
\affiliation{Tsung-Dao Lee Institute, Shanghai Jiao Tong University,
  800 Dongchuan Road, Shanghai 200240, People's Republic of China}
\affiliation{School of Physics and Astronomy, Shanghai Jiao Tong University, 800 Dongchuan Road, Shanghai 200240, People's Republic of China}
\author{R. Paul Butler}
\affiliation{Earth and Planets Laboratory, Carnegie Institution for Science, Washington, DC 20015, USA}
\author{Steven S. Vogt}
\affiliation{UCO/Lick Observatory, University of California, Santa Cruz, CA 95064,USA}
\author[0000-0001-8933-6878]{Matthew S. Clement}
\affiliation{Earth and Planets Laboratory, Carnegie Institution for Science, Washington, DC 20015, USA}
\author{C.G. Tinney}
\affiliation{School of Physics and Australian Centre for Astrobiology, University of New South Wales, Sydney 2052, Australia}
\author{Kaiming Cui}
\affiliation{Tsung-Dao Lee Institute, Shanghai Jiao Tong University,
  800 Dongchuan Road, Shanghai 200240, People's Republic of China}
\author{Masataka Aizawa}
\affiliation{Tsung-Dao Lee Institute, Shanghai Jiao Tong University,
  800 Dongchuan Road, Shanghai 200240, People's Republic of China}
\author{Hugh R. A. Jones}
\affiliation{Centre for Astrophysics Research, University of Hertfordshire, College Lane, AL10 9AB, Hatfield, UK}

\author{J. Bailey}
\affiliation{School of Physics and Australian Centre for Astrobiology, University of New South Wales, Sydney 2052, Australia}
\author{Jennifer Burt}
\affiliation{Jet Propulsion Laboratory, California Institute of
  Technology, 4800 Oak Grove drive, Pasadena CA 91109}
\author{B.D. Carter}
\affiliation{University of Southern Queensland, Centre for Astrophysics, USQ Toowoomba, QLD 4350 Australia}
\author{Jeffrey D. Crane}
\affiliation{Observatories of the Carnegie Institution for Science, 813 Santa Barbara St., Pasadena, CA 91101}
\author[0000-0002-8881-3078]{Francesco Flammini Dotti}
\affiliation{Tsung-Dao Lee Institute, Shanghai Jiao Tong University,
  800 Dongchuan Road, Shanghai 200240, People's Republic of China}
\affiliation{Department of Physics, Xi{'}an Jiaotong-Liverpool University, 111 Ren{'}ai Rd., \\
Suzhou Dushu Lake Science and Education Innovation District, Suzhou Industrial Park, Suzhou 215123, P.R. China}
\affiliation{Department of Mathematical Sciences, University of Liverpool, Liverpool L69 3BX, UK}
\affiliation{Astronomisches Rechen-Institut, Zentrum f\"ur Astronomie, University of Heidelberg, M\"onchhofstrasse 12--14, 69120, Heidelberg, Germany}
\author{Bradford Holden}
\affiliation{UCO/Lick Observatory, University of California, Santa Cruz, CA 95064,USA}
\author{Bo Ma}
\affiliation{Department of Astronomy, Sun Yat-sen University, 74 Zhongshan Er Road, Guangzhou 510085, People's Republic of China}
\author[0000-0002-8300-7990]{Masahiro Ogihara}
\affiliation{Tsung-Dao Lee Institute, Shanghai Jiao Tong University,
  800 Dongchuan Road, Shanghai 200240, People's Republic of China}
\affiliation{School of Physics and Astronomy, Shanghai Jiao Tong University, 800 Dongchuan Road, Shanghai 200240, People's Republic of China}
\author{Rebecca Oppenheimer}
\affiliation{Astrophysics Department, American Museum of Natural History, Central Park West at 79th Street, New York, NY 10024, USA}
\author{S.J. O'Toole}
\affiliation{Australian Astronomical Optics, Faculty of Science and Engineering, Macquarie University, 105 Delhi Rd, North Ryde, NSW 2113, Australia}
\author{Stephen A. Shectman}
\affiliation{Observatories of the Carnegie Institution for Science, 813 Santa Barbara St., Pasadena, CA 91101}
\author{Robert A. Wittenmyer}
\affiliation{University of Southern Queensland, Centre for Astrophysics, USQ Toowoomba, QLD 4350 Australia}
\author[0000-0002-6937-9034]{Sharon X. Wang}
\affiliation{Department of Astronomy, Tsinghua University, Beijing 100084, People's Republic of China}
\author{D.J. Wright}
\affiliation{University of Southern Queensland, Centre for Astrophysics, USQ Toowoomba, QLD 4350 Australia}
\author{Yifan Xuan}
\affiliation{Tsung-Dao Lee Institute, Shanghai Jiao Tong University, 800 Dongchuan Road, Shanghai 200240, People's Republic of China}

\correspondingauthor{Fabo Feng}
\email{ffeng@sjtu.edu.cn}

\begin{abstract}
We analyse 5108 AFGKM stars with at least five
high precision radial velocity points as well as Gaia and Hipparcos
astrometric data utilizing a novel pipeline developed in previous work. We find 914 radial velocity signals with periods longer than 1000\,d. Around these signals, 167 cold giants and 68 other types of companions are identified by combined analyses of radial velocity, astrometry, and imaging data. Without correcting for detection bias, we estimate the minimum occurrence rate of the wide-orbit brown dwarfs to be 1.3\%, and find a significant brown dwarf valley around 40\mj. We also find a power-law distribution in the host binary fraction beyond 3\,au similar to that found for single stars, indicating no preference of multiplicity for brown dwarfs. Our work also reveals nine sub-stellar systems (GJ 234 B, GJ 494 B, HD 13724 b, HD 182488 b, HD 39060 b and c, HD 4113 C, HD 42581 d, HD 7449 B, and HD 984 b) that have previously been directly imaged, and many others that are observable at existing facilities. Depending on their ages we estimate that an additional 10-57 sub-stellar objects within our sample can be detected with current imaging facilities, extending the imaged cold (or old) giants by an order of magnitude. 
\end{abstract}

\section{Introduction}\label{sec:intro}
Brown dwarfs (BDs) are failed stars, which were unable to initiate nuclear fusion of hydrogen and helium. They are not considered planets as they can induce the fusion of other light elements such as Deuterium and, for very massive BDs, the fusion of Lithium \citep[e.g.,][]{burgasser08}. Following the studies of \cite{kumar63a,hayashi63,burrows01,spiegel11,baraffe15,marley21}, we use the hydrogen-burning and deuterium-burning mass limits of 13 and 75\,$M_{\rm Jup}$ to define the range of BD masses, although we are aware of the issue of using such an observationally ambiguous distinction \citep{boss96nat}. While stars are typically formed through core collapse and planets form in circumstellar disks, brown dwarfs could form through both channels. Hence BDs can be considered as an
unique population bridging planets and stars, deserving intensive scientific investigations. 

BDs can have four different spectral types M, L, T, and Y. GJ 229 B \citep{nakajima95} and Teide 1 \citep{rebolo95} are the first two BDs unambiguously discovered through direct imaging in 1995. The former is a T-type companion BD to an M dwarf, while the latter is a free-floating M-type brown dwarf located in the Pleiades open star cluster. Since 1995, astronomers have discovered more than 2500 BDs (e.g., \citealt{kirkpatrick21}) with the vast majority of those classified as individual ``free-floating'' objects (e.g., objects which are gravitationally bounded only to the galaxy central potential rather than as a companion in a binary). \cite{luhman07} found stars and brown dwarfs are mixed homogeneously based on their spatial kinematics being  indistinguishable in Chamaeleon I, a young star forming region, which further supports that both stars and brown dwarfs have the same formation mechanism. 

For directly imaged BDs, whether free-floating or having a stellar companion, the cooling model is typically used to provide an indirect estimate of their mass \citep[e.g.,][]{baraffe15,marley21}. Due to the lack of hydrogen fusion in the core of BDs, they cannot sustain their high temperature and brightness and thus cool down over time. The cooling process as a function of time depends on BD mass, metallicity, cloud coverage, etc. However, cooling models are diverse and sometimes estimate a mass inconsistent with the dynamical mass constrained by direct imaging of BDs. Hence a large sample of BD companions to stars with known masses are essential to test various cooling models. The brown dwarf host star provides a natural reference point where age might potentially be accurately determined for us to better quantify the formation and evolution of BDs. 

While we have few direct mass determinations of the thousands of free-floating BDs, we have a much smaller population of BDs orbiting around stars, whose masses are well constrained from orbital fits. Hereafter, we will refer to these BDs orbiting stars as ``circumstellar BDs''. Radial velocity surveys find the occurrence rate of circumstellar BDs is measured to be about $0.5-2\%$ from samples of thousand stars \citep[e.g.,][]{vogt02,patel07,sahlmann11,grieves17,kiefer19}. The BD desert hypothesis was proposed to explain the low detection rate of circumstellar BDs. This hypothesis was formulated in the late 1980s, when the first precise radial velocity surveys compiled results \citep[e.g.,][]{campbell88,marcy89,marcy95,marcy00}, although the significant observational biases of radial velocity (RV) and latter transit surveys were considered as plausible causes. Through an adaptive optics imaging survey, \citet{metchev09} found that there are more BDs at wide orbits than BDs in the brown dwarf desert. Recently, more and more BDs have been found to reside in this desert \citep[e.g.,][]{persson19,carmichael19,carmichael20,acton21}. As suggested by \citet{shahaf19}, characterization of the shape of the brown dwarf desert in the period-mass diagram by a large sample of circumstellar BDs can improve our understanding of its origin. 

The correlation between inner companion planets
and the wide-orbit companions may be the key to improve our understanding of
planet formation. For example, one of the puzzles is
whether the existence of a companion in a wide orbit affects the
formation of inner companions \citep{fontanive19,ziegler21}. In this
work, we can detect both inner companions and BDs with wide
orbits. Therefore, our sample allows the study of the correlation between the
wide-orbit companion and the inner companion. Another long-lasting
puzzle is whether cold Jupiters (hereafter CJs) were formed by core accretion or
gravitational instability \citep{chabrier14}. In this manner, our work aims to bolster the sample of detected systems hosting CJs in order to provide improved constraints for theoretical formation models. According to \cite{zhu12}, when BDs form due to gravitational instability, multiple CJs and BDs are
expected to form simultaneously. Therefore, it is reasonable to consider that CJs formed inside the orbit of BDs formed by the gravitational instability are both formed through the same process. We expect to be sensitive to such systems in this project.

To constrain the BD cooling models, to test the BD desert hypothesis, to find the correlation between BDs, CJs and other types of planets, and to provide a large sample of candidates for direct
imaging, we need a larger circumstellar sub-stellar sample covering the BD mass range. This requires
the improvement of BD detection sensitivity through the synergy of
all available detection techniques. A good approach is to
combine the RV and the astrometric difference between Hipparcos \citep{perryman97,leeuwen07} and
Gaia \citep{gaia16,gaia18,gaia20} for nearby stars. Various groups have used this approach to
estimate the dynamical mass of directly imaged brown dwarfs \citep{snellen18,brandt19,kervella19,xuan20,kiefer21,kervella22}.

We follow the approach developed by \cite{feng19b} to use both the position and proper motion differences between Gaia and Hipparcos to constrain the orbits of long period companions. Because the positional difference between Gaia and Hipparcos after subtracting the linear stellar motion is proportional to the square of Gaia-Hipparcos time difference, it is more sensitive to companion-induced acceleration of primary star than the proper motion difference, which is a linear function of time. In other words, 
\begin{eqnarray}
\Delta r &=& \frac{1}{2}g\Delta t^2~,\\
\Delta\mu&=& g\Delta t~,
\label{eqn:dr_dmu}
\end{eqnarray}
where $\Delta r$ is the amplitude of the positional change, $\Delta \mu$ is the proper motion change, $\Delta t$ is the difference between the reference times of Gaia EDR3 and Hipparcos catalog, $g$ is the acceleration of the primary star induced by a companion. The combined analysis of both proper motion and positional differences is found to be optimal \citep{feng21} through a comparison of different approaches (e.g., \citealt{brandt21}) to the Gaia data when considering a sample of low mass companions similar to those considered here. Although a global calibration can remove systematics to some extent {\it a priori} (e.g., \citealt{cantat21}), we prefer using astrometric jitters and offsets to model the known and unknown systematics {\it a posteriori} in order to avoid over-fitting or under-fitting problems (e.g., \citealt{foreman-mackey15} and \citealt{feng16}).

This paper is structured as follows. The RV and astrometry data are
introduced in section \ref{sec:data}. The combined modeling of RV and
astrometry is described in section \ref{sec:method}. The BDs
that are discovered and confirmed by our work are listed in
section \ref{sec:candidates}. The following section \ref{sec:stats}
explains the statistics of this sample. The detectability of this
sample by the current imaging facilities is discussed in section 
\ref{sec:imaging}. The dynamical stability of the systems are investigated in section \ref{sec:stability}. Finally, we present conclusion in section \ref{sec:conclusion}. 

\section{Data}\label{sec:data}
In this work, we use the RV data of the University College
London Echelle Spectrograph (UCLES; \citealt{diego90}) mounted on the Anglo-Australian
Telescope (AAT), the Automated Planet Finder (APF; \citealt{vogt14}) and Levy
Spectrometer at the Lick Observatory, the CORALIE spectrometer \citep{udry00}
installed at the Swiss 1.2-metre Leonhard Euler Telescope at ESO’s La
Silla Observatory, the ELODIE spectrograph \citep{baranne96} of Observatoire de
Haute-Provence, the High Accuracy Radial velocity Planet Searcher
(HARPS; \citealt{pepe00}) at the ESO La Silla 3.6m telescope, the HARPS
 for the Northern hemisphere (HARPS-N or HARPN; \citealt{cosentino12}) installed at
the Italian Telescopio Nazionale Galileo (TNG), the HIRES spectrometer \citep{vogt94} at
the Keck observatory, the Lick Observatory Hamilton echelle
spectrometer \citep{vogt87}, the Echelle Spectrograph for Rocky
Exoplanet and Stable Spectroscopic Observations (ESPRESSO; \citealt{pepe10}) installed
on VLT, the Magellan Inamori Kyocera Echelle
(MIKE) spectrograph \citep{bernstein03} and the Carnegie Planet Finder Spectrograph (PFS;
\citealt{crane10}) on the Magellan Clay Telescope, the SOPHIE
spectrograph \citep{perruchot08} at the 1.93-m telescope of
Haute-Provence Observatory, the ESO UV-visual echelle spectrograph (UVES) on the Unit Telescope 2 of the
VLT array, and the high resolution spectrograph (HRS; \citealt{tull98}) mounted on the
Hobby-Eberly Telescope (HET; \citealt{ramsey98}).

The HARPS data is reduced by \cite{trifonov20}, using the SERVAL pipeline \citep{zechmeister18}. There is a known offset in the RV zero point for the
post-2015 dataset \citep{curto15}. Hence we label the pre-2015 data set as ``HARPSpre'', and the post-2015 data as
``HARPSpost''. The AAT, APF, MIKE, PFS, and UVES data are reduced using the pipeline developed
by \cite{butler96}. The APF data for HD 182488 (or GJ 758) is
published by \cite{bowler18}. This data set is denoted by ``APF1''
while the other archived APF data is labeled ``APF2''. For $\beta$ Pic (or HD 39060), the RV data reduced
by \cite{lagrange19} is used and labeled ``AL19''. We use the published RV data from the
Lick Hamilton spectrograph and label the versions due to various
updates by Lick6, Lick8, and Lick13 \citep{fischer14}. Since the first operation of CORALIE at 1998, it had major upgrades in 2007 \citep{segransan10} and in
2014. Hence we use COR98, COR07, COR14 to denote the three versions of
data sets. The ELODIE data for HIP 63762 and the SOPHIE data for HIP 94931, HIP 14729, and
HIP 22203 are downloaded from the SOPHIE/ELODIE archives\footnote{\url{http://atlas.obs-hp.fr/}} and reduced
using the SERVAL pipeline \citep{zechmeister18}. We also use the RVs
for HD 10697, HD 136118, HD 190228, HD 23596, HD 28185, HD 38529, HD 72659, and HD 95128 measured by the 2.7\,m Harlan
J. Smith Telescope (HJS) and/or HRS at the McDonald Observatory
\citep{wittenmyer09}. For HD 14067, we use the data published by
\cite{wang14}, including: the data from the High
Dispersion Spectrograph (HDS; \citealt{noguchi02}) installed on the
Subaru telescope, and the RV data measured by the High Dispersion
Echelle Spectrograph (HIDES) at the Okayama Astrophysical Observatory
(OAO), and the data from the High Resolution Spectrograph attached at the Cassegrain focus of the 2.16\,m telescope at
Xinglong Observatory (XINGLONG). For HD139357, we use the RVs measured by the coud\'e \'echelle
spectrograph mounted on the 2\,m Alfred Jensch Telescope (AJT) of the
Thueringer Landessternwarte Tautenburg \citep{dollinger09}.
For HD 106515A, we use the RV data measured by the
high-resolution spectrograph SARG at TNG \citep{desidera12}. The new RV data for all targets are shown in the figures in the appendix.

For a given target with both RV and revised Hipparcos catalog data
\citep{leeuwen07}, we use the {\small gaiadr2.tmass\_best\_neighbour} cross-matching catalog in the Gaia data archive to find the Gaia DR2 source identity and use the {\small gaiaedr3.dr2\_neighbourhood} cross-matching catalog to find the EDR3 data \footnote{\url{https://www.cosmos.esa.int/web/gaia/earlydr3}}. For a target without Gaia counterparts in the cross-matching catalog, we select the Gaia sources within 0.1 degree from
its Hipparcos ICRS coordinates and with a parallax differing
from the Hipparcos one by less than 10\%. For stars with both DR2 and
EDR3 data, we use the difference between the revised Hipparcos catalog and the Gaia EDR3 to
constrain the orbits of companions. For stars with DR2 but without
EDR3 data, we use the Hipparcos-DR2 difference. The Hipparcos and Gaia data used in this work are shown in Table \ref{tab:hostpar}. The stellar mass for each star is from the TESS input catalog \citep{stassun19} unless it can be derived from the combined analyses of RV, astrometric and imaging data.

To compare the significance of companion-induced proper motion ($\Delta \mu$) and positional ($\Delta r$) differences between Gaia and Hipparcos, we calculate the signal-to-noise ratios (SNRs) for $\Delta \mu$ and $\Delta r$ as follows:
\begin{align}
    {\rm SNR}_\mu&=\Delta\mu/\sigma_\mu~,\\
    {\rm SNR}_r&=\Delta r/\sigma_r~.
    \label{eqn:snr}
\end{align}
In the above equations, the proper motion ($\Delta \mu$) and positional ($\Delta r$) differences and their uncertainties ($\sigma_\mu$ and $\sigma_r$) as well as the error-weighted proper motion ($\mu_\alpha$ and $\mu_\delta$) are 
\begin{align}
\Delta\mu&=\sqrt{(\mu_{\alpha}^{G}-\mu_{\alpha}^{H})^2+(\mu_{\delta}^{G}-\mu_{\delta}^{H})^2}~,\\
\Delta r&=\sqrt{\{(\alpha^G-\alpha^H)\cos{[(\delta^G+\delta^H)/2]}-\mu_\alpha\Delta t\}^2+(\delta^G-\delta^H-\mu_\delta\Delta t)^2}~,\\
\sigma_\mu&=\sqrt{\sigma^2_{\mu_{\alpha}^{H}}+\sigma^2_{\mu_{\delta}^{H}}+\sigma^2_{\mu_{\alpha}^{ G}}+\sigma^2_{\mu_{\delta}^{G}}}~,\\
\sigma_r&=\sqrt{\sigma^2_{\alpha^H}+\sigma^2_{\delta^H}+\sigma^2_{\alpha^G}+\sigma^2_{\delta^G}}~,\\
\mu_\alpha&=\frac{\mu_{\alpha}^{H}/\sigma^2_{\mu_{\alpha}^{H}}+\mu_{\alpha}^{G}/\sigma^2_{\mu_{\alpha}^{G}}}{1/\sigma^2_{\mu_{\alpha}^{H}}+1/\sigma^2_{\mu_{\alpha}^{G}}}~,\\
\mu_\delta&=\frac{\mu_{\delta}^{H}/\sigma^2_{\mu_{\delta,H}}+\mu_{\delta}^{G}/\sigma^2_{\mu_{\delta}^{G}}}{1/\sigma^2_{\mu_{\delta}^{H}}+1/\sigma^2_{\mu_{\delta}^{G}}}~,
\end{align}
where $\{\alpha^G,\delta^G,\mu_{\alpha}^{G},\mu_{\delta}^{G}\}$ and $\{\alpha^H,\delta^H,\mu_{\alpha}^H,\mu_{\delta}^{H}\}$ are respectively the Gaia and Hipparcos astrometry data, including right ascension (RA), declination (DEC), and proper motion in RA and DEC. The uncertainty of the astrometry data is denoted by $\sigma$ with corresponding subscripts. The correlation between various astrometric parameters are not considered here although they are considered in our full modeling of astrometric data. The SNR$_\mu$ and SNR$_r$ for the stellar sample in this work will be presented in section \ref{sec:candidates}. In the calculation of positional difference, the linear motion due to proper motion is subtracted. Although the perspective acceleration is not considered in the calculation of SNR, it is taken into account in our rigorous modeling of astrometry that will be introduced in the following section. 

\section{Method}\label{sec:method}
The past and current RV surveys provide a great legacy for the
discovery of CJs, BDs, and low mass stellar
companions. While RV alone is unable to give
the true mass of a companion, astrometry data can break the
degeneracy between mass and inclination and fully constrain the mass
and orbit. As a successor of the Hipparcos astrometric survey, the  
Gaia Early Data Release (EDR3) is a third epoch release of astrometric data for more than 1 billion stars. The comparison between the Hipparcos and Gaia
catalogs for common stars provides an additional constraint on potential
accelerations induced by companions on primary stars.

Rather than just relying on Gaia and Hipparcos proper motions, the frequently used approach is to compare Gaia and Hipparcos positions to derive a third proper motion to calibrate the Gaia and Hipparcos catalogs, as done by
\cite{michalik15} and \cite{brandt18}. Nevertheless, such a third proper motion
is still biased for targets hosting massive companions with orbital
periods comparable or longer than 24 years. Instead of calibrating
Gaia and Hipparcos catalogs {\it a priori}, we fit the calibration
parameters and signal-model parameters simultaneously to the raw catalog data to avoid
overfitting problems caused by conducting calibration before signal
search \citep{foreman-mackey15,feng17b}. This method is optimal compared
with the previous methods in terms of constraining the masses and orbits of small planets according to \cite{feng21}. 

Considering that the models and numerical methods are introduced by
\cite{feng19b} and \cite{feng21}, we only briefly describe the
methodology in this section. The radial velocity model consists of
Keplerian components and red noise components that are modeled using
the moving average model (MA; \citealt{tuomi12} and \citealt{feng16}). We select the optimal order of the MA model
in the Bayesian framework. Specifically, we calculate the maximum
likelihood, $L_{\rm max}$, for the $q^{\rm th}$ order MA model (or MA(q)) to derive
the Bayesian information criterion (BIC) of a model, i.e. $(q+1)\ln{n}-2\ln{\mathcal{L}_{\rm max}}$, where $n$ is the
number of data points. We select the optimal order $q_{\rm opt}$ if 
BIC of MA$(q_{\rm opt}-1)$ relative to MA$(q_{\rm opt})$ is higher
than 10 and the BIC of MA$(q_{\rm opt})$ relative to MA$(q_{\rm
  opt}+1)$ is lower than 10. A detailed description of noise model
comparison is given by \cite{feng17b}. 

Reflex motion describes the orbital motion of a primary star around the barycenter of the system. The full astrometric model for a star consists of three components: stellar reflex motion, proper motion and parallax of the barycenter of the target system. Though Gaia synthetic data can be generated by GOST\footnote{GOST: \url{https://gaia.esac.esa.int/gost/}}, we cannot access the real intermediate data for a reliable detection of short period companions. Because our aim is to detect long period companions, we treat the position $(\alpha,\delta)$ and proper motion $(\mu_\alpha,\mu_\delta)$ at the reference epochs of Gaia and Hipparcos as the instantaneous astrometry data to be modeled by the combination of the motion of the target system barycenter and the reflex motion. Considering a typical systematic RV of 10\auyr for the barycenter of a system, the change of parallax is about $0.001/d\varpi$, where $d$ is the heliocentric distance of the target system in pc. For a star 10\,pc away from the Sun, the parallax change over 24 years is about 0.01\,mas, which is far below the current precision of Gaia parallaxes. Thus we ignore the change of parallax. 

For systems with both long-period (P$>$1000\,d) and short-period planets (P$<$1000\,d), we only use the Gaia-Hipparcos astrometry to constrain the long period signals while leaving the inclination $I$ and longitude of ascending node $\Omega$ of the short-period companion unconstrained. As is described in \cite{feng19b}, we model the instantaneous proper motion and position of the target star at the Gaia epoch by combining the barycenter proper motion and the stellar reflex motion at the Gaia reference epoch. We then propagate the proper motion and position of the barycenter to the Hipparcos epoch, and add the stellar reflex motion at the Hipparcos reference epoch to model the proper motion and position of the target star at the Hipparcos epoch. In the calculation of likelihood, we use offsets and jitters to account for unknown systematics in the Gaia and Hipparcos catalog data. 

To obtain posterior samples, we use the adaptive and parallel Markov Chain Monte Carlo
(MCMC) developed by \cite{haario01} and \cite{feng19a}. All
time-related parameters such as orbital periods and correlation time
scale follow a log-uniform prior distribution. The inclination $I$ is
sampled from a uniform prior distribution over $\sin{I}$. The other
parameters have a uniform prior distribution. We use BIC$>10$ or its
equivalent ln(Bayes Factor)$>5$ \citep{kass95,feng16} to
determine whether additional companions are necessary to explain the
data. To find signals efficiently, we first conduct an RV-only analysis
following the method developed by \cite{feng20a}. Then we
apply the full modeling of RV and astrometry to the systems with
$P>1000$\,d signals to fully constrain the dynamics of the system. For
companions with shorter orbital periods ($P<1000$\,d), the
Hipparcos and Gaia EDR3 astrometry significantly deviate from the
instantaneous astrometry at reference epochs. Because these
short-period companions do not induce significant astrometric signals, they are only constrained by the RV data. For a system
with both short and long period companions, the short period
companions are constrained only by the RV data while the long period
ones are constrained both by RV and by astrometric data. Therefore, the combined analysis of both the RV and the astrometric data for a
multi-companion system would give reliable orbital solutions for all companions. 

\section{Results of combined analyses}\label{sec:candidates}
Among all of the available RV targets from various RV surveys, we
select 5108 stars with each star having more than five high precision RV data points. Following the
methodology above (with a ln(Bayes Factor)$>5$; \citealt{kass95,feng16})
and using the RV data alone we find 869 of these stars show long
period signals (LPSs; $P>1000$\,d). By our combined analyses of RV and astrometry, 167 of them are confirmed as companions with masses from 5 to 120\mj and with relative mass
uncertainty of less than 100\%. 

\begin{figure}
    \centering
    \includegraphics[scale=0.8]{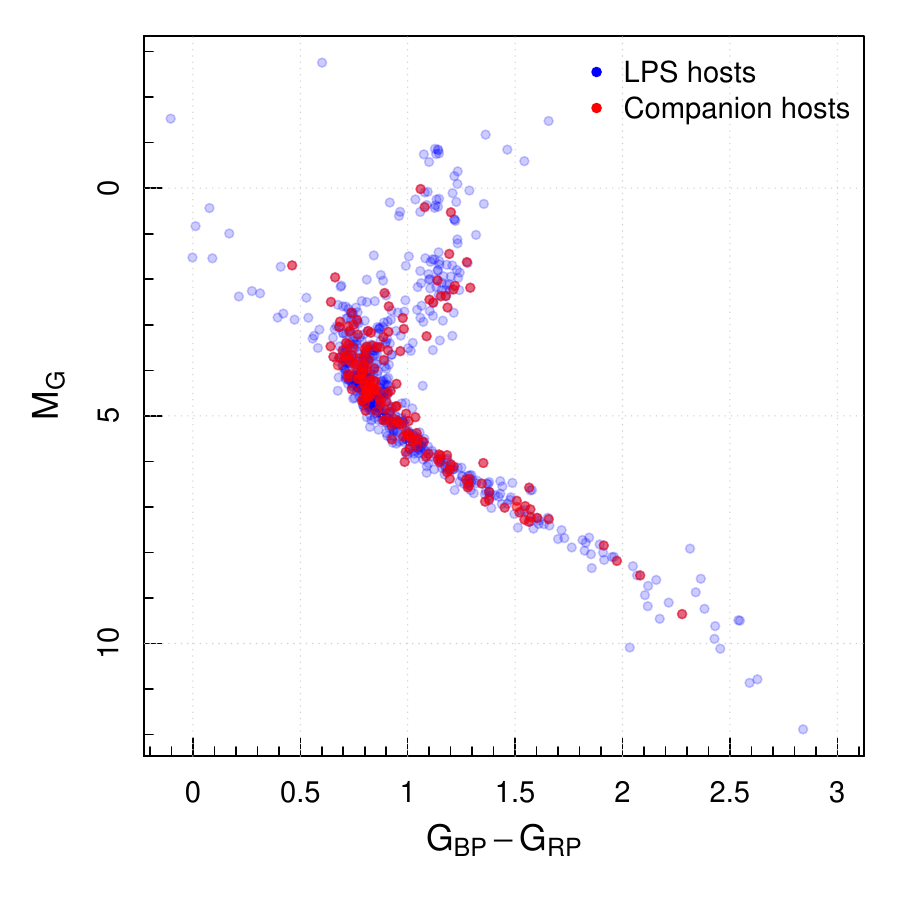}
    \caption{The Hertzsprung-Russell diagram for the stellar sample in this work. The x axis is the BP-RP color while the y axis is the absolute G magnitude derived from the Gaia EDR3 catalog. The 869 stars showing long period RV signals are shown in blue color. Of these 161 of them are found to host companions with mass from 5 to 120\mj and are shown in red color. }
    \label{fig:hr}
\end{figure}

The stellar mass and astrometry data used for the sample of 161 stars that host the 167 companions are shown in Table \ref{tab:hostpar}. For the companions which are directly imaged, the mass of their hosts are inferred together with other parameters {\it a posteriori}. The Hertzsprung-Russell diagram for the sample of stars based on the Gaia BP-RP color is shown in Fig. \ref{fig:hr}. Most of the hosts of companions are main-sequence AFGKM stars while few hosts are sub-giants. 

\begin{figure}
    \centering
    \includegraphics[scale=1]{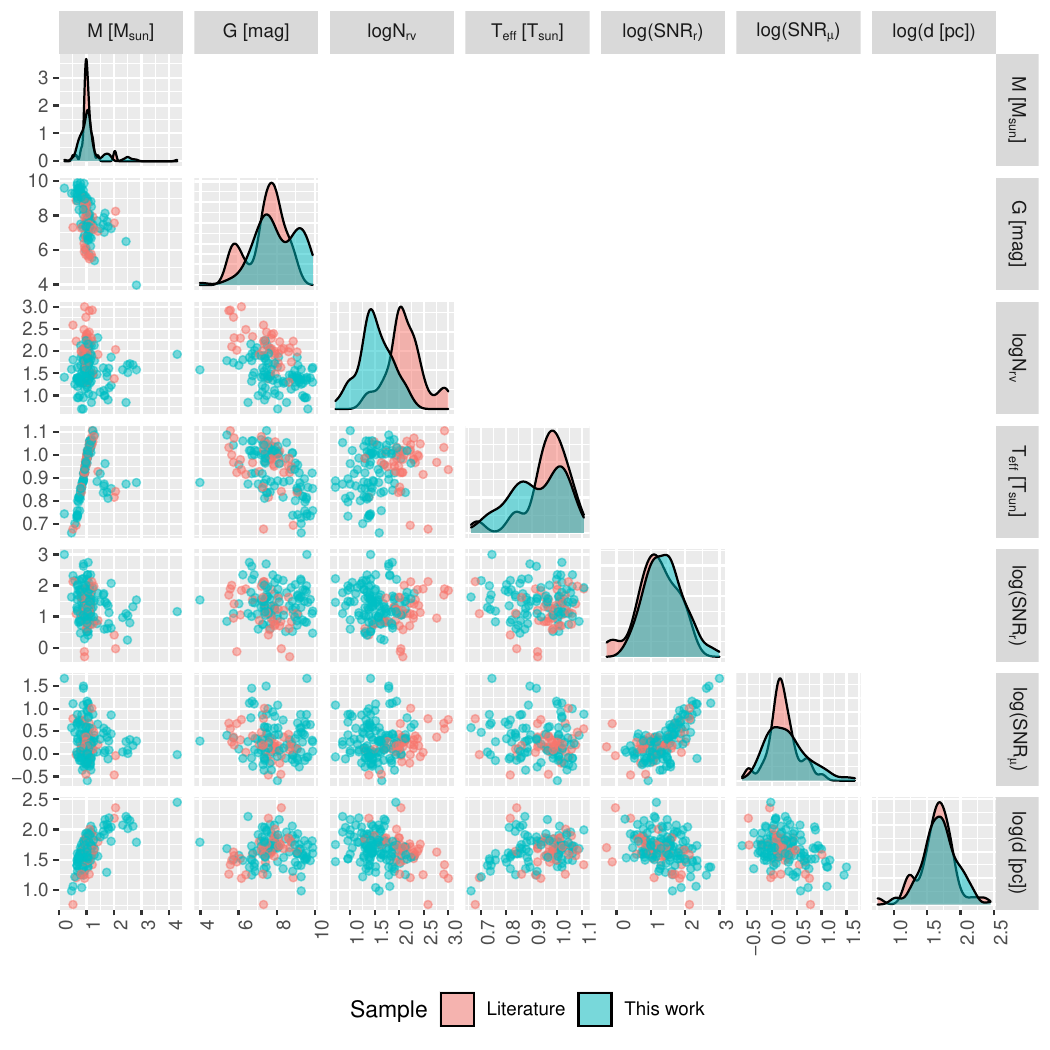}
    \caption{Corner plot for the distribution of the 161 companion hosts over various parameters. The diagonal panels show the histograms of corresponding parameters. Each of the lower panels shows the distribution of the sample over a parameter pair. The red and light green points respectively represent the hosts of known companions and the new companions identified in this work. From left to right on the top, the parameters are stellar mass ($M$), G magnitude, number of RV points ($N_{\rm rv}$), effective temperature ($T_{\rm eff}$), log(SNR$_r$), log(SNR$_\mu$), and heliocentric distance ($d$). The nonlinear positional difference $\Delta r$ is derived by subtracting the positional difference caused by proper motion from the total positional difference. The logarithmic scales are defined using base 10.}
    \label{fig:distr}
\end{figure}

We also show the distribution of stellar mass, G magnitude, number of RV points, effective temperature, SNR$_r$, SNR$_\mu$, and heliocentric distance in Fig. \ref{fig:distr}. In the figure, we divide the total sample of 161 companions into the known ones found in literature and the new ones found in this work. The major difference between these two population is that the new companions are typically identified from fewer RV data points than the known companions. This is expected because the the astrometric data help to constrain the long period signals that the RV data alone cannot confirm. As shown in Fig. \ref{fig:distr}, the majority of our sample are brigher than G$=$8\,mag and less than 100\,pc away from the Sun. It is apparent that the SNR of positional difference is one order of magnitude higher than the SNR of proper motion difference. 

\begin{figure}
  \centering
  \includegraphics[scale=0.35]{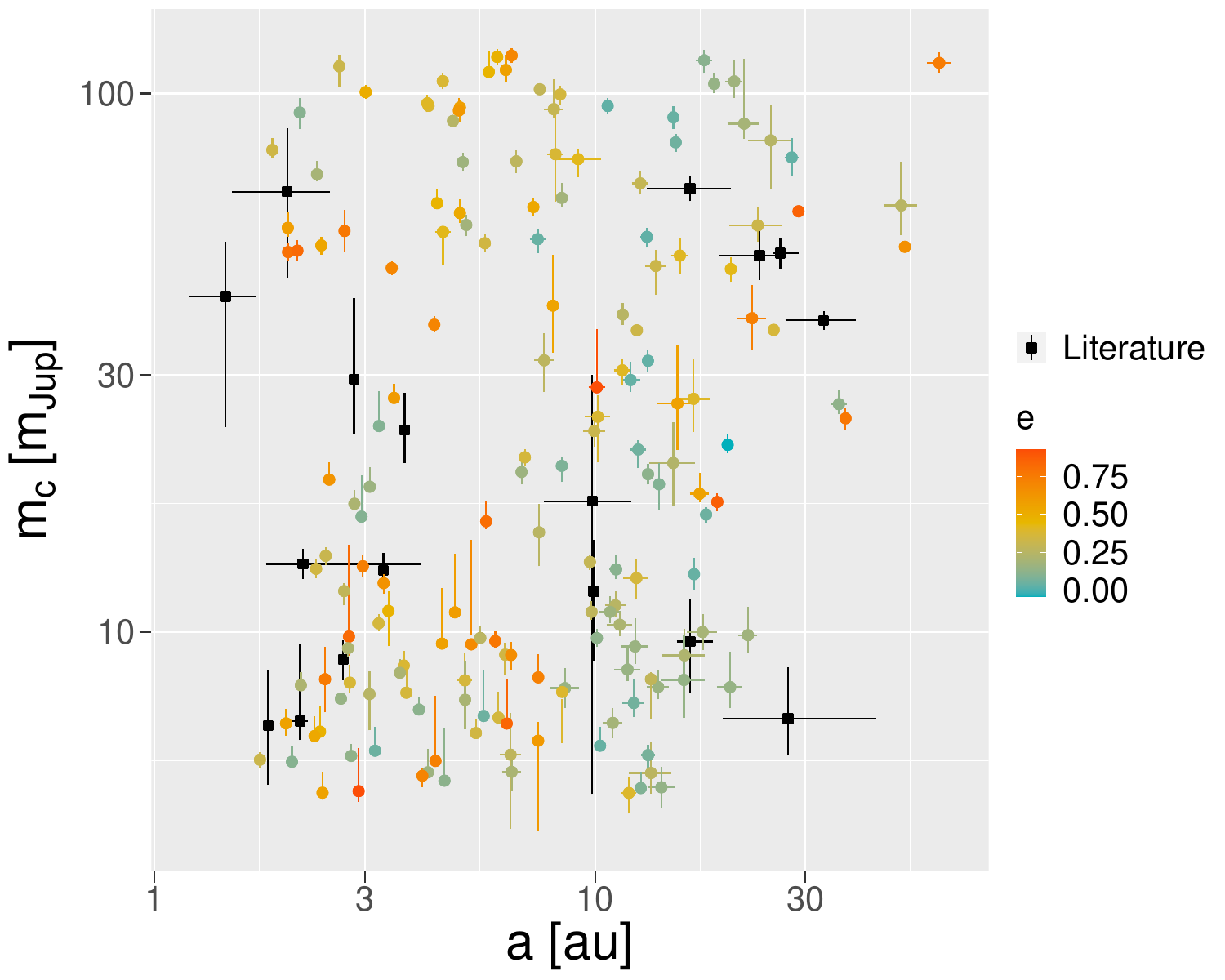}
  \includegraphics[scale=0.35]{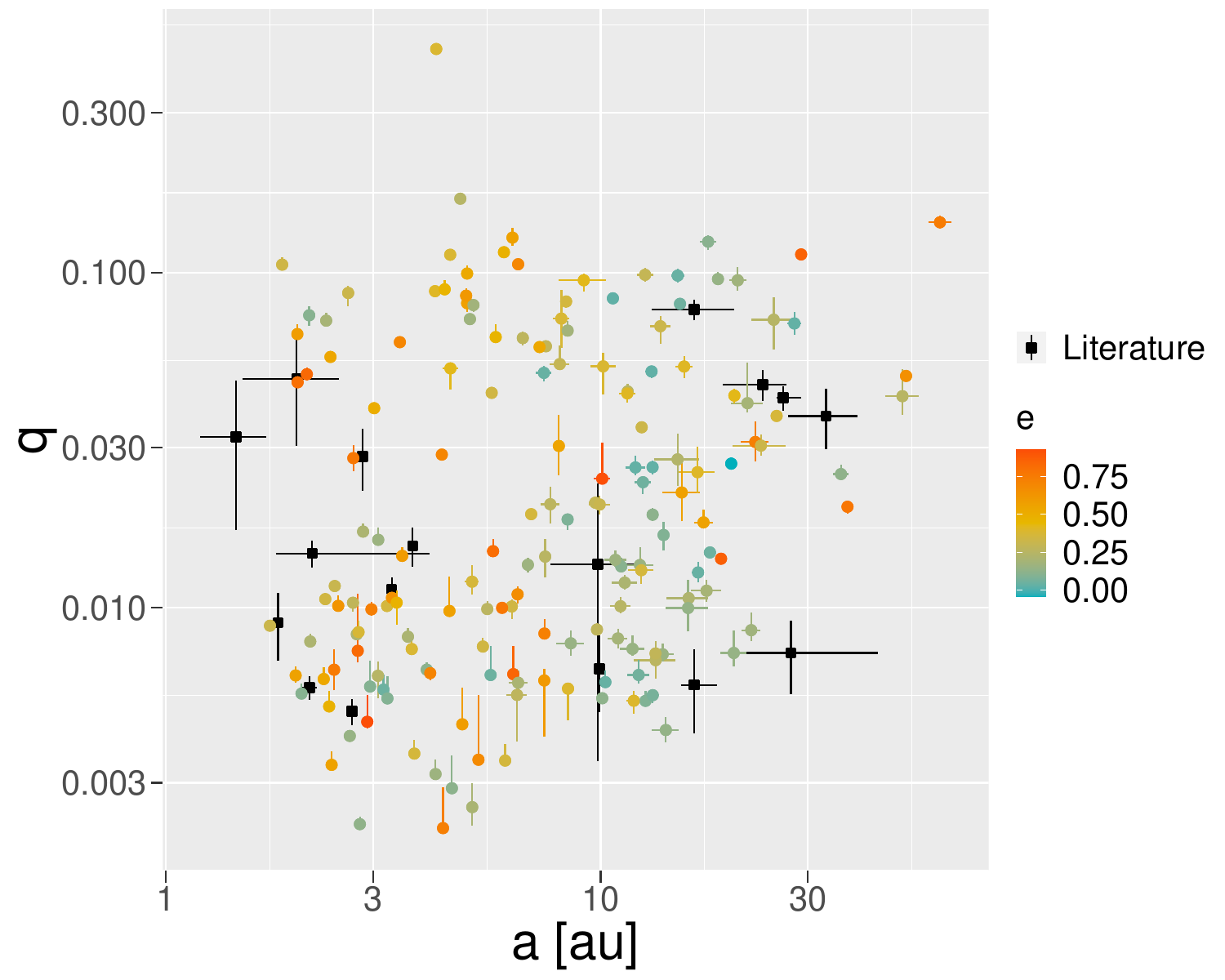}
  \caption{The confirmed companions and detected companions found in this work, successively compared
    with previous companions with relative mass uncertainty less than 100\%. The left panel shows
    the distribution over mass and semi-major axis while the right
    panel shows the distribution over companion-host mass ratio and
    semi-major axis. The companions are selected from this work and
    from the catalog from \url{exoplanet.eu} (denoted by black squares
    and labelled ``Literature'') if their relative mass uncertainties
    are less than 100\%, inclinations and eccentricities are estimated, host-star
    masses are higher than 0.2\mstar, orbital periods are longer than
    1000\,days. The mass ratio for some previous companions are
    missing in the right panel plot because their host-star masses are not
    given. 
  }
  \label{fig:ame}
\end{figure}

Around the 161 hosts of cold sub-stellar companions with a mass from 5
to 120\mj and an orbital period longer than 1000\,d, we also find 63
other types of companions, including 60 planets, 1 sub-stellar
companion and 2 stellar companions. To count the number of different
types of multi-companion systems, we define the mass ranges for
planets, BDs, and stars to be $<13$\mj, $[13,75]$\mj, and $>75$\mj,
respectively. The number of stars and different types of companions
are shown in Table \ref{tab:Ncomp}. Our sample of cold giants is 10
times larger than the current sample of 17 companions with parameters estimated to a similar precision, as shown in Fig. \ref{fig:ame}. 

\begin{deluxetable}{cccccc}
  \tablecaption{Number of companions identified in this work. \label{tab:Ncomp}}
  \tablehead{
    \colhead{Relative uncertainty} & \colhead{Full sample} &
    \colhead{LPS (systems)} & \colhead{Super-Jupiter (systems)} &
    \colhead{BDs (systems)} & \colhead{Stellar companions (systems)} 
    \\
    \colhead{}& \colhead{}& \colhead{}& \colhead{5-13\mj} & \colhead{13-75\mj} & \colhead{$>$75\mj}}
  \startdata
  $\sigma_{m_c}/m_c<100$\%&5108 &  914 (869) & 71 (68) & 67 (66) & 29 (29)\\
  $\sigma_{m_c}/m_c<20$\%&5108& 914& 40 (37) & 49 (48) & 24 (24)\\
  \enddata
  \tablecomments{The column of ``LPS'' shows the number of long period signals. The number of companions is given for two difference cases: the conservative case in brackets defined by $<$20\% relative mass error ($\sigma_{m_c}/m_c<20$\%) and the optimal case defined by $<$100\% relative mass error ($\sigma_{m_c}/m_c<100$\%).}
\end{deluxetable}

\begin{figure}
\hspace*{-0.3in}\includegraphics[scale=0.55]{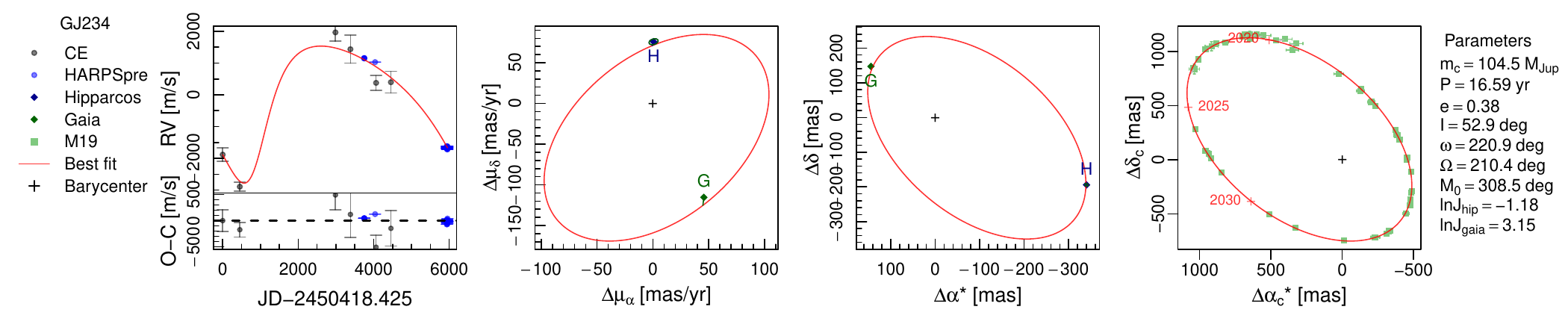}\vspace*{-0.1in}
\hspace*{-0.3in}\includegraphics[scale=0.55]{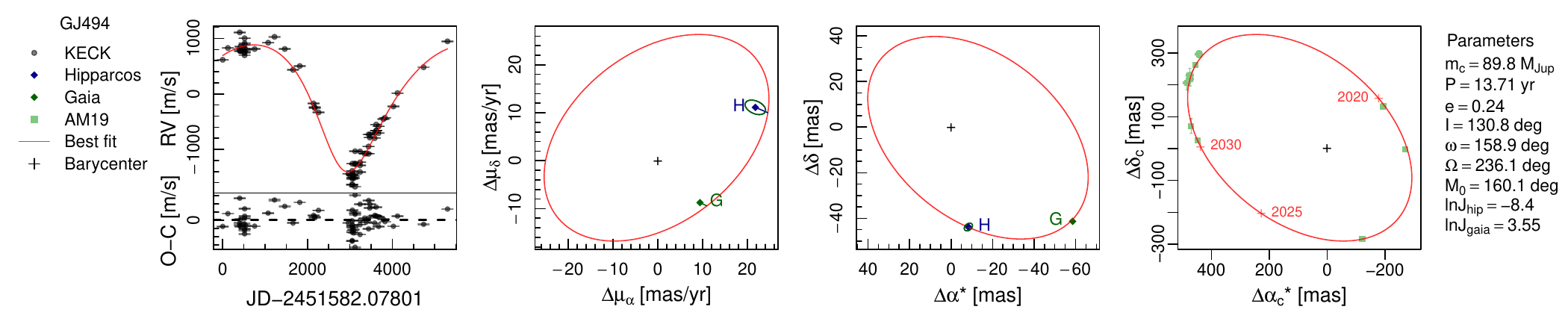}\vspace*{-0.1in}
\hspace*{-0.3in}\includegraphics[scale=0.55]{HD13724_N0_fit}\vspace*{-0.1in}
\hspace*{-0.3in}\includegraphics[scale=0.55]{HD182488_N0_fit}\vspace*{-0.1in}
\caption{Combined fit to RV, Gaia-Hipparcos astrometry, and direct imaging data for all brown dwarf targets. The left panels show the RV fits, the second and third columns respectively show the fit to the proper motion and positional differences between Gaia and Hipparcos, the right panels show the fit to the direct imaging data collected from literature. For each target, the linear motion is subtracted from the proper motion and position to show planet-induced nonlinear motion. The RV data sets as well as Gaia and Hipparcos astrometric data are encoded by the same colors as shown by the left-hand legends. The covariances of the Gaia and Hipparcos proper motions and positions are denoted by error ellipses. The straight line connects the data point to the best fitting model, which represents where the companion is expected to be. The MAP values of parameters of the companion with the longest orbital period are shown on the right side of the figure. For stars with multiple companions, only the astrometry fits for companions with periods longer than 1000 days are shown because the contribution of short-period companions to astrometry is minor. The right panels show the orbits of all companions in a system as well as the relative astrometry derived from imaging data.}
\label{fig:imaging1}
\end{figure}

\begin{figure}
\hspace*{-0.3in}\includegraphics[scale=0.55]{HD39060_N0_fit}\vspace*{-0.1in}
\hspace*{-0.3in}\includegraphics[scale=0.55]{HD4113_N0_fit}\vspace*{-0.1in}
\hspace*{-0.3in}\includegraphics[scale=0.55]{HD42581_N0_fit}\vspace*{-0.1in}
\hspace*{-0.3in}\includegraphics[scale=0.55]{HD7449_N0_fit}\vspace*{-0.1in}
\hspace*{-0.3in}\includegraphics[scale=0.55]{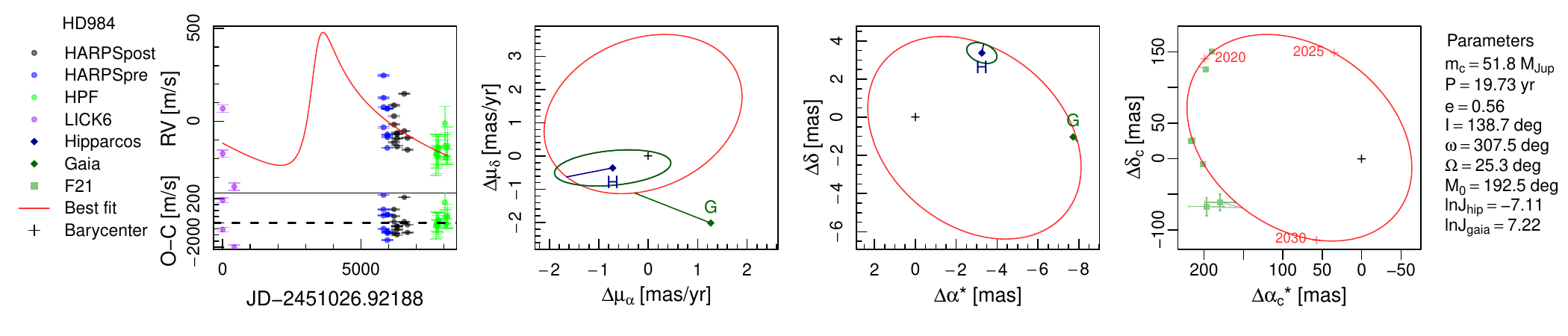}
\caption{Similar to Fig. \ref{fig:imaging1} but for other targets.}
\label{fig:imaging2}
\end{figure}

\begin{figure}
\includegraphics[scale=0.6]{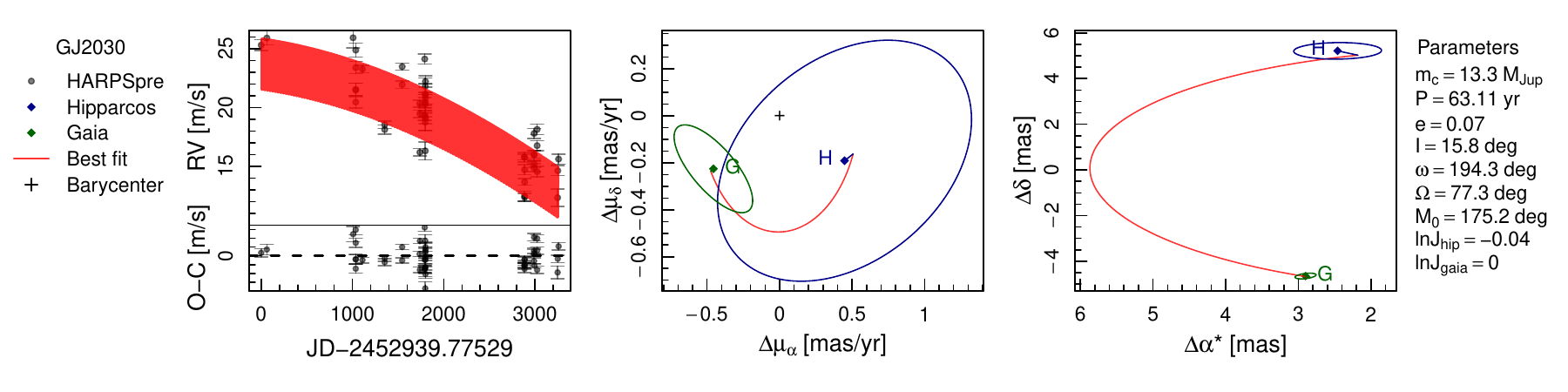}\vspace*{-0.25in}
\includegraphics[scale=0.6]{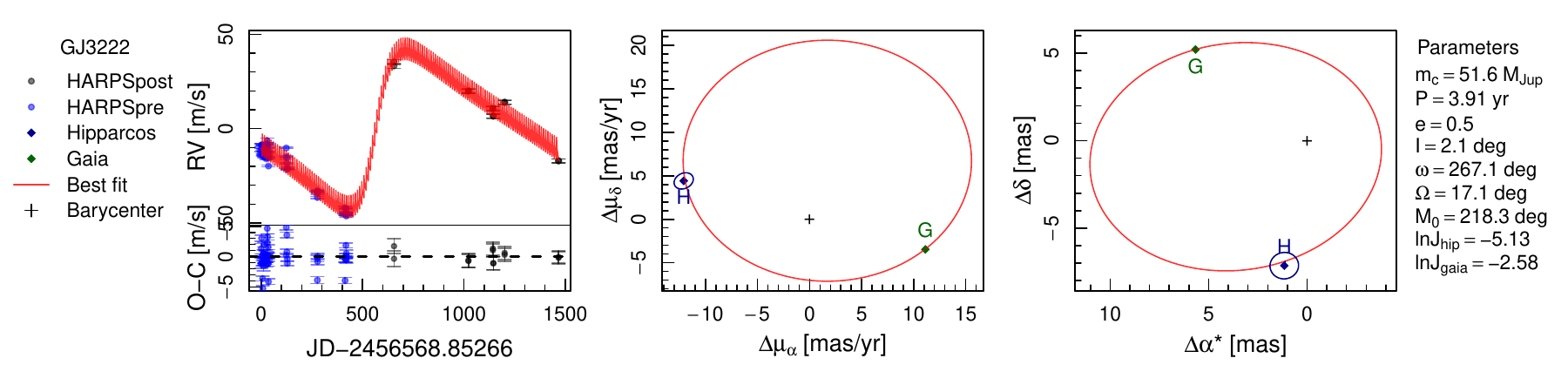}\vspace*{-0.25in}
\includegraphics[scale=0.6]{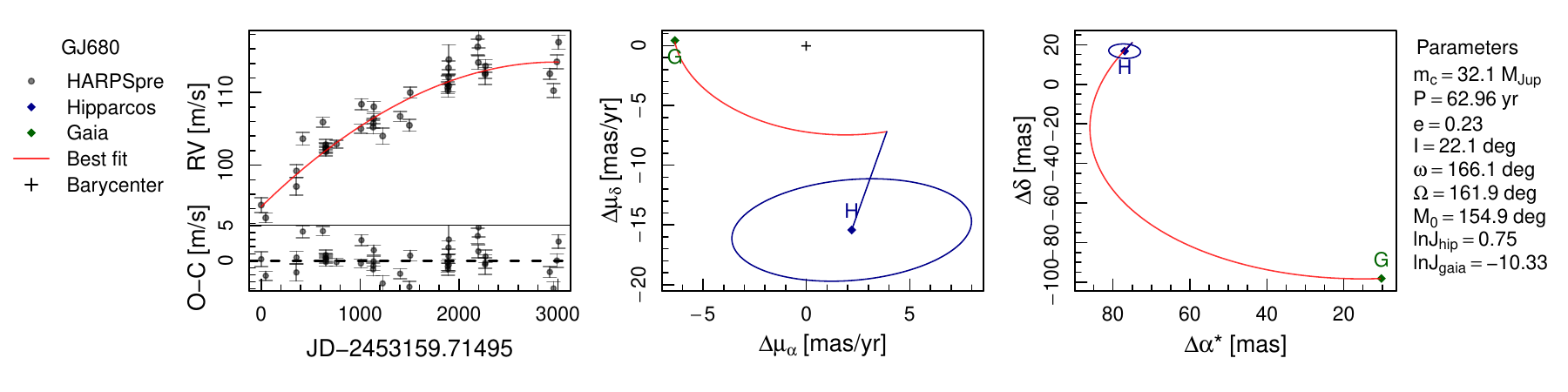}\vspace*{-0.25in}
\includegraphics[scale=0.6]{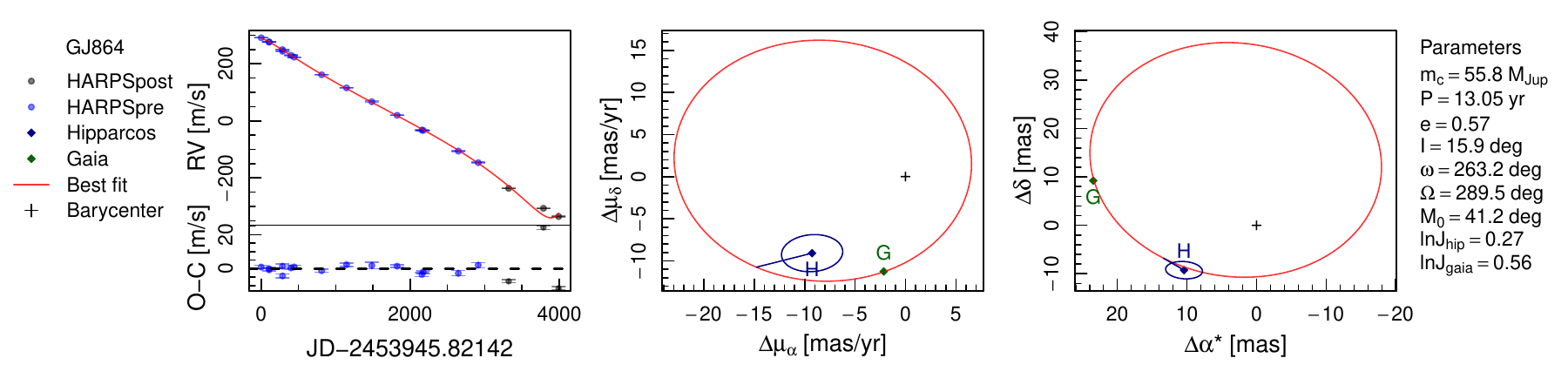}\vspace*{-0.25in}
\includegraphics[scale=0.6]{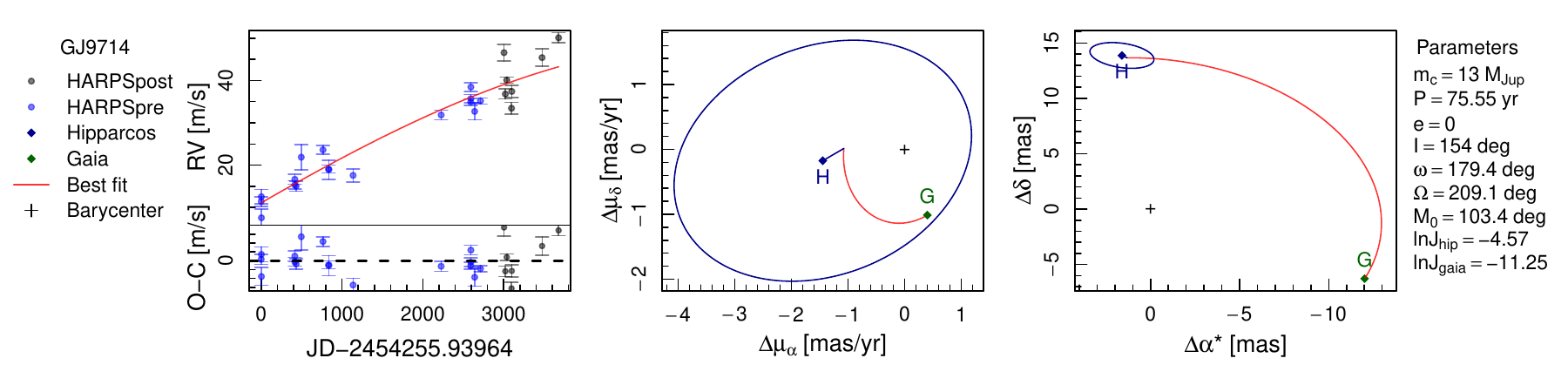}\vspace*{-0.25in}
\caption{Combined fit to the RV and Gaia-Hipparcos data for all brown dwarf targets. The left panels show the RV fits,
    the middle and right ones respectively show the fit to the proper motion and positional differences between Gaia and Hipparcos. For each target, the linear motion is subtracted from the
    proper motion and position to show planet-induced nonlinear
    motion. The RV data sets as well as Gaia and Hipparcos astrometric
    data are encoded by the same colors as shown by the left-hand legends. The
    covariances of the Gaia and Hipparcos proper motions and positions
    are denoted by error ellipses. The straight line connects the data point
    to the best fitting model, which represents where the companion is
    expected to be. The MAP values of parameters of the companion with the longest orbital period are shown on the right side of the figure. For stars with multiple companions, only the astrometry fits for companions with periods longer than 1000 days are shown because the contribution of short-period companions to astrometry is minor.}
\label{fig:fit1}
\end{figure}

\begin{figure}
\includegraphics[scale=0.6]{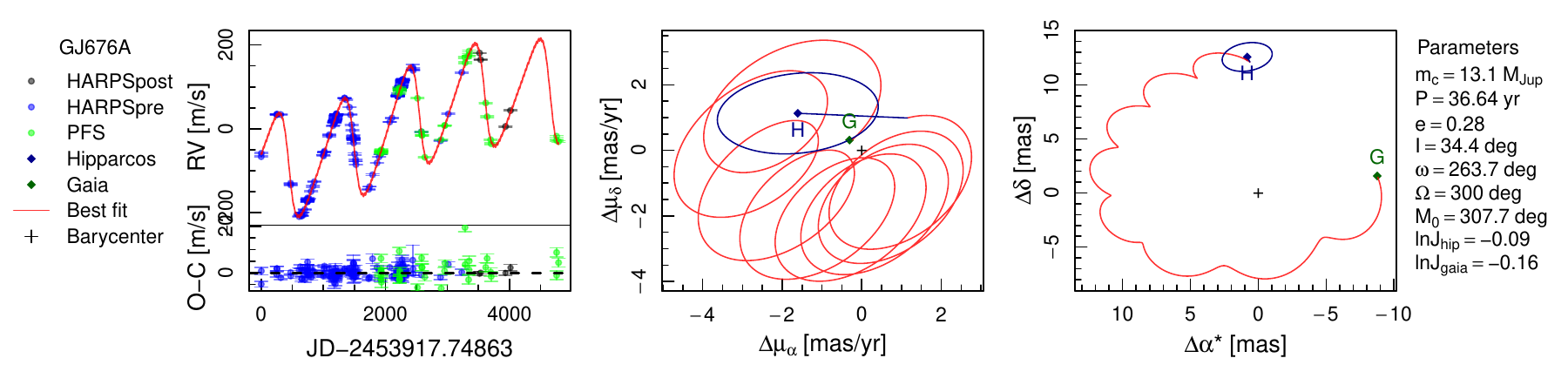}\vspace*{-0.25in}
\includegraphics[scale=0.6]{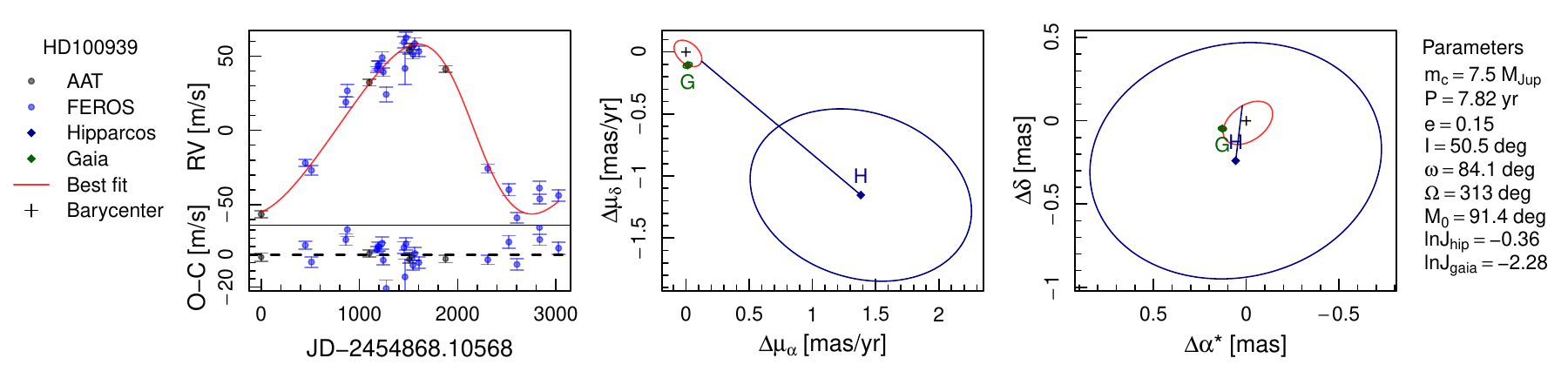}\vspace*{-0.25in}
\includegraphics[scale=0.6]{HD105618_N0_fit}\vspace*{-0.25in}
\includegraphics[scale=0.6]{HD105811_N0_fit}\vspace*{-0.25in}
\includegraphics[scale=0.6]{HD106515A_N0_fit}\vspace*{-0.25in}
\includegraphics[scale=0.6]{HD10697_N0_fit}
\caption{Similar to Fig. \ref{fig:fit1} but for other targets.}
\label{fig:fit2}
\end{figure}

The fitting results for targets with direct imaging data are shown in Fig. \ref{fig:imaging1} and \ref{fig:imaging2} while example fittings for targets without imaging data are shown in Fig. \ref{fig:fit1} and \ref{fig:fit2}. For a target with imaging data and shown in Fig. \ref{fig:imaging1} and \ref{fig:imaging2}, the first panel from left to right shows the optimal fit to the RV data. The second and third panels respectively shows the fit of the reflex motion induced by wide-orbit companions ($P>1000$\,d) to the proper motions and positions at the Hipparcos and Gaia epoch after subtracting the barycentric motion. The fourth panel shows the companion's binary orbit to its relative astrometry derived from imaging data. For targets without imaging data and shown in Fig. \ref{fig:fit1} and \ref{fig:fit2}, the fourth panel is not shown. The fit to the Gaia-Hipparcos difference for a target is based on a prediction of the reflex motion with the optimal parameters over a time-span ranging from the Hipparcos epoch to the Gaia epoch. For a system consist of multiple wide-orbit companions, the reflex motion could be complex (e.g. HD 7449 in Fig. \ref{fig:imaging2}, GJ 676 A and HD 105811 in Fig. \ref{fig:fit2}).

We present Hipparcos and Gaia EDR3 astrometry for the stars in Table \ref{tab:hostpar} and the orbital parameters for stellar planetary companions based on combined RV and astrometric analyses in Table \ref{tab:par}. Because this catalog records dynamical mass and orbital parameters through combined analyses of RV and astrometry data, it is not directly comparable with proper motion anomaly catalogs that are more comprehensive but do not break the degeneracy between companion mass and orbital parameters (e.g., \citealt{kervella22}). We proceed to discuss the individual targets that are found to host companions in previous studies. 
\begin{deluxetable*}{llll llll lllr}
\tablecaption{Hipparcos and Gaia EDR3 catalog astrometry for the sample of stars.\label{tab:hostpar}}
\tablehead{
\colhead{ Star }&\colhead{ Mass }&\colhead{ $\alpha^H$ }&\colhead{ $\delta^H$ }&\colhead{ $\tilde{\omega}^H$ }&\colhead{ $\mu_\alpha^H$ }&\colhead{ $\mu_\delta^H$ }&\colhead{ $\alpha^G$ }&\colhead{ $\delta^G$ }&\colhead{ $\tilde{\omega}^G$ }&\colhead{ $\mu_\alpha^G$ }&\colhead{ $\mu_\delta^G$ } \\
\colhead{  }&\colhead{ $M_\odot$ }&\colhead{ deg }&\colhead{ deg }&\colhead{ mas }&\colhead{ \masyr }&\colhead{ \masyr }&\colhead{ deg }&\colhead{ deg }&\colhead{ mas }&\colhead{ \masyr }&\colhead{ \masyr } }
\startdata
GJ 2030  & 1.0(1) &
50.82375&-7.793565&26.9(5)&2.2(6)&-219.3(3)&50.82376&-7.795073&26.79(9)&1.3(2)&-219.3(1) \\ 
GJ 234 * & 0.195(1) &
97.34581&-2.812477&242(3)&705(3)&-612(2)&97.3507&-2.817014&243.0(9)&750(2)&-803(1) \\ 
GJ 3222  & 0.9(1) &
50.89679&-40.07649&57.4(7)&40.8(6)&43.1(6)&50.89727&-40.07622&56.56(3)&63.99(2)&35.17(4) \\ 
GJ 494 $^\dagger$ & 0.511(4) &
195.1956&12.37576&86(2)&-616(2)&-14(1)&195.1911&12.37559&86.9(1)&-628.7(2)&-33.5(1) \\ 
GJ 676A  & 0.63(6) &
262.5477&-51.63653&61(2)&-260(1)&-184.3(8)&262.5448&-51.6378&62.58(3)&-258.76(3)&-185.12(2) \\ 
GJ 680  & 0.46(5) &
263.8064&-48.68197&103(3)&83(4)&454(3)&263.8072&-48.67878&103.31(2)&74.07(2)&470.17(1) \\ 
GJ 864  & 0.58(6) &
339.0402&-0.8401511&58(2)&49(2)&-628(1)&339.0406&-0.844402&60.38(4)&55.90(4)&-630.31(4) \\ 
GJ 9714  & 0.67(8) &
315.4122&-32.52418&49(1)&244(2)&-121(1)&315.4142&-32.52502&48.79(2)&246.27(2)&-122.03(2) \\ 
HD 100939  & 1.7(1) &
174.2007&-37.03904&8.2(7)&25.2(6)&-15.6(5)&174.2009&-37.03914&8.22(2)&23.86(2)&-14.54(2) \\ 
HD 105618  & 1.0(1) &
182.4028&11.21158&17(1)&-103(1)&-2.3(6)&182.4021&11.21157&14.32(2)&-104.35(2)&-2.02(2) \\ 
$\cdots$&
$\cdots$&
$\cdots$&
$\cdots$&
$\cdots$&
$\cdots$&
$\cdots$&
$\cdots$&
$\cdots$&
$\cdots$&
$\cdots$&
$\cdots$\\
\enddata
\tablecomments{Table \ref{tab:hostpar} is published in its entirety in the electronic edition of the {\it Astrophysical Journal Supplement Series}. A portion is shown here for guidance regarding its form and content.}
\tablecomments{The masses of stars are from the TESS input catalog \citep{stassun19}. The superscripts ``gaia'' and ``hip'' are respectively used to denote the Gaia EDR3 and Hipparcos right ascension ($\alpha$), declination ($\delta$), parallax ($\varpi$), proper motion in right ascension ($\mu_\alpha$), and the proper motion in declination ($\mu_\delta$). The hosts with $\dagger$ superscript have been directly imaged and the stellar mass is infered from the relative astrometry {\it a posteriori}. For a star without EDR3 but with DR2 data, a star symbol is added behind its name.}
\end{deluxetable*}

\begin{longrotatetable}
\begin{deluxetable*}{
llllllll
r}
\tabletypesize{\small}
\tablecaption{\bf Orbital parameters for stellar and planetary companions detected in this work.
      \label{tab:par}}
\tablehead{ \colhead{Companion}&\colhead{Host}&\colhead{$P$}&\colhead{I}&\colhead{$e$}&\colhead{$m_c$}&\colhead{$a$}&\colhead{HZ Stability}&\colhead{Reference} \\
\colhead{Name}&\colhead{Name}&\colhead{yr}&\colhead{deg}&\colhead{}&\colhead{$M_{\rm Jup}$}&\colhead{au}&\colhead{}&\colhead{} }
\startdata
GJ 2030 b&HD 21019&$ 0.007 \pm 4\times 10^{-7} $&---&$ 0.246 \pm 0.073 $&$ 0.014 \pm 0.002 $&$ 0.034 \pm 0.002 $ & Stable &  \\
GJ 2030 c&HD 21019&$ 70.185 \pm 6.819 $&$ 17.21 \pm 2.84 $&$ 0.045 \pm 0.038 $&$ 12.934 \pm 2.361 $&$ 16.752 \pm 1.312 $ & Stable &  \\
GJ 234 B&Ross 614&$ 16.586 \pm 0.004 $&$ 52.92 \pm 0.02 $&$ 0.382 \pm 1\times 10^{-4} $&$ 94.646 \pm 1.144 $&$ 4.186 \pm 0.008 $ & Stable & 1, 2, 3 \\
GJ 3222 b&HD 21175$^\star$&$ 0.029 \pm 6\times 10^{-6} $&---&$ 0.913 \pm 0.061 $&$ 0.038 \pm 0.010 $&$ 0.091 \pm 0.004 $ & Unstable &  \\
GJ 3222 c&HD 21175$^\star$&$ 3.912 \pm 0.024 $&$ 2.08 \pm 0.19 $&$ 0.547 \pm 0.049 $&$ 52.317 \pm 5.132 $&$ 2.385 \pm 0.103 $ & Unstable &  \\
GJ 494 B$^\dagger$&BD+132618&$ 13.677 \pm 0.035 $&$ 130.68 \pm 0.21 $&$ 0.245 \pm 0.001 $&$ 88.492 \pm 2.231 $&$ 4.755 \pm 0.015 $ & Stable & 1, 4, 5, 6, 7, 8 \\
GJ 676 A d&CD-5110924&$ 0.010 \pm 5\times 10^{-7} $&---&$ 0.19 \pm 0.07 $&$ 0.012 \pm 0.001 $&$ 0.039 \pm 0.001 $ & Unstable & 9, 10, 11 \\
GJ 676 A e&CD-5110924&$ 0.097 \pm 4\times 10^{-5} $&---&$ 0.16 \pm 0.08 $&$ 0.021 \pm 0.002 $&$ 0.181 \pm 0.006 $ & Unstable & 9, 10, 11 \\
GJ 676  A b&CD-5110924&$ 2.879 \pm 0.001 $&$ 49.12 \pm 3.19 $&$ 0.319 \pm 0.003 $&$ 5.788 \pm 0.477 $&$ 1.733 \pm 0.058 $ & Unstable & 9, 10, 11 \\
GJ 676  A c&CD-5110924&$ 37.824 \pm 3.458 $&$ 33.71 \pm 1.39 $&$ 0.289 \pm 0.039 $&$ 13.451 \pm 1.091 $&$ 9.655 \pm 0.676 $ & Unstable & 9, 10, 11 \\
GJ 680 b&CD-4811837$^\star$&$ 48.359 \pm 12.404 $&$ 21.85 \pm 5.21 $&$ 0.377 \pm 0.115 $&$ 23.505 \pm 7.716 $&$ 10.197 \pm 1.797 $ & Stable &  \\
GJ 864 b&HD 214100&$ 14.157 \pm 1.026 $&$ 15.48 \pm 0.63 $&$ 0.525 \pm 0.040 $&$ 61.008 \pm 6.971 $&$ 4.904 \pm 0.289 $ & Stable &  \\
GJ 9714 b&HD 199981$^\star$&$ 58.669 \pm 20.864 $&$ 155.37 \pm 4.77 $&$ 0.210 \pm 0.194 $&$ 10.070 \pm 2.483 $&$ 13.048 \pm 3.073 $ & Stable &  \\
HD 100939 b&HD 100939&$ 7.410 \pm 0.429 $&$ 87.05 \pm 38.71 $&$ 0.116 \pm 0.064 $&$ 7.058 \pm 1.563 $&$ 4.581 \pm 0.196 $ & Stable &  \\
HD 105618 b&HD 105618&$ 0.027 \pm 4\times 10^{-6} $&---&$ 0.457 \pm 0.042 $&$ 0.079 \pm 0.008 $&$ 0.091 \pm 0.004 $ & Stable &  \\
HD 105618 c&HD 105618&$ 226.881 \pm 52.684 $&$ 31.73 \pm 3.73 $&$ 0.138 \pm 0.090 $&$ 27.241 \pm 3.681 $&$ 37.210 \pm 5.771 $ & Stable &  \\
HD 105811 b&HD 105811&$ 3.223 \pm 0.017 $&$ 0.50 \pm 0.11 $&$ 0.772 \pm 0.043 $&$ 57.923 \pm 10.085 $&$ 2.700 \pm 0.053 $ & Unstable &  \\
HD 105811 B&HD 105811&$ 6.760 \pm 0.008 $&$ 81.09 \pm 5.72 $&$ 0.609 \pm 0.003 $&$ 207.590 \pm 8.628 $&$ 4.571 \pm 0.085 $ & Unstable &  \\
HD 106515 A b&HD 106515 A$^\star$&$ 9.849 \pm 0.145 $&$ 89.73 \pm 36.43 $&$ 0.564 \pm 0.027 $&$ 12.696 \pm 4.859 $&$ 4.491 \pm 0.188 $ & Stable & 12, 13, 14 \\
HD 10697 b&109 Psc&$ 2.944 \pm 0.002 $&$ 86.03 \pm 19.69 $&$ 0.105 \pm 0.008 $&$ 6.116 \pm 0.733 $&$ 2.047 \pm 0.083 $ & Stable & 15, 16, 17, 18 \\
HD 109988 b&HD 109988$^\star$&$ 47.503 \pm 8.217 $&$ 33.30 \pm 4.12 $&$ 0.061 \pm 0.041 $&$ 20.992 \pm 3.094 $&$ 12.558 \pm 1.546 $ & Stable &  \\
HD 110537 b&HD 110537&$ 71.162 \pm 26.741 $&$ 53.56 \pm 7.26 $&$ 0.386 \pm 0.128 $&$ 28.727 \pm 10.551 $&$ 17.131 \pm 4.325 $ & Stable &  \\
HD 111031 b&HD 111031&$ 46.080 \pm 3.765 $&$ 169.60 \pm 0.52 $&$ 0.037 \pm 0.027 $&$ 53.741 \pm 5.749 $&$ 12.942 \pm 0.895 $ & Stable & 19 \\
$\cdots$&$\cdots$&$\cdots$&$\cdots$&$\cdots$&$\cdots$&$\cdots$&$\cdots$&$\cdots$\\
\enddata
\tablecomments{\bf Table \ref{tab:par} is published in its entirety in the
electronic edition of the {\it Astrophysical Journal Supplement Series}.  A portion is
shown here for guidance regarding its form and content. The systems are sorted by the host names listed in the first column while the companions in each system are sorted by their orbital periods. For companions with orbital period less than 1000 days, the inclination is not given and the mass reported in this table is $m_c\sin{I}$. For companions with mass higher than 75\mj, we use capital letters to label them. Among the parameters, directly inferred parameters are $P$ (orbital period), $K$ (RV semi-amplitude), $e$ (eccentricity), $\omega_\star$ (argument of periastron), $M_0$ (mean anomaly at the minimum epoch of RV data), $I$ (inclination), $\Omega$ (longitude of ascending node). The derived parameters are $T_p$ (periastron epoch), $m_c$ (companion mass), and $a$ (semi-major axis). The dynamical stability of Earth-like planets in the habitable zones of their host stars are determined numerically, as described in section \ref{sec:stability}. The median and the 1-$\sigma$ quantiles (i.e. 16\% and 84\% quantiles) are used to measure the uncertainty of each parameter. The companions with $\dagger$ superscript have been imaged and their relative astrometry data are analyzed in combination with other types of data in the solutions. The targets with $\star$ superscript have wide binary companions that are identified by \cite{elbadry21} or other studies.}
\tablerefs{
(1) \citet{mann19}; (2) \citet{agati15}; (3) \citet{bonavita16}; (4) \citet{burgasser10}; (5) \citet{heintz94}; (6) \citet{beuzit04}; (7) \citet{goldman10}; (8) \citet{dupuy13}; (9) \citet{forveille11}; (10) \citet{stassun17}; (11) \citet{angladaescude12}; (12) \citet{marmier13}; (13) \citet{desidera12}; (14) \citet{li21}; (15) \citet{zucker00}; (16) \citet{simpson10}; (17) \citet{wittenmyer09}; (18) \citet{vogt00}; (19) \citet{hinkel19}; (20) \citet{mayor04}; (21) \citet{borgniet14}; (22) \citet{xuan20}; (23) \citet{borgniet19}; (24) \citet{jones16}; (25) \citet{fischer07}; (26) \citet{ment18}; (27) \citet{giguere15}; (28) \citet{sato13}; (29) \citet{wilson16}; (30) \citet{luhn19}; (31) \citet{locurto10}; (32) \citet{howard10}; (33) \citet{martioli10}; (34) \citet{fischer02}; (35) \citet{rickman19}; (36) \citet{rickman20}; (37) \citet{dollinger09}; (38) \citet{wang14}; (39) \citet{tinney02}; (40) \citet{wittenmyer12}; (41) \citet{wittenmyer20}; (42) \citet{butler03}; (43) \citet{wittenmyer07}; (44) \citet{feng15}; (45) \citet{diaz12}; (46) \citet{patel07}; (47) \citet{marcy99}; (48) \citet{pilyavsky11}; (49) \citet{naef01}; (50) \citet{jones21}; (51) \citet{arriagada10}; (52) \citet{thalmann09}; (53) \citet{vigan16}; (54) \citet{brandt21}; (55) \citet{bowler18}; (56) \citet{marcy05}; (57) \citet{fuhrmann04}; (58) \citet{tokovinin14a}; (59) \citet{roberts15b}; (60) \citet{perrier03}; (61) \citet{sahlmann11}; (62) \citet{han01}; (63) \citet{reffert11}; (64) \citet{diaz16}; (65) \citet{segransan10}; (66) \citet{robertson12}; (67) \citet{moutou17}; (68) \citet{ginski16}; (69) \citet{moutou11}; (70) \citet{kane19}; (71) \citet{melo07}; (72) \citet{venner21}; (73) \citet{jenkins17}; (74) \citet{rosenthal21}; (75) \citet{moutou05}; (76) \citet{trifonov17}; (77) \citet{angladaescude10}; (78) \citet{kurster15}; (79) \citet{santos01}; (80) \citet{tokovinin14b}; (81) \citet{griffin12}; (82) \citet{bouchy16}; (83) \citet{wittenmyer17}; (84) \citet{zurlo18}; (85) \citet{fischer01}; (86) \citet{lagrange09}; (87) \citet{nielsen14}; (88) \citet{lagrange19}; (89) \citet{lacour21}; (90) \citet{jones02}; (91) \citet{huang18}; (92) \citet{alvarez20}; (93) \citet{jenkins15}; (94) \citet{tamuz08}; (95) \citet{cheetham18}; (96) \citet{mugrauer14}; (97) \citet{feng20}; (98) \citet{nakajima95}; (99) \citet{naef10}; (100) \citet{allen12}; (101) \citet{butler06}; (102) \citet{barbato18}; (103) \citet{moutou09}; (104) \citet{naef04}; (105) \citet{dumusque11}; (106) \citet{wittenmyer19}; (107) \citet{rodigas16}; (108) \citet{demangeon21}; (109) \citet{sozzetti06}; (110) \citet{bang20}; (111) \citet{fischer09}; (112) \citet{hartmann10}; (113) \citet{johnsongroh17}; (114) \citet{wagner19}; (115) \citet{franson22}; (116) \citet{jones15}; (117) \citet{jones17}; (118) \citet{jones15b}; (119) \citet{mason01}
.}
\end{deluxetable*}
\end{longrotatetable}


 
For previously published planets, we discuss our new results on an object-by-object basis as follows.
\begin{itemize}
\item GJ 234 A (or Ross 614) is an M-dwarf hosting a companion with a mass of about 100\mj \citep{agati15,bonavita16,mann19}. Based on combined analyses of the RV, Gaia-Hipparcos astrometry and the imaging data collected by \cite{mann19}, we estimate a mass of $94.84_{-1.37}^{+0.88}$\mj. 

\item GJ 494 (or Ross 458) is an M-dwarf binary
  \citep{heintz94,beuzit04} hosting a young BD known as GJ 494 c which
  has a mass of 11.3$\pm$4.5\mj \citep{dupuy13}. Because GJ 494 c is on an extremely wide orbit, it does not change the motion of the inner binary much and is thus not considered in our solution. Based on combined
  analyses of the RV and Gaia-Hipparcos data as well as the relative
  astrometry data from \cite{mann19} (labeled ``AM19''), we estimate a mass of 0.511$\pm$0.004\msun for GJ 494 A and a mass of $88.92_{-2.84}^{+1.84}$\mj for GJ 494 B. According to the mass-luminosity relation derived by \cite{mann19}, GJ 494 B is an M7.4-type dwarf while GJ 494 A is an M1.5-type dwarf, differing from previous classifications based on less accurate mass estimation \citep{heintz94,beuzit04}. 

\item GJ 676 is a binary and the primary hosts four planets \citep{forveille11,anglada12}. Our solution constrains the inclinations of GJ 676 A b and c to be $48.92_{-2.78}^{+3.31}$ and $33.69_{-1.32}^{+1.36}$\deg. Our solutions determines their masses as 5.79$_{-0.48}^{+0.47}$\mj and $13.49_{-1.13}^{+1.05}$1\mj, respectively. With a projected separation of 761\,au from GJ 676 A \citep{elbadry21}, GJ 676 B may play a role to warp the planetary disk such that the orbits of GJ 676 A d and e are slightly misaligned. 

\item HD 100939 (HIP 56640) is a K-type giant hosting a companion with a minimum mass of $3.67\pm 0.14$\mj \citep{jones21}. Our combined analyses estimate a dynamical mass of $5.30_{-0.00}^{+3.309}$\mj, putting it into the super-Jupiter category. 

\item HD 106515 A (GJ 9398) is solar type star hosting a planet with a minimum mass of $9.33\pm 0.16$\mj \citep{mayor11,desidera12}. Our combined analyses give a dynamical mass of $9.52_{-0.13}^{+6.39}$\mj. 

\item HD 10697 (109 Psc) is a G star hosting a companion with a minimum mass of 6.38$\pm$0.08\mj \citep{vogt00,butler06,wittenmyer09,luhn19}. Our analyses estimate a mass of $5.74_{-0.29}^{+1.01}$\mj. 

\item HD 111031 (GJ 3746) is a G-type star hosting a planet candidate with a minimum mass of 3.7\mj \citep{hinkel19}. Based on combined analyses of the RV and astrometry data, we estimate a mass of $54.17_{-6.15}^{+5.32}$\mj. 

\item HD 111232 (HIP 62534) is a G star hosting a companion with a minimum mass of 6.80\mj \citep{mayor04,stassun17}. Our analyses estimate a mass of $7.97_{-0.48}^{+1.13}$\mj. We also find a BD companion with a mass of $18.06_{-1.61}^{+4.21}$\mj on an 17\,au-wide orbit. The HD 111232 system is one of few systems hosting multiple sub-stellar companions in our sample. 

\item HD 113337 hosts a disc with an inclination of about 13$_{-9}^{+10}$\,deg \citep{xuan19}. \cite{xuan19} estimate an inclination of 31$_{-4}^{+5}$\,deg based on analyses of Gaia DR2 and Hipparcos data. However, our solution determines an inclination of $57.48_{-3.65}^{+4.55}$\,deg. Such discrepancy is probably due to our use of the EDR3 catalog. 

\item HD 11343 hosts a super-Jupiter \citep{jones16} and a wide-orbit M-dwarf companion \citep{elbadry21}. We find a new Jupiter analog, HD 11343 c, with a minimum mass of about 0.5\mj. The astrometric signal induced by this HD 11343 c is weak and thus the astrometric data does not constrain its inclination well. The mass of HD 11343 b is $7.71_{-1.19}^{+0.73}$\mj. 

\item HD 11506 is a G0V-type star hosting a warm Saturn and a CJ \citep{fischer07}. Thanks to the long baseline between Hipparcos and Gaia, we detect an extremely cold super-Jupiter, HD 11506 d, with a period of $40.31_{-7.5}^{+7.7}$\,yr and a mass of $7.38_{-1.09}^{+2.02}$\mj. This makes the system unique in terms of a solar analog hosting a warm Saturn and two CJs.

\item HD 120084 is a G star hosting a companion with a minimum mass of 4.5\mj \citep{sato13}. Our combined analyses estimate a mass of $5.76_{-0.31}^{+4.62}$\mj. 

\item HD 122562 is a G star hosting a companion with a minimum mass of 24$\pm$2\mj \citep{wilson16}. Our solution gives a dynamical mass of 38$\pm$4\mj. In addition to this companion, we find a warm super-Neptune, HD 122562 c, with an orbital period of 84\,d. On the basis that there seem to be other planets in the system we remain consistent with previous naming of the more massive companion and label it HD 122562 b. Its mass is $37.20_{-2.60}^{+3.67}$\mj. 

\item HD 125390 hosts a companion with a minimum mass of 22.16$\pm$0.96\mj \citep{luhn19}. Our combined analyses estimate a mass of $27.20_{-0.31}^{+4.13}$\mj. 

\item HD 125612 is a G star hosting three companions with minimum masses of 3.0, 0.058, and 7.2\mj and with orbital periods of 502, 4.1547, and 3008\,d, respectively \citep{fischer07,locurto10,ment18}. Through combined analyses of RV and astrometric data, we are able to constrain the mass of the Jupiter-like planet, HD 125612 d, to be $7.18_{-0.45}^{+0.93}$\mj. 

\item HD 126614 is a binary consisting of a G-type primary star and a M dwarf companion with the G star hosting a planet with a minimum mass of 0.38$\pm$0.04\mj \citep{howard10,stassun17}. With combined analyses, we estimate a mass of $0.34_{-0.02}^{+0.20}$\mj for the planetary companion and $81.13_{-7.92}^{+7.78}$\mj for the M-dwarf companion. 

\item HD 127506 (GJ 554) is a K-type star hosting a brown dwarf HD 127506 B \citep{halbwachs00,zucker01}. The BD is further constrained by \cite{sozzetti10} and \cite{reffert11} using Hipparcos intermediate astrometric data. A mass of 45$\pm$21\mj is estimated by \cite{sozzetti10} while a minimum mass of 36\mj is given by \cite{halbwachs00}. We estimate a mass of $47.41_{-3.53}^{+3.70}$\mj, much more precise than previous estimations. This system also hosts a Neptune-sized planet with a minimum mass of about 27\me. Its inclination is consistent with the inclination of HD 127506 B, supporting a coplanar architecture.

\item HD 129191 (BD-04 37333 or HIP 71803) is a star hosting a companion candidate with a minimum mass of 6.8\mj \citep{hinkel19}. The use of Gaia-Hipparcos data allow us to break the degeneracy between mass and inclination and confirm this candidate as a BD with a mass of $76.06_{-14.60}^{+16.22}$\mj. 

\item HD 136118 (HIP 74948) is an F-type star hosting a BD with a mass of 42$_{-18}^{+11}$\,mj \citep{fischer02,wittenmyer09}. Our solution estimates a mass of $13.10_{-1.27}^{+1.35}$\mj. 

\item HD 13724 (HIP 10278) is a G-type star hosting HD 13724 b or (HD
  13724 B), a brown dwarf with a mass of 50.5$_{-3.5}^{+3.3}$\mj
  \citep{rickman20}. We estimate a mass of $36.32_{-1.60}^{+1.48}$\mj after
  considering the 10\% uncertainty of the mass of the primary
  star. Without using direct imaging data, we constrain the BD mass to
  a precision similar to that given by \citep{rickman20} (labeled ``ER20'') based on combined RV and direct imaging data analysis. We also find a warm Saturn with a minimum mass of about 0.2\mj or 72\me. Without a well constrained inclination for HD 13724 b, we cannot determine the mutual inclination between HD 13724 b and B. 

\item HD 139357 is a K-type star hosting a companion with a minimum mass of 9.76$\pm$2.15\mj \citep{dollinger09}. With combined analyses, we confirm this companion to be a BD with a mass of $16.38_{-0.00}^{+7.88}$\mj. 

\item HD 14067 is a G9III star hosting a sub-stellar companion with a minimum mass of 9.0\mj \citep{wang14}. Our combined analyses constrain the mass of the companion to be $9.49_{-0.00}^{+13.30}$\mj. 

\item HD 142 is a wide binary consist of a F-type star and an M dwarf
  \citep{tokovinin16}. The primary hosts at least two planets
  \citep{tinney02,wittenmyer12}. Our analyses confirm HD 142 A d, a warm Saturn
  identified by \cite{wittenmyer20}. The astrometric data put a strong
  constraint on the inclination of HD 142 A c, leading to a mass
  estimation of $10.90_{-0.94}^{+1.28}$\mj. 

\item HD 145675 (14 Her) hosts two companions \citep{butler03,wittenmyer07,gagliuffi21}. Our combined analyses estimate $m_b=8.05_{-0.88}^{+1.63}$\mj, $m_c=5.03_{-1.07}^{+0.87}$\mj, $I_b=144.65_{-3.24}^{+6.28}$\,deg and $I_c=120.10_{-29.05}^{+6.26}$\,deg for HD 145675 b and c, consistent with an aligned orbital configuration. However, \cite{gagliuffi21} give a solution of $I_b=32.7_{-3.2}^{+5.3}$\mj and $I_c=101_{-33}^{+31}$\,deg, indicating a strong misalignment. Such a discrepancy may be due to the parameter degeneracy in their constraint of four orbital parameters ($\Omega_b$, $\Omega_c$, $I_b$, and $I_c$) by using only the proper motion difference between Gaia and Hipparcos (equivalent to 2 data points). 

\item HD 145934 hosts a Jupiter analog with a minimum mass of 2.28$\pm$0.6\mj \citep{feng15}. We confirm the outer companion identified by \cite{feng15} as a (sub-)stellar companion, HD 145934 B, which is on a wide orbit and has a mass of $87.87_{-13.93}^{+70.23}$\mj. Though well separated from the primary star, this companion cannot be resolved by Gaia due to the small parallax of the system (about 4.36\,mas). 

\item HD 167665 (HIP 89620) is an F-type star hosting a BD \citep{patel07} with a minimum mass of 50.6$\pm$1.7\mj \citep{wilson16}. Based on our combined RV and astrometry analysis, HD 167665 B has a mass of $52.71_{-4.40}^{+5.11}$\mj after considering stellar mass uncertainty. Its inclination is $95.20_{-9.86}^{+7.47}$\,deg, suggesting a non-zero transit probability.

\item HD 170707 (HIP 90988) is a K giant hosting a planet with a minimum mass of  1.96$\pm$0.11\mj \citep{jones21}. Our combined analyses estimate a minimum mass of $2.10_{-0.09}^{+0.08}$\mj for this companion. We also find a stellar companion with a mass of $101.79_{-6.11}^{+5.91}$\mj. 

\item HD 181234 (HIP 95015) is a G-type star hosting a companion with a minimum mass of 8.37$_{-0.36}^{+0.34}$\mj \citep{rickman19}. Based on combined RV and astrometry analysis, we constrain the mass of this companion to be $8.24_{-0.00}^{+2.15}$\mj and the inclination to be $98.73_{-41.61}^{+24.47}$\,deg. 

\item HD 182488 hosts a BD companion named ``HD 182488 B'' or ``GJ 758 B''
  \cite{thalmann09} and imaged by \cite{vigan16}. It is further
  studied by multiple groups (e.g., \citealt{bowler18} and \citealt{brandt21}) through combined analyses of astrometry,
  imaging data (labeled ``BB18'') and RV data. In particular, \cite{brandt21} estimate a mass of $38.0\pm 0.8$\mj and an eccentricity of 0.24$\pm$0.11. By analyzing the same data set as that used by \cite{brandt21} as well as the Gaia-Hipparcos positional difference, we constrain the mass to be $36.39_{-1.08}^{+1.21}$\mj and the eccentricity to be $0.37_{-0.08}^{+0.08}$, consistent with the solution given by \cite{brandt21}. With astrometry for both components of the binary, we also estimate a mass of 0.93$\pm$0.03\msun for the primary {\it a posteriori}. 

\item HD 185414 (HIP 96395; WDS 19359+5659) is a spectroscopic binary hosting a wide-orbit companion detected by 2MASS and Gaia \citep{mason01,fuhrmann04,tokovinin14b,roberts15,elbadry21}. The spectroscopic binary is unresolved. Our solution estimates a mass of $117.59_{-9.55}^{+9.03}$\mj for this unresolved companion. 

\item HD 190228 is a G star hosting a Jupiter-like companion, HD 190228 b \citep{perrier03,wittenmyer09}. The inclination of HD 190228 b is 4.5$\pm$2.1\,deg based on the analyses of the Hipparcos intermediate astrometric data \citep{reffert11}. Our solution does not put a strong constraint on the inclination because the orbital period is comparable with the Hipparcos or Gaia EDR3 baselines. A detailed analyses of intermediate data in a future Gaia data release will strongly constrain the mass.

\item HD 191806 is a K star hosting a companion with a minimum mass of 8.52$\pm$0.63\mj \citep{diaz16a}. Our combined analyses constrain the mass to be $9.33_{-0.85}^{+0.92}$\mj. 

\item HD 203473 B was discovered by \cite{ment18} based on analysis of KECK RV data. With both KECK and HARPS data as well as the  Hipparcos-Gaia astrometry, we estimate a mass of about 96\mj, much higher than the minimum mass of 7.84$\pm$1.15\mj estimated by \cite{ment18}. Our solution estimates an orbital period of $8.10_{-0.02}^{+0.01}$\,yr, much longer than the period of 4.25\,yr estimated by \cite{ment18}, who may have identified a half-period harmonic of the true signal due to limited RV baseline. 

\item HD 204313 is claimed to host three planets based on the RV data collected by the 2.7 m Harlan J. Smith Telescope and CORALIE spectrometer \citep{segransan10,robertson12}. However, \cite{diaz16} do not confirm HD 204313 d based on their analyses of previous and new HARPS data. Our solution agrees with \cite{diaz16} that there is no significant signal around a period of 2800\,d as claimed by \cite{robertson12}. However, we confirm the trend identified by \cite{diaz16} as a new long period signal with a period of $20.06_{-1.01}^{+1.10}$\,yr based on combined RV and astrometry analysis. This new companion is dubbed ``HD 204313 e'' to be distinguished from the dubious candidate HD 204313 d. It corresponds to a BD with a mass of $15.32_{-5.18}^{+4.89}$\mj. 

\item HD 211847 is a G star hosting a sub-stellar companion with a minimum mass of 19.2$\pm$1.2\mj \citep{sahlmann11, moutou17}. Our analyses constrain the mass to be $55.32_{-18.49}^{+1.34}$\mj. 

\item HD 214823 (HIP 111928) is a G-type star hosting a companion \citep{diaz16,ment18} with a minimum mass of 20.56$\pm$0.32\mj and an orbital period of 1853.9$\pm$1.6\,d \citep{luhn19}. With Gaia and Hipparcos data, we are able to break the degeneracy between mass and inclination, and to constrain the absolute mass to $18.61_{-1.07}^{+4.14}$\mj and the period to be $5.078_{-0.004}^{+0.004}$\,yr. 

\item HD 217786 (HIP 113834) is a F star hosting a companion with a minimum mass of 13$\pm$0.8\mj \citep{moutou11} and a stellar companion on an extremely wide orbit \citep{ginski16,elbadry21}. Because the RV and astrometric variation of the primary star caused by the wide stellar companion is insignificant, we only model the reflex motion due to the substellar companion and estimate a mass of $13.85_{-1.31}^{+1.27}$\mj. We also find a hot super-Earth with an orbital period of 2.5\,d. 

\item HD 217958 (HIP 113948) is a G star hosting a possible stellar companion \citep{kane19}. Our combined analyses of the RV, Gaia-Hipparcos astrometry and the imaging data given by \citep{kane19} do not show strong evidence for this candidate companion on a wide orbit. However, we discover a companion with a mass of $81.82_{-38.20}^{+34.17}$\mj and a planet with a mass of $0.52_{-0.06}^{+0.35}$\mj on a Jupiter-like orbit. 

\item HD 219077 (HIP 114699) is a G star hosting a companion with a minimum mass of 13.4$\pm$0.78\mj \citep{marmier13,kane19}. Our combined analyses gives a dynamical mass of $9.62_{-0.73}^{+1.00}$\mj.

\item HD 219828 (HIP 115100) is a G star hosting two companions with minimum masses of 15.1 and 0.0661\mj \citep{melo07,ment18}. Our combined analyses estimate a mass of $16.05_{-1.27}^{+3.52}$\mj for the bigger companion on a Jupiter-like orbit. 

\item HD 221420 (GJ 4340) is a G-type star hosting a BD \citep{kane19} with a wide-orbit M-dwarf companion \citep{venner21,elbadry21}. Based on combined RV and the HGCA catalog of calibrated Gaia DR2 and Hipparcos proper motions \citep{brandt18}, \cite{venner21} estimated a mass of $20.35_{-3.43}^{+1.99}$\mj, consistent with 20$\pm$3\mj estimated in this work. 

\item HD 224538 (HIP 118228) is an F star hosting a companion with a minimum mass of 5.97$\pm$0.42\mj \citep{jenkins17}. Our combined analyses estimate a dynamical mass of $6.53_{-0.35}^{+1.84}$\mj. 

\item HD 23596 (HIP 17747) is an F star hosting a sub-stellar companion with a mass of 8.1\mj \citep{perrier03, wittenmyer09, stassun17} and a wide-binary companion \citep{elbadry21}. Our combined analyses show a dynamical mass of $11.91_{-1.77}^{+0.99}$\mj. 

\item HD 25015 (HIP 18527) is a K star hosting a companion with a minimum mass of 4.5$\pm$0.3\mj \citep{rickman19}. Our combined analyses show a mass of $9.08_{-1.81}^{+1.16}$\mj. 

\item HD 26161 (HIP 19428) is a G-type star hosting a companion with a minimum mass of 13.5$_{-3.7}^{+8.5}$\mj \citep{rosenthal21}. Based on the combined RV and astrometry analysis, the dynamical mass of this companion is $28.46_{-0.22}^{+20.05}$\mj and the orbital period is about 39\,years. HD 26161 also hosts an M dwarf companion with a projected separation of 561\,au \citep{elbadry21}. 

\item HD 27894 (HIP 20277) is a K star hosting three companions with masses of 5.42, 0.16, and 0.67\mj \citep{moutou05,angladaescude10,kurster15,trifonov17}. However, we only find strong evidence for the biggest two companions. By constraining the inclination of the biggest companion using astrometry, we find a dynamical mass of $6.49_{-0.35}^{+0.99}$\mj. 

\item HD 28185 (HIP 20723) is a G-type star hosting a companion \citep{santos01,wittenmyer09} with a minimum mass of 6.7\mj and an orbital period of 379$\pm$2\,d \citep{minniti09}. With both RV and astrometry data, we constrain the mass of HD 28185 b to be $7.07_{-0.79}^{+1.29}$\mj. Moreover, we detect a BD companion with a mass of $19.64_{-2.14}^{+2.27}$\mj.

\item HD 28192 (HIP 20752) is a G star hosting a stellar companion on an extremely wide orbit and with a mass of about 0.4\msun \citep{tokovinin14b,elbadry21}. This companion at most induces a reflex motion of 0.1\msyr and is thus not considered in our combined analyses. We find another stellar companion with a mass of $94.78_{-7.52}^{+8.51}$\mj on a 10\,au-wide orbit and a planet with a minimum mass of $0.31_{-0.03}^{+0.02}$\mj and an orbital period of about 14\,d. 

\item HD 29461 (HIP 21654) is a G star hosting a companion with a minimum mass of about 0.08\msun \citep{griffin12,bouchy16}. Our combined analyses estimate a mass of $92.96_{-4.93}^{+12.61}$\mj on a 5\,au-wide orbit. The small separation between this companion and the primary star explains null detection of it by previous imaging surveys.

\item HD 30177 (HIP 21850) is a G star hosting two companions with minimum masses of 3$\pm$0.3 and 8.07$\pm$0.12\mj \citep{tinney02,wittenmyer17,barbato18}. The null detection of these companions in the direct imaging survey conducted by \cite{zurlo18} leads to an upper limit of 28-30\mj and a minimum inclination of 15\deg. Our combined analyses estimate masses of $8.40_{-0.49}^{+1.24}$\mj and $6.15_{-0.34}^{+1.31}$\mj, consistent with the constraints given by previous imaging and RV data analyses. This system is one of few systems hosting two super-Jupiters in our sample. 

\item HD 38529 (HIP 27253) is a G star hosting two companions with masses of 18$\pm 3$ and 0.17\mj \citep{fischer03,wittenmyer09, barbato18,benedict10,xuan20}. Our combined analyses estimate a mass of $10.38_{-0.88}^{+1.03}$\mj and an inclination of $104.56_{-8.72}^{+6.39}$\,deg for HD 38529 c, consistent with the the solution given by \cite{xuan20} but with higher precision. 

\item HD 39060 ($\beta$ Pic) is a well-studied system, hosting two giant planets \citep{lagrange09,lagrange19}. We present the parameters of $\beta$ Pic b and c given by the most recent studies in Table \ref{tab:betaPic}. Thanks to the valuable data collected by previous studies, we analyze the relative RV data for b \citep{snellen14}, the updated relative astrometry data from \cite{lacour21}, the recently released Gaia EDR3, and the RV data used by \cite{lagrange20} and  \cite{vandal20} with different reprocessing to remove stellar activity noise. We model the relative astrometry for multiple companions by (1) calculating the reflex motion of the host due to the innermost companion; (2) calculating the reflex motion of the barycenter of the host and the innermost companion due to the outer companion; (3) repeating the above steps until the reflex motion due to all companions are modeled; (4) calculating the position of an outer companion relative to the host star by converting the barycenter of the host and inner companions to the host position. This procedure is introduced by \cite{lacour21} in detail and is also used by \cite{brandt21b}.

We find that our solution is quite sensitive to the RV data. Although both the RV datasets used are from the same HARPS observations different corrections are made to stellar activity particularly from pulsations. We use the \cite{lagrange20} dataset (dubbed ``AL20'') to find $\beta$ Pic b to be $7.56_{-1.69}^{+1.35}$\mj and that of c to be $8.94_{-0.78}^{+0.75}$\mj while using \cite{vandal20} dataset (dubbed ``TV20'') we find a mass of $11.75_{-2.15}^{+2.34}$\mj and $10.15_{-1.07}^{+1.20}$. In Table \ref{tab:betaPic} we present our solutions along with the range of solutions from the literature and note the considerable scatter and discrepancy from the masses of 3.2\mj and 5.6$\pm$1.5\mj measured by \cite{lagrange20} and \cite{nowak20} respectively for $\beta$ Pic b using the AL20 RVs when they adopt an uninformative prior. For Table \ref{tab:par} we adopt our TV20 solution since it agrees better with the independent astrometric solution for $\beta$ Pic c based on interferometric data by \cite{lacour21} and provides a solution more consistent with the higher mass predicted by various cooling models (e.g. \citealt{baraffe03}, \citealt{spiegel12}). We can anticipate that further evolution in the processing of the RV data for stellar activity as well as the incorporation of Gaia intermediate data into analysis solutions will be useful to resolve the discrepancies that we find in mass measurement from different RV data sets.

\begin{deluxetable}{lllllllll}
\tablecaption{Parameters for the $\beta$ Pic system based on analyses of various data sets in literature and in this work \label{tab:betaPic}}
\tablehead{
\colhead{Reference\tablenotemark{a}}&\colhead{EN20\tablenotemark{b}}&\colhead{AL20\tablenotemark{c}}& \colhead{TV20\tablenotemark{d}} & \colhead{MN20\tablenotemark{e}}&\colhead{TB21} & \colhead{SL21} & \colhead{This paper (AL20)} & \colhead{This paper (TV20)}}
\startdata
         $m_\star$ [$m_\odot$]&1.76$_{-0.02}^{+0.03}$&1.77$\pm$0.03&1.80$_{-0.04}^{+0.03}$ &1.82$\pm$0.03&1.83$\pm$0.04&1.75$_{-0.02}^{+0.03}$&1.76$\pm$0.02&1.80$\pm$0.03\\
         $m_b$ [$m_{\rm Jup}$]&8.03$_{-2.62}^{+2.61}$&11.1$\pm$0.8&$11.7\pm 1.4$&$9.0\pm 1.6$&9.3$_{-2.5}^{+2.6}$&11.90$_{-3.04}^{+2.93}$&$7.56_{-1.69}^{+1.35}$&$11.75_{-2.15}^{+2.34}$\\
         $m_c$ [$m_{\rm Jup}$]&$9.18_{-0.87}^{+0.96}$&7.8$\pm$0.4&$8.5\pm 0.5$&$8.2\pm 0.8$&8.3$\pm$1.0&8.89$_{-0.75}^{+0.75}$&$8.94_{-0.78}^{+0.75}$&$10.15_{-1.07}^{+1.20}$\\
         $I_b$ [deg]&88.82$_{-0.01}^{+0.01}$&89.01$\pm$0.01&$88.88_{-0.03}^{+0.04}$&88.99$\pm$0.01&88.94$\pm$0.02&$88.93_{-0.01}^{+0.00}$&88.93$_{-0.09}^{+0.09}$&$89.01_{-0.01}^{+0.01}$\\
         $I_c$ [deg]&88.85$_{-0.71}^{+0.72}$&89.01$\pm$0.01&--&89.17$\pm$0.50&89.1$\pm$0.66&$88.95_{-0.10}^{+0.09}$&89.00$_{-0.01}^{+0.01}$&$88.95_{-0.09}^{+0.08}$\\\hline
         AL20 RV&&\checkmark&&\checkmark&&&\checkmark&\\ 
        TV20 RV&&&\checkmark&&\checkmark&\checkmark&&\checkmark\\ 
         b RV &&&&&\checkmark&&\checkmark&\checkmark\\
         b astrometry&\checkmark&\checkmark&\checkmark&\checkmark&\checkmark&\checkmark&\checkmark&\checkmark\\ 
         c astrometry &&\checkmark&&\checkmark&\checkmark&\checkmark&\checkmark&\checkmark\\ 
         HDR2\tablenotemark{f}&\checkmark&&&&\checkmark&&&\\ 
         HEDR3\tablenotemark{g}&&&&&&&\checkmark&\checkmark\\ 
\enddata
\tablenotetext{a}{EN20: \cite{nielsen20}; AL20: \cite{lagrange20}; TV20: \cite{vandal20}; MN20: \cite{nowak20}; TB21: \cite{brandt21b}; SL21: \cite{lacour21}}
\tablenotetext{b}{The solution shown in this table is the so-called "coplanar fit" by EN20 who adopt a Gaussian prior centered on zero and with a standard deviation of 1\deg. Without such assumption, the inclination of c has an error of 13\deg, leading to significant mass uncertainty for c. }
\tablenotetext{c}{In AL20, the inclination of $\beta$ Pic c is assumed to be equal to $\beta$ Pic b because the relative astrometry of $\beta$ Pic c is not used to constrain it. The mass of b reported here is estimated by AL20 using a Gaussian prior of $14\pm 1$\mj. Without such informative prior, the mass is 3.2\mj.}
\tablenotetext{d}{In TV20, the stellar mass is given {\it a priori} by \cite{wang16}.}
\tablenotetext{e}{The mass of b reported here is estimated by MN20 using a Gaussian prior of $15\pm 3$\mj. Without such informative prior, the mass is 5.6$\pm$1.5\mj.}
\tablenotetext{f}{HDR2 represents the data of proper motion difference beween Hipparcos and Gaia DR2.}
\tablenotetext{g}{HEDR3 represents the data of both proper motion and positional difference between Hipparcos and Gaia EDR3.}
\end{deluxetable}

\item HD 39091 ($\pi$ Men) is a G star hosting at least two companions with masses of 0.015 and 13\mj \citep{jones02,huang18,damasso20} and a possible third companion found recently by \cite{hatzes22}. The smaller companion is a transiting planet while the bigger companion is found to be significantly misaligned with the inner one \citep{damasso20,derosa20,xuan20,kunovac21}. With combined analyses of the RV data from AAT, CORALIE, ESPRESSO, HARPS, and PFS as well as the Gaia-Hipparcos data, we are able to constrain the mass of this companion to be $12.33_{-1.38}^{+1.19}$\mj and the inclination to be $54.44_{-3.72}^{+5.94}$\,deg. This inclination of $\pi$ Men b differs from the $~$90\,deg inclination of the $\pi$ Men c by 6$\sigma$, suggesting significant misalignment as proposed by previous studies. However, we fail to confirm the third companion found by \cite{hatzes22}.

\item HD 39213 (HIP 27491) is a K star hosting a BD with a minimum mass of 0.07$\pm$0.01\msun \citep{jenkins15}. Our combined analyses determine a dynamical mass of $70.77_{-5.43}^{+10.21}$\mj, putting it around the boundary between BD and stellar object. Follow-up direct imaging of this object is needed to further characterize the companion.

\item HD 4113 A (HIP 3391) is a Sun-like star hosting a BD (HD 4113 A
  b or HD 4113 b) and a sub-stellar companion (HD 4113 C)
  \citep{tamuz08,cheetham18} as well as HD 4113 B, an M dwarf on an
  extremely wide orbit \citep{mugrauer14}. HD 4113 A b has a minimum
  mass of 1.602$_{-0.075}^{+0.076}$\mj based on RV analysis
  \citep{cheetham18}. HD 4113 C has a dynamical mass of
  65.8$_{-4.4}^{+5.0}$\mj based on analyses of direct imaging data
  (labeled by ``AC18'') and an isochronal mass of 36$\pm$5\mj based on cooling models \citep{cheetham18}. Our combined analysis of the RV, astrometry, and imaging data constrains the mass of HD 4113 C to be $51.91_{-0.46}^{+0.60}$\mj, relaxing the previous tension between dynamical and isochronal masses without invoking binarity in HD 4113 C.

\item HD 42581 (GJ 229 A) hosts the first imaged BD, GJ 229 B
  \citep{nakajima95} as well as two planets \citep{feng20a}. Recently,
  \cite{brandt21} estimated a mass of $71.4\pm 0.6$\mj for GJ 229 B
  based on combined analyses of RV, imaging data (labeled ``MB21''), Gaia EDR3 and Hipparcos data. This mass is in tension with the mass predicted by cooling models. However, with nearly the same data but with additional constraint from Hipparcos-Gaia positional difference, our combined analyses estimate a mass of $60.42_{-2.38}^{+2.34}$\mj, consistent with the $64.8\pm 0.1$\mj predicted by cooling models \citep{brandt21}. This suggests that the use of positional difference between Hipparcos and Gaia might be important to avoid potential bias by using proper motion difference alone.

\item HD 43197 is a Sun-like star hosting a warm Jupiter \citep{naef10}. We detect a cold super-Jupiter, HD 43197 c, with a mass of $7.9\pm 1.7$\mj on a wide orbit with a period of 27$\pm$9\,yr and a nearly face-on inclination ($11.42_{-3.07}^{+5.39}$\,deg). However, the inclination of the inner companion is not well constrained. Assuming a coplanar configuration, HD 43197 b would have a mass of about 4\mj. 

\item HD 65430 (HIP 39064) is a spectroscopic binary with an orbital period of 3138\,d \citep{allen12}. Our combined analyses estimate a mass of $105.40_{-8.95}^{+8.37}$\mj, confirming its stellar origin.

\item HD 66428 b and c were discovered by \cite{butler06} and \cite{rosenthal21}, respectively. In addition to the Keck data used by \cite{rosenthal21}, we analyze the HARPS data reduced by \cite{trifonov20} and estimate a mass of 10.6$\pm$2.8\mj for HD 66428 b and 3.1$\pm$1.7\mj for HD 66428 c. Compared with an orbital period of $107_{-49}^{+153}$\,yr and a minimum mass of $0.085_{-0.053}^{+0.069}$\mj estimated for HD 66428 c by \cite{rosenthal21}, our estimation of a period of $28.69_{-5.35}^{+9.21}$\,yr and $1.76_{-0.04}^{+3.40}$\mj is much more precise. However, further constraint on the inclination of HD 66428 c is needed to determine whether the orbits of the two companions are misaligned. 

\item HD 72659 (HIP 42030) is a G-type star hosting a companion \citep{butler03,wittenmyer09} with a minimum mass of 3.15$\pm$0.14\mj \citep{moutou11}. Based on the combined analysis of RV and astrometry, HD 72659 b is found to have a mass of $2.99_{-0.10}^{+2.59}$\mj. We also find a new companion HD 72659 c with a mass of $18.81_{-4.80}^{+4.44}$\mj. Assuming a coplanar configuration between b and c, HD 72659 b would have a mass of about 15\mj. 

\item HD 72892 is a Sun-like star hosting a super-Jupiter
  \citep{jenkins17}. In addition to this companion, we also detect
  another companion HD 72892 B with a mass of $77.12_{-35.48}^{+41.76}$\mj on an edge-on and eccentric orbit. The high eccentricity of HD 72892 b ($e=0.419\pm 0.003$)
  might be caused by the strong perturbations from HD 72892 B, which is on an orbit with an eccentricity of 0.38$\pm$0.06.

\item HD 73267 is a solar type star hosting a Jupiter-like planet, HD 73267 b \citep{moutou09}. Through combined RV and astrometry data analyses, we identify an additional companion named ``HD 73267 c''. It is a super-Jupiter with a mass of $5.13_{-0.28}^{+0.91}$\mj and with an orbital period of $46.74_{-2.98}^{+2.15}$\,yr. The orbits of the two companions are probably misaligned though with large uncertainty. 

\item HD 74014 (HIP 42634) is a star hosting a BD companion \citep{patel07} with a minimum mass of 49.0$\pm$1.7\mj \citep{sahlmann11}. Our combined RV and astrometry analyses constrain the mass to $61.54_{-5.77}^{+5.61}$\mj. 

\item HD 74156 (HIP 42723) is a G star hosting two sub-stellar companions with masses of 1.78 and 8.00\mj \citep{naef04,wittenmyer09}. Our combined analyses lead to estimated masses of $1.71_{-0.06}^{+0.96}$ and $8.67_{-0.47}^{+1.39}$\mj. 
\item HD 7449 A b and HD 7449 A c (or HD 7449 B) are detected by
  \cite{dumusque11} and \cite{rodigas16}, respectively. Through
  combined analyses of RV, Gaia-Hipparcos astrometry, and the relative
  astrometry derived from the imaging data collected by
  \cite{rodigas16} (labeled ``TR16''), we estimate inclinations of $171.63_{-3.74}^{+2.61}$\,deg and $68.40_{-3.89}^{+4.10}$\,deg for Ab and B, indicating significant misalignment between the two companions. The mass of HD 7449 A b is $8.17_{-2.70}^{+3.06}$\mj. The mass of HD 7449 B is $178.15_{-13.66}^{+16.61}$\mj, consistent with the value of 0.23$_{-0.05}^{+0.22}$\msun estimated by \cite{rodigas16} based on photometry. 

\item HD 80869 (HIP 46022) is a G star hosting a planetary companion with a minimum mass of 4.86$_{-0.29}^{+0.65}$\mj \citep{demangeon21}. The use of Gaia-Hipparcos astrometry allows us to break the inclination-mass degeneracy and estimate a mass of $5.07_{-0.56}^{+2.54}$\mj. 

\item HD 81040 (HIP 46076) is a G star hosting a Jupiter-like companion with a mass of $7.24^{+1.0}_{-0.37}$\mj \citep{stassun17,sozzetti06,li21}. Our combined analyses give a mass of $6.77_{-0.87}^{+1.10}$\mj. Though we use the same RV data sets as \cite{li21}, we model the correlated RV noise using the MA(1) model. This makes our estimation of the dynamical mass more uncertain but more conservative than the values given by \cite{li21}.

\item HD 81817 is a K-type star hosting a substellar companion (HD 81817 b) with a minimum mass of 27.1\mj \citep{bang20}. With both RV and astrometry data, we are able to constrain its mass to $24.13_{-0.71}^{+9.83}$\mj. We also find another BD in this system (HD 81817 c) although it was diagnosed as an activity signal by \cite{bang20} due to a dubious overlap with powers in the periodograms of H$\alpha$. However, we confirm HD 81817 c as a BD because this signal shows a unique power (see Fig. \ref{fig:HD81817}) in the BFP and is strictly periodic and quite circular based on MCMC posterior samplings. The evidence strongly support a Keplerian origin instead of an activity origin though its inclination is not well constrained due to its short orbital period.
\begin{figure}
    \centering
    \hspace{-0.1in}
    \includegraphics[scale=0.67]{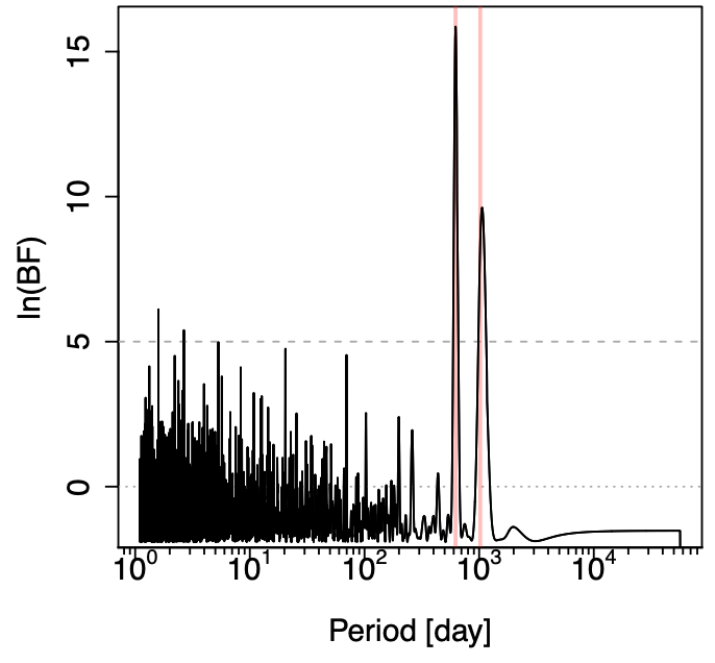}
    \includegraphics[scale=0.67]{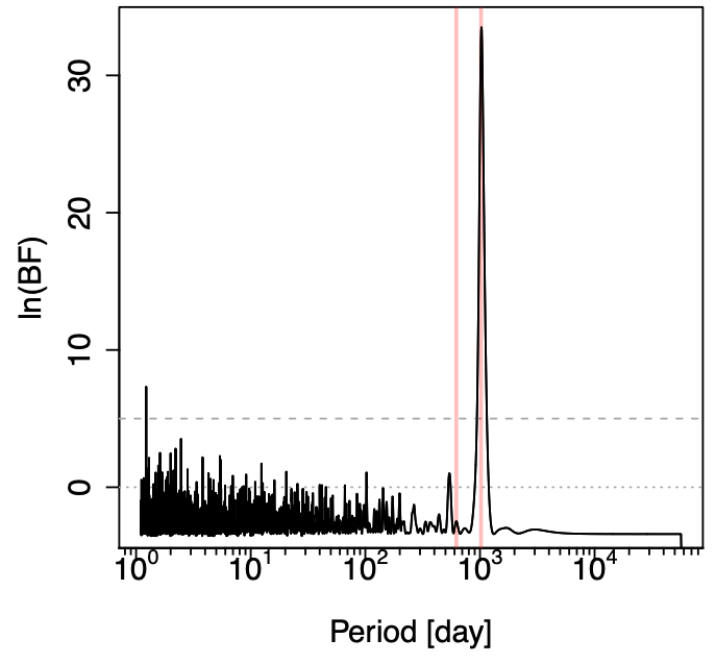}
    \includegraphics[scale=0.55]{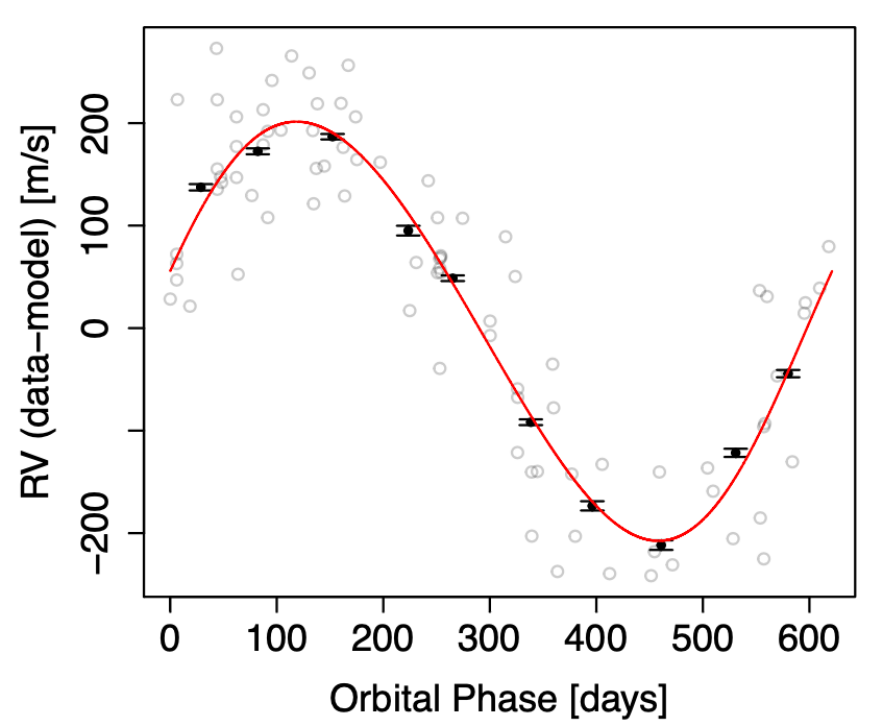}
    \includegraphics[scale=0.55]{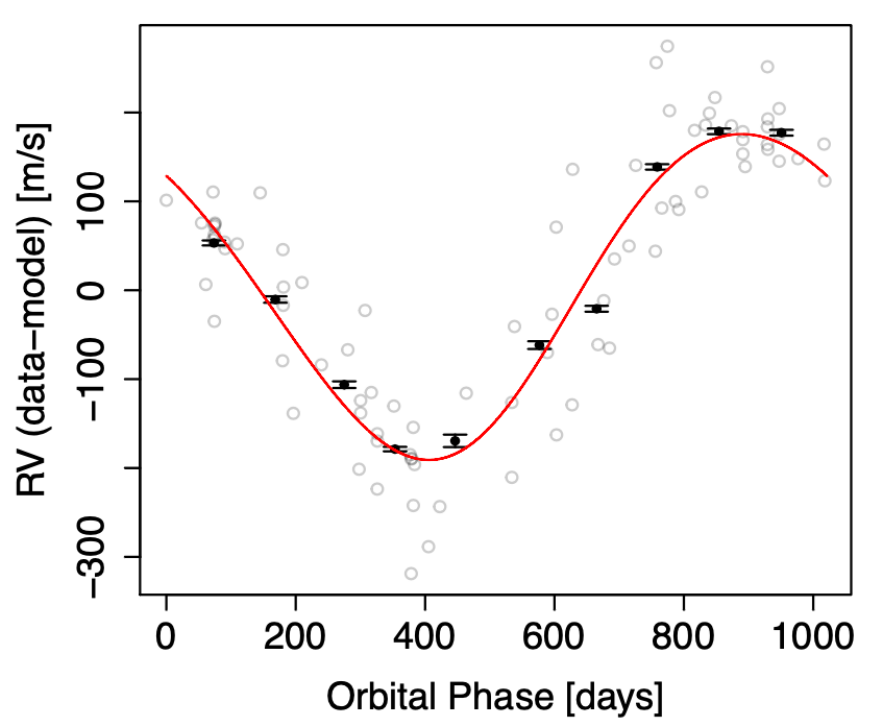}    
    \caption{BFPs and phase curves for the 600 and 1000d signals
      detected in the RV data for HD 81817. Upper panels: the left
      panel shows the BFP for the raw data while the right panel shows
      the BFP for the residual with the 600-day signal subtracted. The
      red lines in the upper panels indicate the periods of the two
      signals. The dashed and dotted lines show the thresholds of lnBF$=$5 and 0, respectively. Lower panels: the red curves in the lower panels show the best fit corresponding to the parameter values at the maximum a posteriori. The black error bars represent 10 binned data points.}
    \label{fig:HD81817}
\end{figure}

\item HD 86264 (HIP 48680) is a K star hosting a companion with a minimum mass of $7\pm 1.6$\mj \citep{fischer09}. According to the solution based on our combined analyses, the dynamical mass of the companion is $9.81_{-1.95}^{+11.71}$\mj. 

\item HD 8673 (HIP 6702) is a double star system including a F star and an early M-dwarf with a mass of 0.33–0.45\msun \citep{roberts15}. The primary F star hosts a sub-stellar companion with a minimum mass of 14.2$\pm$1.6\mj. Our combined analyses constrain the mass of the sub-stellar companion to be $13.25_{-1.42}^{+1.70}$\mj. 

\item HD 87883 (HIP 49699) is a K star hosting a companion with a minimum mass of $6.31_{-0.32}^{+0.31}$\mj \citep{fischer09,stassun17,li21}. Our solution estimates a mass of 5.3$\pm$0.7\mj. Compared with the bi-modal posterior distribution of inclination given by \cite{li21} based on RV and Gaia-Hipparcos proper motion difference, we use both proper motion and position differences between Gaia and Hipparcos so that we are able to break the degeneracy between $I$ and $I+\pi$, and to constrain the inclination to be $25.45_{-1.05}^{+1.61}$\,deg.

\item HD 95544 (HIP 54203) is a G star hosting a companion with a minimum mass of 6.84$\pm$0.31 \citep{demangeon21}. According to our combined analyses of RV and astrometric data, the dynamical mass is $6.02_{-0.26}^{+1.62}$\mj and the inclination is $86.50_{-24.60}^{+31.19}$\,deg, indicating an edge-on configuration.

\item HD 984 (or HIP 1134) is an F star hosting a brown dwarf with a
  mass of $61.0\pm 4.0$\mj \citep{johnson17,franson22}. Through
  combined analyses of the RV, Gaia-Hipparcos, and the imaging data
  collected by \cite{franson22}  and is labeled ``KF22'', we find a
  dynamical mass of $40.37_{-18.27}^{+24.33}$\mj.
  
\item HD 98649 (HIP 55409) is a G type star hosting a companion with a minimum mass of 6.79$_{-0.3}^{+0.5}$\mj \citep{marmier13,rickman19}. By combined analyses of RV and Gaia-Hipparcos astrometry, \cite{li21} estimate a mass of $9.7_{-1.9}^{+2.3}$\mj. Using both proper motion and position differences between Gaia and Hipparcos, we estimate a mass of $6.76_{-0.00}^{+3.61}$\mj, consistent with and more precise than the mass given by \cite{li21}. 

\item HIP 22203 (HD 30246) is a G star hosting a brown dwarf with a minimum mass of $55.1^{+20.3}_{-8.2}$\mj \citep{diaz12}. Our combined analyses constrain the mass to be 51$\pm$5\mj and the inclination to be $50.76_{-4.54}^{+4.73}$\,deg. 

\item HIP 65891 (HD 117253) is a red giant branch K star hosting a planet with a minimum mass of about 6\mj \citep{jones15}. Using both RV and Gaia astrometric excess noise, \cite{kiefer21} estimate a mass ranging from 168.4 to 713.7\mj for HD 117253 b. However, based on Gaia-Hipparcos data, we find a mass of $5.89_{-0.28}^{+0.76}$\mj and an inclination of $88.97_{-17.33}^{+16.21}$\,deg. Although the inclination is not well constrained, the companion orbit is unlikely face-on as \cite{kiefer21} conclude. Considering that the astrometric signal of this companion is insignificant, we cannot be sure about whether this companion is sub-stellar or stellar. 

\item HIP 67537 (HD 120457) is a red giant branch star hosting a planet with a minimum masses of $11.1^{+0.4}_{-1.1}$\mj \citep{jones17}. Our combined analyses constrain the companion's mass to be $10.88_{-0.00}^{+7.78}$\mj. 

\item HIP 67851 (HD 121056) is a K-type giant star hosting two companions with minimum masses of 5.98$\pm$0.76\mj and 1.38$\pm$0.15\mj \citep{jones14, jones15,wittenmyer15}. Our combined analyses constrain the mass of the bigger companion to be $6.94_{-0.52}^{+2.06}$\mj. 

\item HIP 78395 (WDS 16003-0148) is a K star hosting two companions \citep{mason01}. By combined analyses of the RV, Gaia-Hipparcos astrometry, and the relative astrometric data provided by \cite{mason01}, we constrain the mass to be $68.09_{-8.06}^{+8.65}$\mj, putting the companion around the boundary between sub-stellar and stellar categories.

\item HIP 97233 (HD 186641) is a K star hosting a companion with a minimum mass of 20$\pm$0.4\mj \citep{jones15}. Based on our analyses, the mass of this companion is $19.19_{-0.32}^{+3.67}$\mj and is on a nearly edge-on orbit. 
\end{itemize}

\section{Statistics of the companion sample}\label{sec:stats}
\subsection{Mass distribution and occurrence rate}
Through our analysis of the 5108 stars with each star having more than 5 high precision RV data point, we find 869 stars with 914 long period
signals ($>$1000\,d). Of these 167 of them are confirmed as companions
with masses from 5 to 120\mj by our combined analyses of RV and
astrometry. The relative mass uncertainty of this sample is less than
100\%. The masses of 113 companions are constrained to a precision of
better than 20\%. Without correcting for detection bias, the occurrence rate of the wide-orbit BDs is about 1.3\%, consistent with previous estimation (e.g. \citealt{grieves17, kiefer19}). 

\begin{figure}
  \centering
   \includegraphics[scale=0.5]{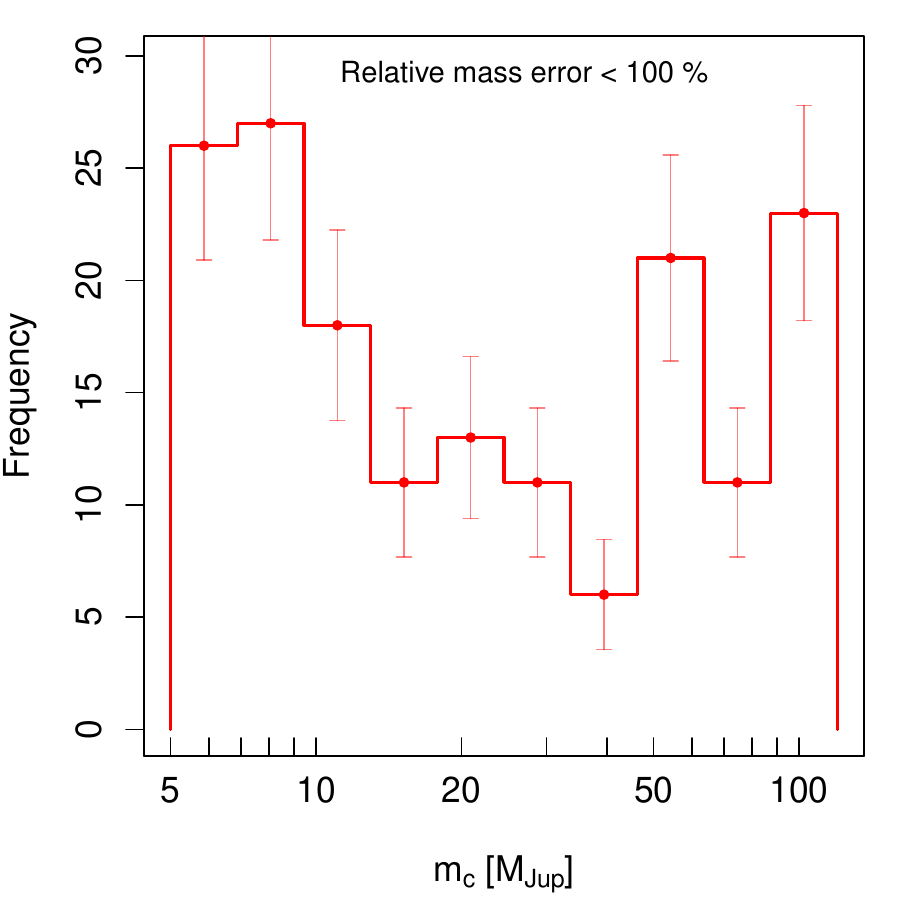}
    \includegraphics[scale=0.5]{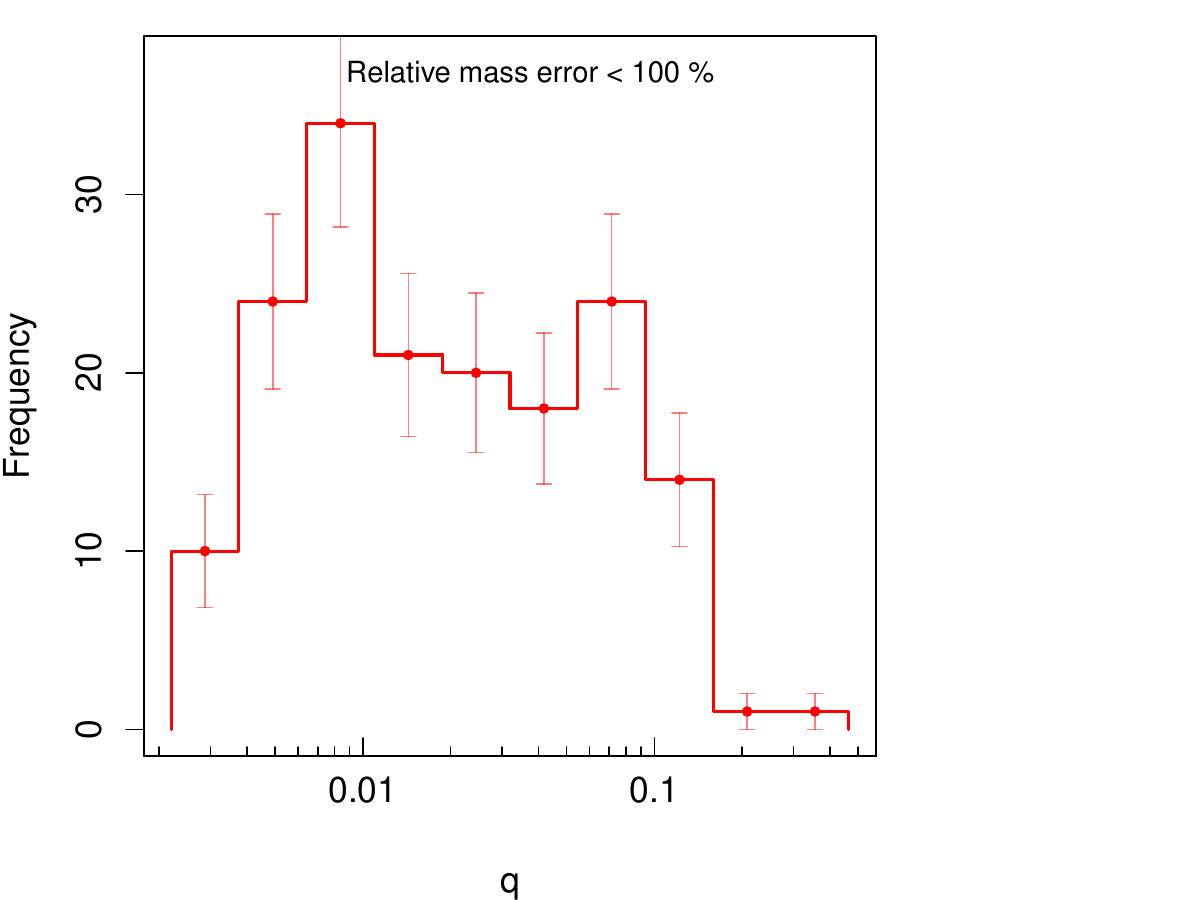}
    \includegraphics[scale=0.5]{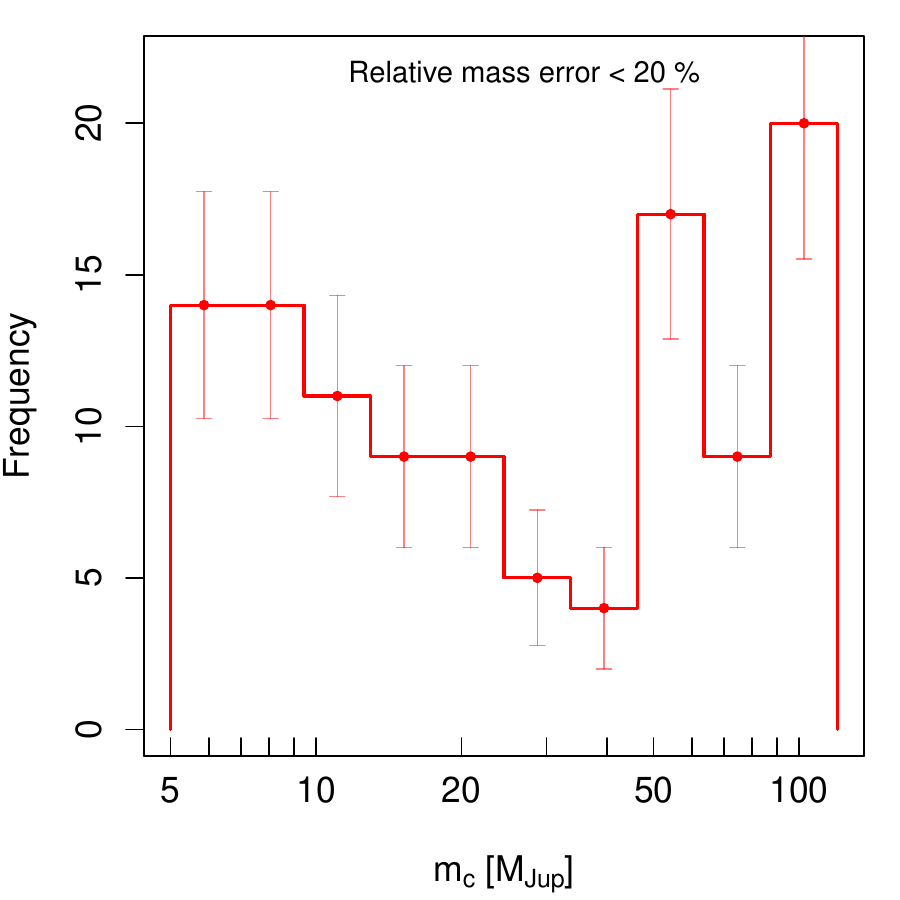}
    \includegraphics[scale=0.5]{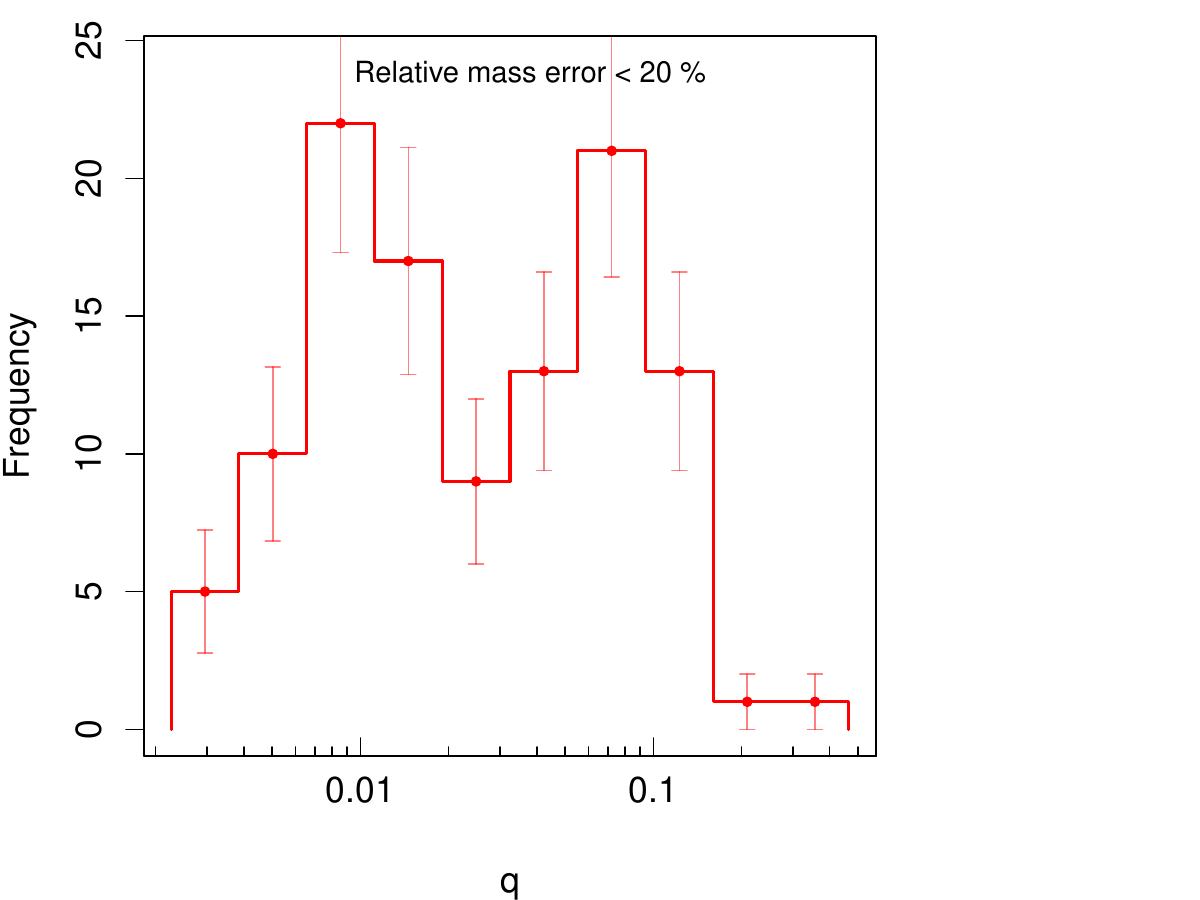}
    \caption{Mass distributions of the sub-stellar companions detected in this work. The samples are divided into 10 bins. The upper and lower panels respectively show the distributions of the companions with relative mass uncertainty of less than 100\% and 20\% . The left and right panels respectively show the distributions over companion mass and companion-host mass ratio. The error bars are determined by assuming that the number of detected companions follows a Poisson distribution. }
\label{fig:mass}
\end{figure}

We define the sample with relative mass uncertainty less than 100\% as the ``optimistic sample'' and the sample with relative mass error less than 20\% as the ``conservative sample''. We show the distribution of the sample over mass and mass ratio in Fig. \ref{fig:mass}. There are at least three features seen in the mass distribution of the optimistic sample: (1) there is a lack of BDs with a mass around 40\mj, consistent with the so-called low-mass and high-mass BD boundary identified by \cite{ma14}; (2) there is also a 2-$\sigma$ valley around the 75\mj boundary between stars and BDs; (3) a sharp decrease of companions around the 13\mj planet-BD boundary is followed by a shallow decrease from 13\mj to 40\mj. While the first two features remain in the conservative sample, the third feature becomes insignificant in the conservative sample. In the distribution of mass ratio (right panels of Fig. \ref{fig:mass}), we see a valley around 0.3-0.4 but fail to find any significant features around the star-BD boundary (0.07 if assuming the host mass to be unit solar mass). Because the detection bias is only significant for cold super-Jupiters, some of the features seen above are not likely to disappear after considering detection bias. In particular, the valley around 40\mj is robust to the choice of sample size and the normalization of companion mass. By investigating the distribution over mass (or mass ratio) and semi-major axis (Fig. \ref{fig:ame}), we observe that the 40\mj valley gradually disappear beyond 10\,au. Nevertheless, a bias-corrected distribution of sub-stellar companions over mass and semi-major axis is necessary to confirm the above patterns in the sample. Such an investigation will be left to a subsequent study of this sample while this paper is focused on companion detection. 

\subsection{Multiplicity} 
In the sample of 161 wide-companion hosts, 61 hosts have multiple planet or stellar companions. Among them, 29 have stellar companions from the EDR3 wide-binary catalog given by \cite{elbadry21}, 6 of them have both companions identified in this work and companions from the wide-binary catalog. Without referring to the EDR3 wide-binary catalog, there are 38 multi-companion systems. Among the 61 multi-companion systems, there are 3 systems that contain planets, BDs, and stellar companions, 12 contain planets and BDs, 21 contain planets and stellar companions, 8 contain BDs and stellar companions, 12 contain multiple planets, 1 contains multiple BDs, and 4 contain multiple stellar companions. 

\begin{figure}
     \centering
     \includegraphics[scale=0.7]{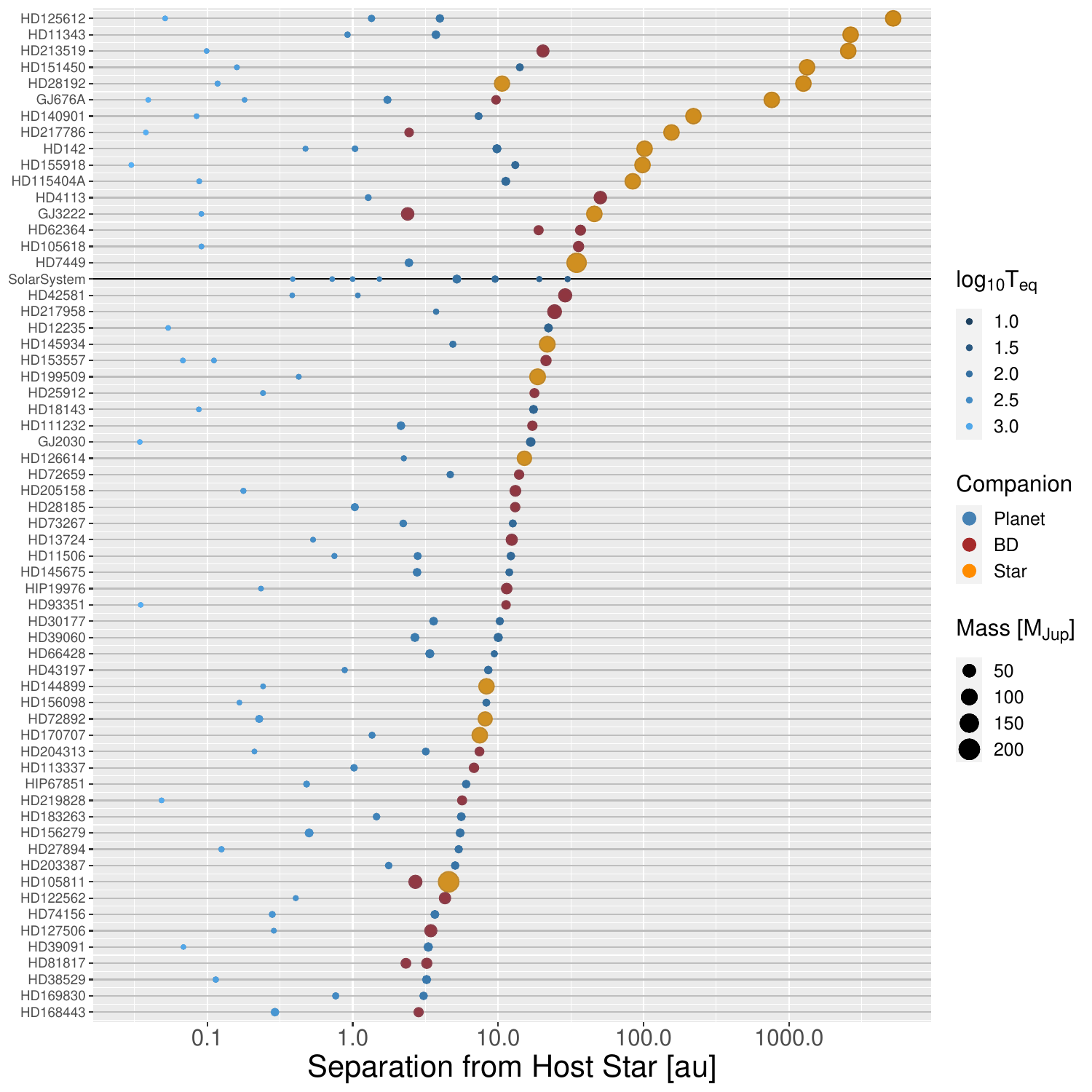}
     \caption{The 61 multi-companion systems identified in this work. The yellow dots represent stellar companions with mass higher than 75\mj, the brown dots represent BDs with mass range from 13 to 75\mj, and the blue dots represent planet companions with mass lower than 13\mj. The semi-major axis is used as a proxy for the separation from host star for the companions identified in this work while the binary separation is used for the EDR3 wide-binary catalog given by \cite{elbadry21}. The Solar System planets are put on the black horizontal line for reference. The masses of the Solar System planets are amplified by 10 times for better visualization. The size of dots represents companion mass and the size for stellar companions is truncated to the largest size.}
     \label{fig:architecture}
 \end{figure}

All multi-companion systems are shown in Fig. \ref{fig:architecture}. The apparent impression is that the widest companion in a system that is wider than the Neptune's orbit tend to be stellar. This is due to the incompleteness of sub-stellar companions on extremely wide orbits (e.g. $>$100\,au). On the other hand, the architecture of the inner system seems to be insensitive to the separation between the outer companion and the primary star. This is either due to the incompleteness of the inner companions or due to the insignificant impact of extremely wide companion on inner system. 

There is a controversy over whether hosts of hot Jupiters or BDs tend to have high rates of widely separated companions \citep{fontanive19,ziegler21,moe21}. While close binaries definitely suppress S-type planets, it is unclear whether wide companions could influence the formation of inner planets significantly. Because our sample is mainly massive companions on wide orbits, we would like to assess the influence of wide-companions on inner planets. By cross-matching the EDR3-based wide binary sample given by \cite{elbadry21} and our sample of massive companions, we find 29 out of the 161 companion hosts have wide stellar companions, indicating a stellar multiplicity rate of at least 18$\pm$3\%. Considering that only two host stars in our sample have distances larger than 200\,pc, the incompleteness of the identified wide companions is mainly caused by decreasing Gaia completeness below an angular resolution of 2$''$ \citep{elbadry21}. 

The detection rate of a wide binary around a companion host is P(DWB,$s|$CH), where $s$ is binary separation. It is derived from the occurrence rate of the wide binary around companion host P(WB,$s|$CH) and the incompleteness of the EDR3 wide binary sample, P(DWB,$s|$WB,$s$), according to P(DWB,$s|$CH)=P(DWB,$s|$WB,s)P(WB,$s|$CH). To calculate the detection rate, P(DWB,$s|$WB,$s$), we sample $s$ from 20 to 10,000\,au. For each $s$, we repetitively draw 100,000 samples from the parallaxes $\tilde\omega$ of all companion hosts and select the ones that could be resolved by Gaia, i.e. $s\tilde\omega>2''$. The sample becomes significantly incomplete for $s<150$\,au or $ln(s/au)<5$. After correcting for this incompleteness, the occurrence rate of wide binaries for companion hosts or P(WB$|$CH) is 32$\pm$6\%, consistent with the binary fraction of 36$\pm$2\% for binary separation from 20 to 10,000\,au \citep{fontanive19}. Hence we do not find any preference of multiplicity for massive companions on wide orbits. 

Based on the calculation of P(DWB,$s|$CH) and P(DWB,$s|$WB,$s$), we derive the wide-binary occurrence rate as a function of separation by using P(WB,$s|$CH)=P(DWB,$s|$CH)/P(DWB,$s|$WB,$s$). We find that the identified wide binaries approximately follow a log-normal distribution centered around $\ln{(s/au)}=5.8$ or $s=330$\,au. A similar peak around 250\,au is also found by \cite{fontanive19} for wide binaries hosting hot-Jupiters. After considering the detection bias, we find a power-law distribution of $P({\rm WB},s|{\rm CH})\propto s^{-1.4}$ to be optimal to model the distribution of the occurrence rate over binary separation. This power-law distribution over separation is consistent with the monotonic decreasing of binary fraction with separation beyond 3\,au for the whole EDR3 wide binary sample \citep{elbadry21}. Considering that the power-law distribution is found after considering detection bias, it is probably intrinsic to the wide binary sample and thus the peak around 250\,au in the separation distribution found by \cite{fontanive19} is probably due to detection bias. A detailed analysis of the whole EDR3 wide binary sample is needed to understand the intrinsic distribution of wide binaries and is beyond the scope of this paper. 

\section{Candidates for direct imaging}\label{sec:imaging}
\begin{figure}
  \centering
  \includegraphics[scale=0.4]{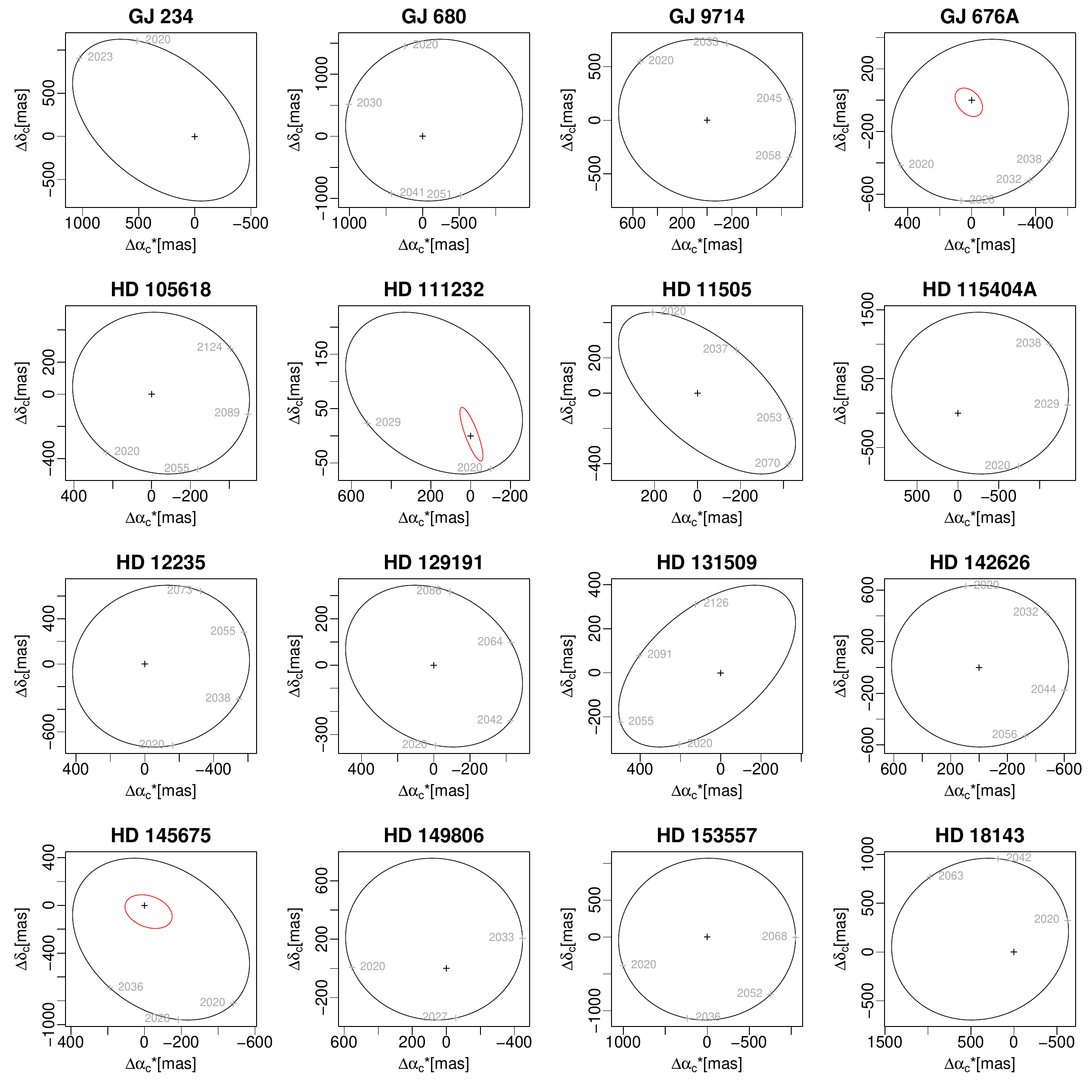}
  \caption{Orbits of 16 companions with separation larger than
    500\,mas from their host stars. For a multi-giant system, the black line represents the giant companion on the widest orbit while the red line represents the other companions with orbital period longer than 1000\,d in the system. The positions of the companions at some epochs are
    also denoted by the corresponding years. The black cross
    in each panel represents the host star. 
  }
  \label{fig:orbit}
\end{figure}

\begin{figure}
  \centering
  \includegraphics[scale=0.55]{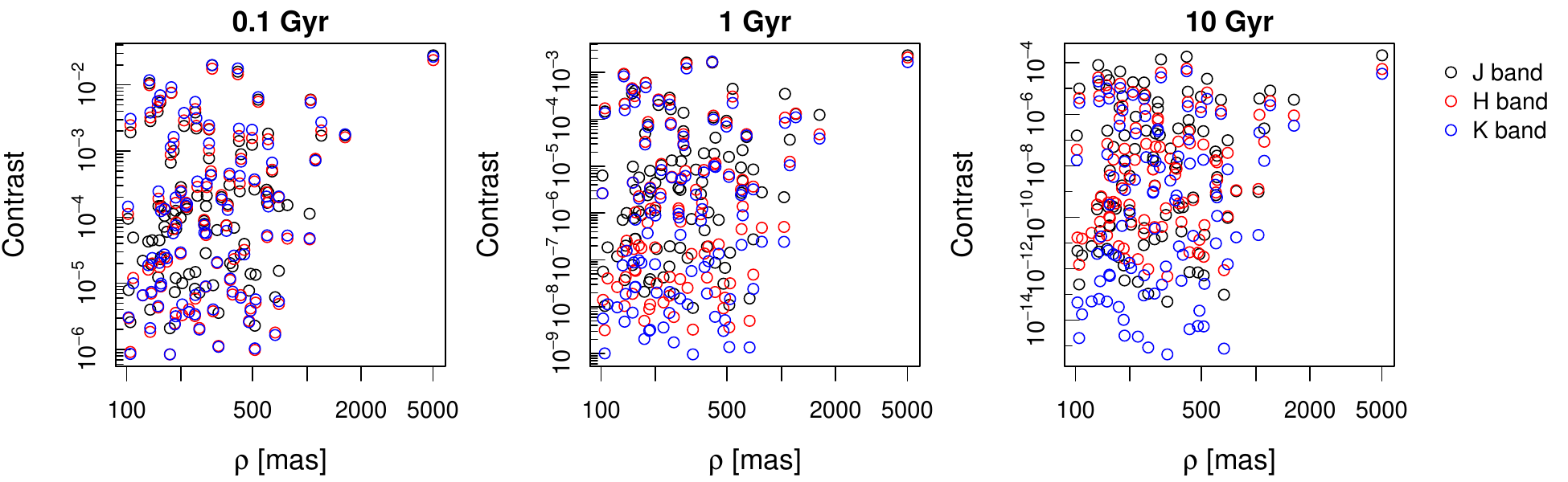}
  \caption{Companion-host contrast in J, H, and K bands by assuming ages of 0.1, 1, and 10 Gyr for the systems. }
  \label{fig:contrast}
\end{figure}

Because the companions identified in this work are massive, nearby,
and on wide orbits, they are appropriate targets for direct imaging by
the currently available facilities and the next-generation
instruments. We find 30 super-Jupiters and BDs with average separation from
their hosts larger than $0.5''$. The orbits of 16 of them are shown in
Fig. \ref{fig:orbit}. 
Considering that the ages of the host stars are typically not well
constrained, we assume ages of 0.1, 1, and 10\,Gyr for each star to
predict the J, H, K magnitudes and the total luminosity (L) of wide
super-Jupiters and BDs (less than 75\mj) based on the cooling models introduced by \cite{phillips20}. The contrast ratio between the companion and their host stars are shown in Fig. \ref{fig:contrast}. For the modern coronagraphs such as SCExAO/CHARIS installed on the Subaru telescope \citep{jovanovic15}, the typical inner
working angle is around 0.2$''$ and the contrast limit is about
$10^{-6}$ \citep{currie20}. Assuming such a detection sensitivity and an age of
1\,Gyr for all host stars, 41\%, 35\%, and 33\% companions are
detectable in the J, H and K bands. The proportions of detectable
companions increase to 62\%, 61\%, and 61\% for all bands if the stars have an age of 0.1\,Gyr, and decrease to 16\% for J band, 12\% for H band, and 11\% for K band if the stars have an age of 10\,Gyr. Considering 0.1 and 10\,Gyr as the
lower and upper limits of the ages of host stars respectively, there are 10-57 sub-stellar objects detectable by the current imaging facilities. 

\startlongtable
\begin{deluxetable*}{lllllllllllr}
\tablecaption{Mean separation and base-10 logarithmic flux ratio of companions to host stars. \label{tab:contrast}}
\tablehead{\colhead{Companion}& \colhead{$\rho$ [mas]}&\colhead{$m_c$ [\mj]}&\colhead{J0.1}&\colhead{H0.1}&\colhead{K0.1}&\colhead{J1}&\colhead{H1}&\colhead{K1}&\colhead{J10}&\colhead{H10}&\colhead{K10} }
\startdata
GJ 2030 c & 449 &$ 12.8_{-2.14}^{+2.34 }$& -4.4&-4.5&-4.6&-6.1&-6.7&-6.9&-9.2&-9.3&-10.7 \\
GJ 3222 c & 135 &$ 52.21_{-4.95}^{+5.11 }$& -2.5&-2.5&-2.4&-3.7&-3.7&-3.8&-4.9&-5.3&-5.5 \\
GJ 680 b & 1047 &$ 25.1_{-11.15}^{+6.16 }$& -2.2&-2.2&-2.3&-3.5&-4.0&-4.1&-5.4&-6.0&-6.7 \\
GJ 864 b & 298 &$ 59.96_{-6.08}^{+8.86 }$& -1.7&-1.8&-1.7&-2.8&-2.8&-2.9&-3.9&-4.4&-4.6 \\
GJ 9714 b & 601 &$ 9.4_{-1.69}^{+3.05 }$& -3.8&-4.3&-4.3&-5.6&-6.3&-6.7&-9.1&-9.0&-11.0 \\
GJ 676  A c & 609 &$ 13.49_{-1.13}^{+1.05 }$& -2.7&-2.8&-2.9&-4.7&-5.3&-5.5&-7.6&-7.9&-9.2 \\
GJ 676  A b & 109 &$ 5.79_{-0.48}^{+0.47 }$& -4.3&-4.9&-5.0&-6.7&-7.4&-7.9&-11.5&-10.8&-13.8 \\
HD 105618 c & 512 &$ 26.49_{-2.83}^{+4.34 }$& -3.6&-3.5&-3.4&-4.8&-5.2&-5.2&-6.7&-7.2&-7.8 \\
HD 106515 A b & 132 &$ 9.52_{-0.13}^{+6.38 }$& -4.4&-4.8&-4.7&-6.1&-6.8&-7.1&-9.7&-9.5&-11.4 \\
HD 109988 b & 293 &$ 21.82_{-4.15}^{+2.26 }$& -3.4&-3.3&-3.4&-4.7&-5.1&-5.2&-6.8&-7.3&-8.1 \\
...&...&...&...&...&...&...&...&...&...&...&... \\
\enddata
\tablecomments{Table \ref{tab:contrast} is published in its entirety in the
electronic edition of the {\it Astrophysical Journal Supplement Series}.  A portion is
shown here for guidance regarding its form and content. The contrasts in different bands are denoted by "BandAge" where "Band" is H, J, or K and "Age" is 0.1, 1, or 10 Gyrs. The contrast or flux ratio is in a base-10 log scale. This table only lists the mean values of the separation $\rho$ and the contrast in different bands for each companion. The separation of the companion to the host vary over time and the exact values for different epochs are shown in Fig. \ref{fig:orbit}. The luminosity is either given by the Gaia DR2 or by the mass-luminosity relation provided by \cite{eker15}. The J, H, and K magntiudes of each star are obtained from the Simbad database \citep{wenger00}. The J, H, and K magnitudes as well as the luminosity for each companion  are derived by the ATMO 2020 cooling models \citep{phillips20}. }
\end{deluxetable*}

\section{Dynamical Stability}\label{sec:stability}
We performed a large number of N-body simulations to study the
dynamical stability of both the planets themselves, as well as the
systems' habitable zones (HZs).  The majority of our computations utilize the $Mercury6$ hybrid integrator \citep{chambers99}, however we also employed a more direct, Bulirsch-Stoer methodology to accurately simulate systems with high eccentricity planets that make excessively low perihelia passages around their host stars.  In general, our simulations are designed to perform a broad, first-order analysis of the long-term dynamical evolution of each planets' orbit given the uncertainties reported in Table \ref{tab:Ncomp}.  Therefore, our work should be viewed as a reasonable measure of the stability of our sample of systems (and a validation of the orbital determinations described earlier in this manuscript), rather than a comprehensive and detailed interrogation of all possible trajectories.  While we consider variations within the determined values of the most dynamically significant properties of each planets' orbit (namely their eccentricities and semi-major axes), we do not investigate the possibility of perturbations from other, undetected massive bodies in each system that might perturb the orbital paths of the detected bodies.  Thus, while our simulations cannot definitively prove that our sample of systems are stable, we can argue with a high degree of confidence that planets exhibiting regular behavior regardless of the orbital parameters being varied are stable.

\subsection{Multi-planet system stability}

\begin{figure}
	\centering
	\includegraphics[width=.5\textwidth]{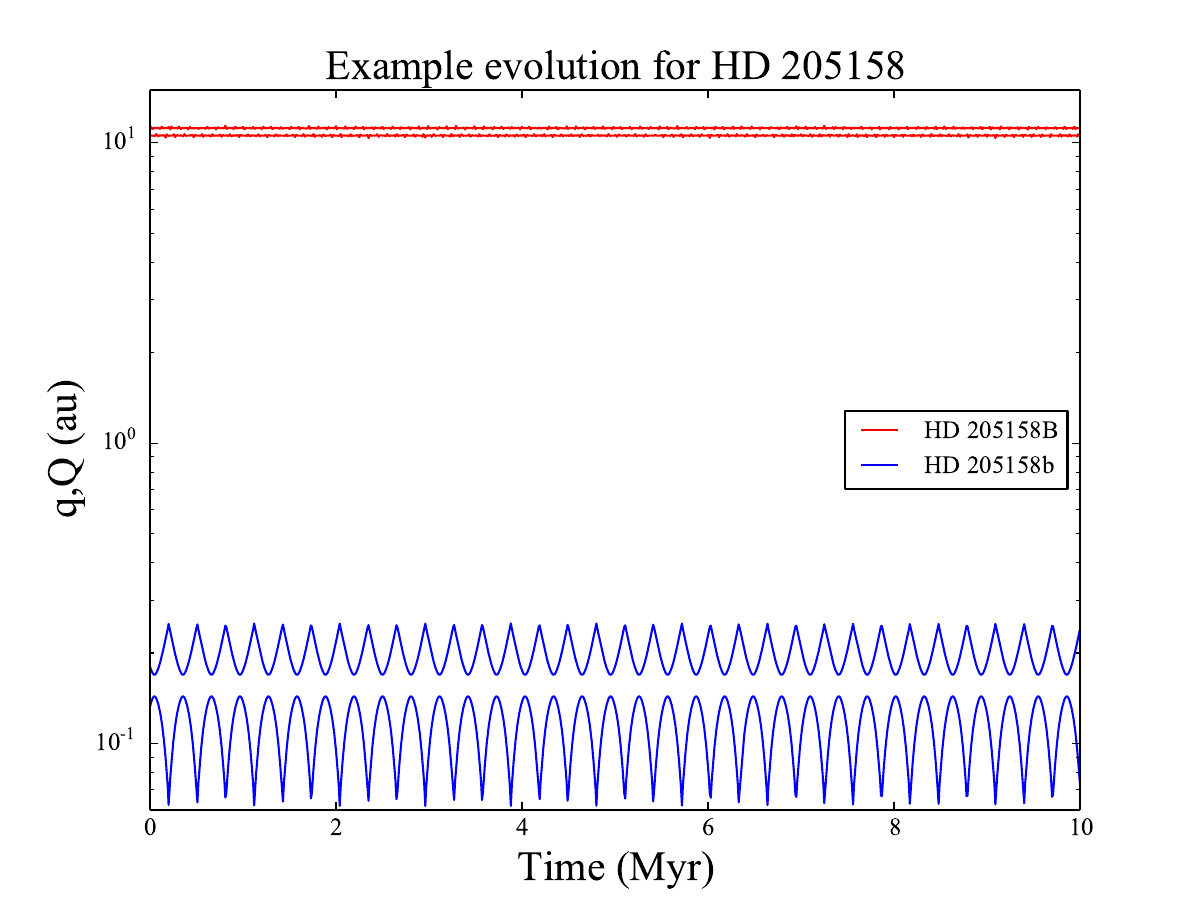}
	\caption{An example evolutionary scheme from one of our dynamical simulations studying the low mass companion and Jupiter analog in HD 205158.  The pericenter and apocenter for each body is plotted with red (HD 205158 B) and blue (HD 205158 b) lines.}
	\label{fig:qaq}
\end{figure}

We ran a series of $\geq$225, 1 Myr dynamical simulations to gauge the orbital stability of each multi-planet system in Table \ref{tab:Ncomp}.  In all cases we consider a grid of five eccentricities and three semi-major axes that span the range of uncertainties reported in Table \ref{tab:Ncomp} for each planet (angular orbital elements not determined through our orbit fitting are determined by sampling from uniform distributions of angles).  Each simulation leverages a time-step of $\sim$5$\%$ the orbital period of the innermost planet.  Through this analysis, we determine that each planet reported in this manuscript are stable in at least 90$\%$ of our numerical integrations (the unstable cases typically occur at larger eccentricities). For some  multi-companion systems we performed extended (10 Myr) simulations. As an  example, the two detected bodies in the HD 205158 system are plotted in Fig. \ref{fig:qaq}.  While the brown dwarf companion HD 205158B drives large secular oscillations in the inner Jupiter analog's eccentricity, the pair evolve on stable orbits for the duration of the simulation in spite of the system's uncharacteristically large mutual inclination.

\subsection{Stability of an Earth analog in the HZ}
\begin{figure}
    \centering
	\includegraphics[scale=0.5]{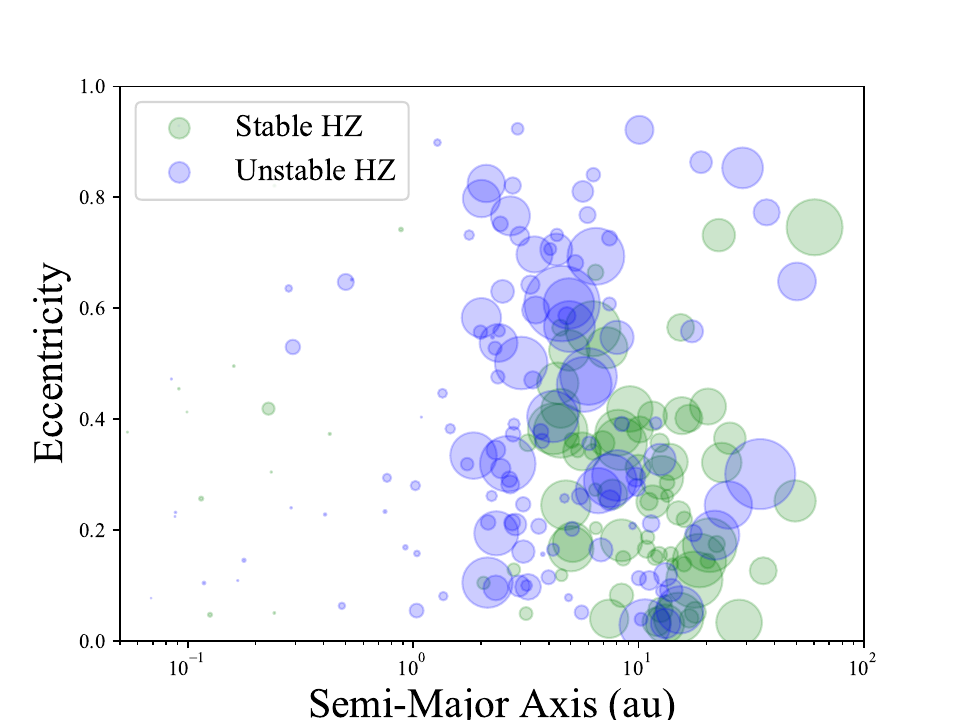}
	\caption{Summary of our simulations studying the stability of an Earth-mass planet in the HZ of our various systems identified in Table 2.  The color of each point identifies whether the Earth analog survived the simulation (green points), or was lost via ejection or collision (blue points).  The size of each point corresponds to the mass of each planet in our various simulations.}
	\label{fig:hz_stab}
\end{figure}

For each system reported in Table \ref{tab:Ncomp}, we investigated the stability of an Earth-mass planet on a nearly circular, co-planar orbit situated in the center of the HZ \citep[as determined via relations from][]{kopparapu13}.  Each of these models utilized the nominal orbital parameters for all planets, the Mercury package's hybrid symplectic integrator, and a total simulation time of 1.0 Gyr.  Fig. \ref{fig:hz_stab} and Table \ref{tab:par} summarize the results from this batch of lengthy integrations.  Unsurprisingly, systems with stable HZs tend to possess planets on longer-period orbits with lower eccentricities.  Additionally, the majority of the detected companion bodies in our sample tend to destabilize the HZ of their host stars.  While these findings are by no means novel, we present these simulations to demonstrate the reasonable feasibility of habitability in the majority of our new systems.

\section{Conclusion}\label{sec:conclusion}
Based on analyses of the proper motion and positional difference
between Gaia and Hipparcos data as well as the RV data, we find 167
circumstellar giants in 161 systems. The occurrence rate of wide-orbit BDs is at least 1.3\%, consistent with previous studies. There are 61 stars hosting multiple companions, with 12 hosting both planets and BDs, 21 hosting both planets and stellar companions, 8 hosting both BDs and stellar companions, and 3 hosting all types of companions. 

Without correction of detection bias, we observe a monotonic decrease of occurrence rate with mass from 5 to 40\mj, a valley around 40\mj and 75\mj. The investigation of these features by reliable correction of observation bias will be left to a subsequent study of this sample. 
We do not find significant preference of multiplicity for wide-orbit companions with mass from 5 to 120\mj. For a system hosting a 5-120\mj companion, the probability of finding a stellar companion on a wide orbit ($>$3\,au) is about 30$\pm$5\% after
accounting for detection bias.  

Because our sample of cold giants are nearby and well separated from
their hosts, they are good targets for direct imaging by the current
facilities. By adopting the ``ATMO 2020'' model to predict the BD
temperature and assuming different ages for the systems, we find that
10-57 super-Jupiters and BDs could be directly imaged by the current
facilities. According to the \url{exoplanet.eu} archive, 161
exoplanets and BDs are imaged, 16 of them are older than
1\,Gyr, 20 of them have dynamical mass, and only 4 of them are older than 1\,Gyr and with dynamical mass. Because our sample of stars are selected from various RV surveys and typically show stable RV variation, most of the
detectable super-Jupiters and BDs are likely to be old. If imaged by
follow-up observations, the substellar companions found in this work will
extend the imaged cold (or old) giants by an order of magnitude. 

We conduct numerical N-body simulations to investigate the
dynamical stability of the systems. Our simulations show a stability
probability of at least 90\% for single-companion systems over 1\,Myr
and for multi-companion systems over 10\,Myr. Our simulations also
show the majority of the massive companions identified in this work
tend to destabilize the HZ of their host stars. Hence
massive companions on wide and eccentric orbits are not friendly to life. 

In addition to the above findings, we also expect the following application of this catalog of sub-stellar companions. 
\begin{itemize}
\item {\bf Sub-stellar mass function:} By accounting for the RV and astrometric detection bias, the occurrence rate of objects of different mass can be corrected to derive the mass function as well as the occurrence rate as a function of mass and other orbital parameters. This will quantify the boundary of BD desert in the parameter space and test various scenarios for the formation of cold giants.   
\item {\bf Misalignment in multi-companion system:} Unlike previous RV
 surveys, our synergistic survey of nearby sub-stellar companions
 fully constrain the orbits and masses of companions. By investigating
 the orbital misalignment between different companions in a system and the
 correlation between misalignment and other orbital parameters, we can discover the causes of this misalignment and constrain 
  evolutionary models of sub-stellar objects.
\item {\bf Constraint of cooling models for BDs and super-Jupiters:}
 Many of the companions in our sample could be imaged by
 the current facilities. Imaging data will be used to derive the
 luminosity, color, and effective temperature of the
 companions. Because the host stars of these companions are nearby
 and bright, their ages are typically well constrained. Assuming a
 co-natal and coeval formation for companions and their hosts, the
 age, effective temperature, and mass could be used to constrain
 the parameters of various cooling models such as cloud coverage.
\item {\bf Correlation between cold giants and other types of
   planets:} Because our sample significantly extends the
 well-constrained cold-giant sample, it makes the statistical study
 of the correlation between these cold giants and inner planets more
 robust. Unlike the previous work based on RV samples of exoplanets
 with only a minimum mass available, the study of planet correlations
 based on our sample will avoid the mass ambiguity caused by the
 limitation of the RV-only method.
\end{itemize}

In summary, we detect and confirm 167 sub-stellar companions on wide
orbits with well constrained mass, extending the current sample of
similar objects by more than one order of magnitudes. This
catalog is used to study the correlation between substellar companion
and wide binaries, to provide dozens of candidates for direct imaging
by the current facility, and to investigate the influence of cold
giants on the HZs of their hosts. Future works based on our sample
would test the brown dwarf desert hypothesis and shed light on the formation and evolution of planets and sub-stellar companions. 

In future, we will detect short period companions through combined analyses of Hipparcos intermediate data and the synthetic Gaia data generated by GOST. By pushing the detection limit of our combined approach towards the low mass regime, we
will detect and characterize Jupiter-mass and Saturn-mass planets on wide orbits for the study of population synthesis, dynamical origin of orbital misalignment, and provide a unique sample of CJs and Saturns
for direct imaging by James-Webb Space Telescope (JWST; \citealt{danielski18}) and Chinese Space Station Telescope (CSST; \citealt{gong19}). 

\section*{Acknowledgements}
This work has made use of data from the European Space Agency (ESA)
mission {\it Gaia} (https://www.cosmos.esa.int/gaia), processed by the
{\it Gaia} Data Processing and Analysis Consortium (DPAC,
https://www.cosmos.esa.int/web/gaia/dpac/consortium). Funding for the
DPAC has been provided by national institutions, in particular the
institutions participating in the {\it Gaia} Multilateral
Agreement. This research has also made use of the Keck Observatory Archive (KOA), which is operated by the W. M. Keck Observatory and the
NASA Exoplanet Science Institute (NExScI), under contract with
the National Aeronautics and Space Administration.  This
research has also made use of the services of the ESO Science
Archive Facility, NASA's Astrophysics Data System Bibliographic
Service, and the SIMBAD database, operated at CDS, Strasbourg, France.
Support for this work was provided by Chinese Academy of Sciences President’s International Fellowship Initiative Grant No. 2020VMA0033.
The authors acknowledge the years of technical support
from LCO staff in the successful operation of PFS, enabling the
collection of the data presented in this paper. We would also like to
acknowledge the many years of technical support from the UCO/Lick
staff for the commissioning and operation of the APF facility atop
Mt. Hamilton. Part of this research was carried out at the Jet
Propulsion Laboratory, California Institute of Technology, under a
contract with the National Aeronautics and Space Administration
(NASA). This work used the Extreme Science and Engineering Discovery
Environment (XSEDE), which is supported by National Science Foundation
grant number ACI-1548562. Specifically, it used the Bridges2 system,
which is supported by NSF award number ACI-1445606, at the Pittsburgh
Supercomputing Center \citep[PSC:][]{xsede}. F.F.D. was supported by
the Research Development Fund (grant RDF-16–01–16) of Xi’an
Jiaotong-Liverpool University (XJTLU) and was supported by the SPP
1992 exoplanet diversity program and Berlin Technical University. B.M. acknowledges financial support from NSFC grant 12073092 and the
  science research grants from the China Manned Space Project
  (No. CMS-CSST-2021-B09). This paper is partly based on observations collected at the European Organisation for Astronomical Research in the Southern Hemisphere under ESO programmes: 0100.C-0097,0100.C-0414,0101.C-0232,0101.C-0379,072.C-0488,074.C-0364,075.C-0332,075.D-0800,076.C-0155,077.C-0101,077.C-0364,079.C-0657,079.C-0927,081.C-0148,081.C-0802,082.C-0212,082.C-0427,084.C-0228,085.C-0019,085.C-0063,086.C-0284,087.C-0368,087.C-0831,088.C-0662,089.C-0497,089.C-0732,090.C-0395,090.C-0421,091.C-0034,091.C-0866,091.C-0936,091.D-0469,092.C-0721,093.C-0409,094.C-0901,095.C-0551,095.D-0026,096.C-0460,097.C-0090,098.C-0366,098.C-0739,099.C-0458,180.C-0886,183.C-0437,183.C-0972,183.D-0729,188.C-0265,190.C-0027,191.C-0873,192.C-0224,192.C-0852,196.C-1006,198.C-0836,60.A-9036, and 60.A-9700.
\software{Astropy \citep{astropy13, astropy18}, Astroquery \citep{ginsburg19}, Numpy \citep{walt11,harris20}, Mercury \citep{chambers99}, magicaxis \citep{robotham16}, fields \citep{fields}, MASS \citep{ripley13}, minpack.lm
\citep{elzhov16}}

\bibliographystyle{aasjournal}
\bibliography{nm}  

\begin{thebibliography}{}
\expandafter\ifx\csname natexlab\endcsname\relax\def\natexlab#1{#1}\fi
\providecommand{\url}[1]{\href{#1}{#1}}
\providecommand{\dodoi}[1]{doi:~\href{http://doi.org/#1}{\nolinkurl{#1}}}
\providecommand{\doeprint}[1]{\href{http://ascl.net/#1}{\nolinkurl{http://ascl.net/#1}}}
\providecommand{\doarXiv}[1]{\href{https://arxiv.org/abs/#1}{\nolinkurl{https://arxiv.org/abs/#1}}}

\bibitem[{{Acton} {et~al.}(2021){Acton}, {Goad}, {Burleigh}, {Casewell},
  {Breytenbach}, {Nielsen}, {Smith}, {Anderson}, {Battley}, {Bayliss},
  {Bouchy}, {Bryant}, {Csizmadia}, {Eigm{\"u}ller}, {Gill}, {Gillen},
  {Grieves}, {G{\"u}nther}, {Henderson}, {Hodgkin}, {Jackman}, {Jenkins},
  {Lendl}, {McCormac}, {Moyano}, {Nelson}, {Sefako}, {Smith}, {Stalport},
  {Thomas}, {Tilbrook}, {Udry}, {West}, {Wheatley}, {Worters}, {Vines}, \&
  {Alves}}]{acton21}
{Acton}, J.~S., {Goad}, M.~R., {Burleigh}, M.~R., {et~al.} 2021, \mnras, 505,
  2741, \dodoi{10.1093/mnras/stab1459}

\bibitem[{{Agati} {et~al.}(2015){Agati}, {Bonneau}, {Jorissen}, {Souli{\'e}},
  {Udry}, {Verhas}, \& {Dommanget}}]{agati15}
{Agati}, J.~L., {Bonneau}, D., {Jorissen}, A., {et~al.} 2015, \aap, 574, A6,
  \dodoi{10.1051/0004-6361/201323056}

\bibitem[{{Allen} {et~al.}(2012){Allen}, {Burgasser}, {Faherty}, \&
  {Kirkpatrick}}]{allen12}
{Allen}, P.~R., {Burgasser}, A.~J., {Faherty}, J.~K., \& {Kirkpatrick}, J.~D.
  2012, \aj, 144, 62, \dodoi{10.1088/0004-6256/144/2/62}

\bibitem[{{Anglada-Escud{\'e}} \& {Butler}(2012)}]{anglada12}
{Anglada-Escud{\'e}}, G., \& {Butler}, R.~P. 2012, \apjs, 200, 15,
  \dodoi{10.1088/0067-0049/200/2/15}

\bibitem[{{Anglada-Escud{\'e}} {et~al.}(2010){Anglada-Escud{\'e}},
  {L{\'o}pez-Morales}, \& {Chambers}}]{angladaescude10}
{Anglada-Escud{\'e}}, G., {L{\'o}pez-Morales}, M., \& {Chambers}, J.~E. 2010,
  \apj, 709, 168, \dodoi{10.1088/0004-637X/709/1/168}

\bibitem[{{Anglada-Escud{\'e}} \& {Tuomi}(2012)}]{angladaescude12}
{Anglada-Escud{\'e}}, G., \& {Tuomi}, M. 2012, \aap, 548, A58,
  \dodoi{10.1051/0004-6361/201219910}

\bibitem[{{Arriagada} {et~al.}(2010){Arriagada}, {Butler}, {Minniti},
  {L{\'o}pez-Morales}, {Shectman}, {Adams}, {Boss}, \&
  {Chambers}}]{arriagada10}
{Arriagada}, P., {Butler}, R.~P., {Minniti}, D., {et~al.} 2010, \apj, 711,
  1229, \dodoi{10.1088/0004-637X/711/2/1229}

\bibitem[{{Astropy Collaboration} {et~al.}(2013){Astropy Collaboration},
  {Robitaille}, {Tollerud}, {Greenfield}, {Droettboom}, {Bray}, {Aldcroft},
  {Davis}, {Ginsburg}, {Price-Whelan}, {Kerzendorf}, {Conley}, {Crighton},
  {Barbary}, {Muna}, {Ferguson}, {Grollier}, {Parikh}, {Nair}, {Unther},
  {Deil}, {Woillez}, {Conseil}, {Kramer}, {Turner}, {Singer}, {Fox}, {Weaver},
  {Zabalza}, {Edwards}, {Azalee Bostroem}, {Burke}, {Casey}, {Crawford},
  {Dencheva}, {Ely}, {Jenness}, {Labrie}, {Lim}, {Pierfederici}, {Pontzen},
  {Ptak}, {Refsdal}, {Servillat}, \& {Streicher}}]{astropy13}
{Astropy Collaboration}, {Robitaille}, T.~P., {Tollerud}, E.~J., {et~al.} 2013,
  \aap, 558, A33, \dodoi{10.1051/0004-6361/201322068}

\bibitem[{{Astropy Collaboration} {et~al.}(2018){Astropy Collaboration},
  {Price-Whelan}, {Sip{\H{o}}cz}, {G{\"u}nther}, {Lim}, {Crawford}, {Conseil},
  {Shupe}, {Craig}, {Dencheva}, {Ginsburg}, {VanderPlas}, {Bradley},
  {P{\'e}rez-Su{\'a}rez}, {de Val-Borro}, {Aldcroft}, {Cruz}, {Robitaille},
  {Tollerud}, {Ardelean}, {Babej}, {Bach}, {Bachetti}, {Bakanov}, {Bamford},
  {Barentsen}, {Barmby}, {Baumbach}, {Berry}, {Biscani}, {Boquien}, {Bostroem},
  {Bouma}, {Brammer}, {Bray}, {Breytenbach}, {Buddelmeijer}, {Burke},
  {Calderone}, {Cano Rodr{\'\i}guez}, {Cara}, {Cardoso}, {Cheedella}, {Copin},
  {Corrales}, {Crichton}, {D'Avella}, {Deil}, {Depagne}, {Dietrich}, {Donath},
  {Droettboom}, {Earl}, {Erben}, {Fabbro}, {Ferreira}, {Finethy}, {Fox},
  {Garrison}, {Gibbons}, {Goldstein}, {Gommers}, {Greco}, {Greenfield},
  {Groener}, {Grollier}, {Hagen}, {Hirst}, {Homeier}, {Horton}, {Hosseinzadeh},
  {Hu}, {Hunkeler}, {Ivezi{\'c}}, {Jain}, {Jenness}, {Kanarek}, {Kendrew},
  {Kern}, {Kerzendorf}, {Khvalko}, {King}, {Kirkby}, {Kulkarni}, {Kumar},
  {Lee}, {Lenz}, {Littlefair}, {Ma}, {Macleod}, {Mastropietro}, {McCully},
  {Montagnac}, {Morris}, {Mueller}, {Mumford}, {Muna}, {Murphy}, {Nelson},
  {Nguyen}, {Ninan}, {N{\"o}the}, {Ogaz}, {Oh}, {Parejko}, {Parley}, {Pascual},
  {Patil}, {Patil}, {Plunkett}, {Prochaska}, {Rastogi}, {Reddy Janga},
  {Sabater}, {Sakurikar}, {Seifert}, {Sherbert}, {Sherwood-Taylor}, {Shih},
  {Sick}, {Silbiger}, {Singanamalla}, {Singer}, {Sladen}, {Sooley},
  {Sornarajah}, {Streicher}, {Teuben}, {Thomas}, {Tremblay}, {Turner},
  {Terr{\'o}n}, {van Kerkwijk}, {de la Vega}, {Watkins}, {Weaver}, {Whitmore},
  {Woillez}, {Zabalza}, \& {Astropy Contributors}}]{astropy18}
{Astropy Collaboration}, {Price-Whelan}, A.~M., {Sip{\H{o}}cz}, B.~M., {et~al.}
  2018, \aj, 156, 123, \dodoi{10.3847/1538-3881/aabc4f}

\bibitem[{{Bang} {et~al.}(2020){Bang}, {Lee}, {Perdelwitz}, {Jeong}, {Han},
  {Oh}, \& {Park}}]{bang20}
{Bang}, T.-Y., {Lee}, B.-C., {Perdelwitz}, V., {et~al.} 2020, \aap, 638, A148,
  \dodoi{10.1051/0004-6361/201936613}

\bibitem[{{Baraffe} {et~al.}(2003){Baraffe}, {Chabrier}, {Barman}, {Allard}, \&
  {Hauschildt}}]{baraffe03}
{Baraffe}, I., {Chabrier}, G., {Barman}, T.~S., {Allard}, F., \& {Hauschildt},
  P.~H. 2003, \aap, 402, 701, \dodoi{10.1051/0004-6361:20030252}

\bibitem[{{Baraffe} {et~al.}(2015){Baraffe}, {Homeier}, {Allard}, \&
  {Chabrier}}]{baraffe15}
{Baraffe}, I., {Homeier}, D., {Allard}, F., \& {Chabrier}, G. 2015, \aap, 577,
  A42, \dodoi{10.1051/0004-6361/201425481}

\bibitem[{{Baranne} {et~al.}(1996){Baranne}, {Queloz}, {Mayor}, {Adrianzyk},
  {Knispel}, {Kohler}, {Lacroix}, {Meunier}, {Rimbaud}, \& {Vin}}]{baranne96}
{Baranne}, A., {Queloz}, D., {Mayor}, M., {et~al.} 1996, \aaps, 119, 373

\bibitem[{{Barbato} {et~al.}(2018){Barbato}, {Sozzetti}, {Desidera}, {Damasso},
  {Bonomo}, {Giacobbe}, {Colombo}, {Lazzoni}, {Claudi}, {Gratton}, {LoCurto},
  {Marzari}, \& {Mordasini}}]{barbato18}
{Barbato}, D., {Sozzetti}, A., {Desidera}, S., {et~al.} 2018, \aap, 615, A175,
  \dodoi{10.1051/0004-6361/201832791}

\bibitem[{{Bardalez Gagliuffi} {et~al.}(2021){Bardalez Gagliuffi}, {Faherty},
  {Li}, {Brandt}, {Williams}, {Brandt}, \& {Gelino}}]{gagliuffi21}
{Bardalez Gagliuffi}, D.~C., {Faherty}, J.~K., {Li}, Y., {et~al.} 2021, \apjl,
  922, L43, \dodoi{10.3847/2041-8213/ac382c}

\bibitem[{{Benedict} {et~al.}(2010){Benedict}, {McArthur}, {Bean}, {Barnes},
  {Harrison}, {Hatzes}, {Martioli}, \& {Nelan}}]{benedict10}
{Benedict}, G.~F., {McArthur}, B.~E., {Bean}, J.~L., {et~al.} 2010, \aj, 139,
  1844, \dodoi{10.1088/0004-6256/139/5/1844}

\bibitem[{{Bernstein} {et~al.}(2003){Bernstein}, {Shectman}, {Gunnels},
  {Mochnacki}, \& {Athey}}]{bernstein03}
{Bernstein}, R., {Shectman}, S.~A., {Gunnels}, S.~M., {Mochnacki}, S., \&
  {Athey}, A.~E. 2003, in Society of Photo-Optical Instrumentation Engineers
  (SPIE) Conference Series, Vol. 4841, Instrument Design and Performance for
  Optical/Infrared Ground-based Telescopes, ed. M.~{Iye} \& A.~F.~M.
  {Moorwood}, 1694--1704

\bibitem[{{Beuzit} {et~al.}(2004){Beuzit}, {S{\'e}gransan}, {Forveille},
  {Udry}, {Delfosse}, {Mayor}, {Perrier}, {Hainaut}, {Roddier}, {Roddier}, \&
  {Mart{\'\i}n}}]{beuzit04}
{Beuzit}, J.~L., {S{\'e}gransan}, D., {Forveille}, T., {et~al.} 2004, \aap,
  425, 997, \dodoi{10.1051/0004-6361:20048006}

\bibitem[{{Bonavita} {et~al.}(2016){Bonavita}, {Desidera}, {Thalmann},
  {Janson}, {Vigan}, {Chauvin}, \& {Lannier}}]{bonavita16}
{Bonavita}, M., {Desidera}, S., {Thalmann}, C., {et~al.} 2016, \aap, 593, A38,
  \dodoi{10.1051/0004-6361/201628231}

\bibitem[{{Borgniet} {et~al.}(2014){Borgniet}, {Boisse}, {Lagrange}, {Bouchy},
  {Arnold}, {D{\'\i}az}, {Galland}, {Delorme}, {H{\'e}brard}, {Santerne},
  {Ehrenreich}, {S{\'e}gransan}, {Bonfils}, {Delfosse}, {Santos}, {Forveille},
  {Moutou}, {Udry}, {Eggenberger}, {Pepe}, {Astudillo}, \&
  {Montagnier}}]{borgniet14}
{Borgniet}, S., {Boisse}, I., {Lagrange}, A.~M., {et~al.} 2014, \aap, 561, A65,
  \dodoi{10.1051/0004-6361/201321783}

\bibitem[{{Borgniet} {et~al.}(2019){Borgniet}, {Perraut}, {Su}, {Bonnefoy},
  {Delorme}, {Lagrange}, {Bailey}, {Buenzli}, {Defr{\`e}re}, {Henning}, {Hinz},
  {Leisenring}, {Meunier}, {Mourard}, {Nardetto}, {Skemer}, \&
  {Spalding}}]{borgniet19}
{Borgniet}, S., {Perraut}, K., {Su}, K., {et~al.} 2019, \aap, 627, A44,
  \dodoi{10.1051/0004-6361/201935494}

\bibitem[{{Boss}(1996)}]{boss96nat}
{Boss}, A.~P. 1996, \nat, 379, 397, \dodoi{10.1038/379397a0}

\bibitem[{{Bouchy} {et~al.}(2016){Bouchy}, {S{\'e}gransan}, {D{\'\i}az},
  {Forveille}, {Boisse}, {Arnold}, {Astudillo-Defru}, {Beuzit}, {Bonfils},
  {Borgniet}, {Bourrier}, {Courcol}, {Delfosse}, {Demangeon}, {Delorme},
  {Ehrenreich}, {H{\'e}brard}, {Lagrange}, {Mayor}, {Montagnier}, {Moutou},
  {Naef}, {Pepe}, {Perrier}, {Queloz}, {Rey}, {Sahlmann}, {Santerne}, {Santos},
  {Sivan}, {Udry}, \& {Wilson}}]{bouchy16}
{Bouchy}, F., {S{\'e}gransan}, D., {D{\'\i}az}, R.~F., {et~al.} 2016, \aap,
  585, A46, \dodoi{10.1051/0004-6361/201526347}

\bibitem[{{Bowler} {et~al.}(2018){Bowler}, {Dupuy}, {Endl}, {Cochran},
  {MacQueen}, {Fulton}, {Petigura}, {Howard}, {Hirsch}, {Kratter}, {Crepp},
  {Biller}, {Johnson}, \& {Wittenmyer}}]{bowler18}
{Bowler}, B.~P., {Dupuy}, T.~J., {Endl}, M., {et~al.} 2018, \aj, 155, 159,
  \dodoi{10.3847/1538-3881/aab2a6}

\bibitem[{{Brandt} {et~al.}(2021{\natexlab{a}}){Brandt}, {Brandt}, {Dupuy},
  {Li}, \& {Michalik}}]{brandt21b}
{Brandt}, G.~M., {Brandt}, T.~D., {Dupuy}, T.~J., {Li}, Y., \& {Michalik}, D.
  2021{\natexlab{a}}, \aj, 161, 179, \dodoi{10.3847/1538-3881/abdc2e}

\bibitem[{{Brandt} {et~al.}(2021{\natexlab{b}}){Brandt}, {Dupuy}, {Li}, {Chen},
  {Brandt}, {Wong}, {Currie}, {Bowler}, {Liu}, {Best}, \&
  {Phillips}}]{brandt21}
{Brandt}, G.~M., {Dupuy}, T.~J., {Li}, Y., {et~al.} 2021{\natexlab{b}}, \aj,
  162, 301, \dodoi{10.3847/1538-3881/ac273e}

\bibitem[{{Brandt}(2018)}]{brandt18}
{Brandt}, T.~D. 2018, \apjs, 239, 31, \dodoi{10.3847/1538-4365/aaec06}

\bibitem[{{Brandt} {et~al.}(2019){Brandt}, {Dupuy}, \& {Bowler}}]{brandt19}
{Brandt}, T.~D., {Dupuy}, T.~J., \& {Bowler}, B.~P. 2019, \aj, 158, 140,
  \dodoi{10.3847/1538-3881/ab04a8}

\bibitem[{{Burgasser}(2008)}]{burgasser08}
{Burgasser}, A.~J. 2008, Physics Today, 61, 70, \dodoi{10.1063/1.2947658}

\bibitem[{{Burgasser} {et~al.}(2010){Burgasser}, {Simcoe}, {Bochanski},
  {Saumon}, {Mamajek}, {Cushing}, {Marley}, {McMurtry}, {Pipher}, \&
  {Forrest}}]{burgasser10}
{Burgasser}, A.~J., {Simcoe}, R.~A., {Bochanski}, J.~J., {et~al.} 2010, \apj,
  725, 1405, \dodoi{10.1088/0004-637X/725/2/1405}

\bibitem[{{Burrows} {et~al.}(2001){Burrows}, {Hubbard}, {Lunine}, \&
  {Liebert}}]{burrows01}
{Burrows}, A., {Hubbard}, W.~B., {Lunine}, J.~I., \& {Liebert}, J. 2001,
  Reviews of Modern Physics, 73, 719, \dodoi{10.1103/RevModPhys.73.719}

\bibitem[{{Butler} {et~al.}(2003){Butler}, {Marcy}, {Vogt}, {Fischer}, {Henry},
  {Laughlin}, \& {Wright}}]{butler03}
{Butler}, R.~P., {Marcy}, G.~W., {Vogt}, S.~S., {et~al.} 2003, \apj, 582, 455,
  \dodoi{10.1086/344570}

\bibitem[{{Butler} {et~al.}(1996){Butler}, {Marcy}, {Williams}, {McCarthy},
  {Dosanjh}, \& {Vogt}}]{butler96}
{Butler}, R.~P., {Marcy}, G.~W., {Williams}, E., {et~al.} 1996, \pasp, 108,
  500, \dodoi{10.1086/133755}

\bibitem[{{Butler} {et~al.}(2006){Butler}, {Wright}, {Marcy}, {Fischer},
  {Vogt}, {Tinney}, {Jones}, {Carter}, {Johnson}, {McCarthy}, \&
  {Penny}}]{butler06}
{Butler}, R.~P., {Wright}, J.~T., {Marcy}, G.~W., {et~al.} 2006, \apj, 646,
  505, \dodoi{10.1086/504701}

\bibitem[{{Campbell} {et~al.}(1988){Campbell}, {Walker}, \&
  {Yang}}]{campbell88}
{Campbell}, B., {Walker}, G.~A.~H., \& {Yang}, S. 1988, \apj, 331, 902,
  \dodoi{10.1086/166608}

\bibitem[{{Cantat-Gaudin} \& {Brandt}(2021)}]{cantat21}
{Cantat-Gaudin}, T., \& {Brandt}, T.~D. 2021, \aap, 649, A124,
  \dodoi{10.1051/0004-6361/202140807}

\bibitem[{{Carmichael} {et~al.}(2019){Carmichael}, {Latham}, \&
  {Vanderburg}}]{carmichael19}
{Carmichael}, T.~W., {Latham}, D.~W., \& {Vanderburg}, A.~M. 2019, \aj, 158,
  38, \dodoi{10.3847/1538-3881/ab245e}

\bibitem[{{Carmichael} {et~al.}(2020){Carmichael}, {Quinn}, {Mustill}, {Huang},
  {Zhou}, {Persson}, {Nielsen}, {Collins}, {Ziegler}, {Collins}, {Rodriguez},
  {Shporer}, {Brahm}, {Mann}, {Bouchy}, {Fridlund}, {Stassun}, {Hellier},
  {Seidel}, {Stalport}, {Udry}, {Pepe}, {Ireland}, {{\v{Z}}erjal},
  {Brice{\~n}o}, {Law}, {Jord{\'a}n}, {Espinoza}, {Henning}, {Sarkis}, \&
  {Latham}}]{carmichael20}
{Carmichael}, T.~W., {Quinn}, S.~N., {Mustill}, A.~J., {et~al.} 2020, \aj, 160,
  53, \dodoi{10.3847/1538-3881/ab9b84}

\bibitem[{{Chabrier} {et~al.}(2014){Chabrier}, {Johansen}, {Janson}, \&
  {Rafikov}}]{chabrier14}
{Chabrier}, G., {Johansen}, A., {Janson}, M., \& {Rafikov}, R. 2014, in
  Protostars and Planets VI, ed. H.~{Beuther}, R.~S. {Klessen}, C.~P.
  {Dullemond}, \& T.~{Henning}, 619

\bibitem[{{Chambers}(1999)}]{chambers99}
{Chambers}, J.~E. 1999, \mnras, 304, 793,
  \dodoi{10.1046/j.1365-8711.1999.02379.x}

\bibitem[{{Cheetham} {et~al.}(2018){Cheetham}, {S{\'e}gransan}, {Peretti},
  {Delisle}, {Hagelberg}, {Beuzit}, {Forveille}, {Marmier}, {Udry}, \&
  {Wildi}}]{cheetham18}
{Cheetham}, A., {S{\'e}gransan}, D., {Peretti}, S., {et~al.} 2018, \aap, 614,
  A16, \dodoi{10.1051/0004-6361/201630136}

\bibitem[{{Cosentino} {et~al.}(2012){Cosentino}, {Lovis}, {Pepe}, {Collier
  Cameron}, {Latham}, {Molinari}, {Udry}, {Bezawada}, {Black}, {Born},
  {Buchschacher}, {Charbonneau}, {Figueira}, {Fleury}, {Galli}, {Gallie},
  {Gao}, {Ghedina}, {Gonzalez}, {Gonzalez}, {Guerra}, {Henry}, {Horne},
  {Hughes}, {Kelly}, {Lodi}, {Lunney}, {Maire}, {Mayor}, {Micela}, {Ordway},
  {Peacock}, {Phillips}, {Piotto}, {Pollacco}, {Queloz}, {Rice}, {Riverol},
  {Riverol}, {San Juan}, {Sasselov}, {Segransan}, {Sozzetti}, {Sosnowska},
  {Stobie}, {Szentgyorgyi}, {Vick}, \& {Weber}}]{cosentino12}
{Cosentino}, R., {Lovis}, C., {Pepe}, F., {et~al.} 2012, in Society of
  Photo-Optical Instrumentation Engineers (SPIE) Conference Series, Vol. 8446,
  Ground-based and Airborne Instrumentation for Astronomy IV, ed. I.~S.
  {McLean}, S.~K. {Ramsay}, \& H.~{Takami}, 84461V

\bibitem[{{Crane} {et~al.}(2010){Crane}, {Shectman}, {Butler}, {Thompson},
  {Birk}, {Jones}, \& {Burley}}]{crane10}
{Crane}, J.~D., {Shectman}, S.~A., {Butler}, R.~P., {et~al.} 2010, in
  \procspie, Vol. 7735, Ground-based and Airborne Instrumentation for Astronomy
  III, 773553

\bibitem[{{Currie} {et~al.}(2020){Currie}, {Guyon}, {Lozi}, {Sahoo}, {Vievard},
  {Deo}, {Chilcote}, {Groff}, {Brandt}, {Lawson}, {Skaf}, {Martinache}, \&
  {Kasdin}}]{currie20}
{Currie}, T., {Guyon}, O., {Lozi}, J., {et~al.} 2020, in Society of
  Photo-Optical Instrumentation Engineers (SPIE) Conference Series, Vol. 11448,
  Society of Photo-Optical Instrumentation Engineers (SPIE) Conference Series,
  114487H

\bibitem[{{Damasso} {et~al.}(2020{\natexlab{a}}){Damasso}, {Sozzetti}, {Lovis},
  {Barros}, {Sousa}, {Demangeon}, {Faria}, {Lillo-Box}, {Cristiani}, {Pepe},
  {Rebolo}, {Santos}, {Zapatero Osorio}, {Gonz{\'a}lez Hern{\'a}ndez}, {Amate},
  {Pasquini}, {Zerbi}, {Adibekyan}, {Abreu}, {Affolter}, {Alibert}, {Aliverti},
  {Allart}, {Allende Prieto}, {{\'A}lvarez}, {Alves}, {Avila}, {Baldini},
  {Bandy}, {Benz}, {Bianco}, {Borsa}, {Bossini}, {Bourrier}, {Bouchy}, {Broeg},
  {Cabral}, {Calderone}, {Cirami}, {Coelho}, {Conconi}, {Coretti}, {Cumani},
  {Cupani}, {D'Odorico}, {Deiries}, {Dekker}, {Delabre}, {Di Marcantonio},
  {Dumusque}, {Ehrenreich}, {Figueira}, {Fragoso}, {Genolet}, {Genoni},
  {G{\'e}nova Santos}, {Hughes}, {Iwert}, {Kerber}, {Knudstrup}, {Landoni},
  {Lavie}, {Lizon}, {Lo Curto}, {Maire}, {Martins}, {M{\'e}gevand}, {Mehner},
  {Micela}, {Modigliani}, {Molaro}, {Monteiro}, {Monteiro}, {Moschetti},
  {Mueller}, {Murphy}, {Nunes}, {Oggioni}, {Oliveira}, {Oshagh}, {Pall{\'e}},
  {Pariani}, {Poretti}, {Rasilla}, {Rebord{\~a}o}, {Redaelli}, {Riva}, {Santana
  Tschudi}, {Santin}, {Santos}, {S{\'e}gransan}, {Schmidt}, {Segovia},
  {Sosnowska}, {Span{\`o}}, {Su{\'a}rez Mascare{\~n}o}, {Tabernero}, {Tenegi},
  {Udry}, \& {Zanutta}}]{alvarez20}
{Damasso}, M., {Sozzetti}, A., {Lovis}, C., {et~al.} 2020{\natexlab{a}}, \aap,
  642, A31, \dodoi{10.1051/0004-6361/202038416}

\bibitem[{{Damasso} {et~al.}(2020{\natexlab{b}}){Damasso}, {Sozzetti}, {Lovis},
  {Barros}, {Sousa}, {Demangeon}, {Faria}, {Lillo-Box}, {Cristiani}, {Pepe},
  {Rebolo}, {Santos}, {Zapatero Osorio}, {Gonz{\'a}lez Hern{\'a}ndez}, {Amate},
  {Pasquini}, {Zerbi}, {Adibekyan}, {Abreu}, {Affolter}, {Alibert}, {Aliverti},
  {Allart}, {Allende Prieto}, {{\'A}lvarez}, {Alves}, {Avila}, {Baldini},
  {Bandy}, {Benz}, {Bianco}, {Borsa}, {Bossini}, {Bourrier}, {Bouchy}, {Broeg},
  {Cabral}, {Calderone}, {Cirami}, {Coelho}, {Conconi}, {Coretti}, {Cumani},
  {Cupani}, {D'Odorico}, {Deiries}, {Dekker}, {Delabre}, {Di Marcantonio},
  {Dumusque}, {Ehrenreich}, {Figueira}, {Fragoso}, {Genolet}, {Genoni},
  {G{\'e}nova Santos}, {Hughes}, {Iwert}, {Kerber}, {Knudstrup}, {Landoni},
  {Lavie}, {Lizon}, {Lo Curto}, {Maire}, {Martins}, {M{\'e}gevand}, {Mehner},
  {Micela}, {Modigliani}, {Molaro}, {Monteiro}, {Monteiro}, {Moschetti},
  {Mueller}, {Murphy}, {Nunes}, {Oggioni}, {Oliveira}, {Oshagh}, {Pall{\'e}},
  {Pariani}, {Poretti}, {Rasilla}, {Rebord{\~a}o}, {Redaelli}, {Riva}, {Santana
  Tschudi}, {Santin}, {Santos}, {S{\'e}gransan}, {Schmidt}, {Segovia},
  {Sosnowska}, {Span{\`o}}, {Su{\'a}rez Mascare{\~n}o}, {Tabernero}, {Tenegi},
  {Udry}, \& {Zanutta}}]{damasso20}
---. 2020{\natexlab{b}}, \aap, 642, A31, \dodoi{10.1051/0004-6361/202038416}

\bibitem[{{Danielski} {et~al.}(2018){Danielski}, {Baudino}, {Lagage},
  {Boccaletti}, {Gastaud}, {Coulais}, \& {B{\'e}zard}}]{danielski18}
{Danielski}, C., {Baudino}, J.-L., {Lagage}, P.-O., {et~al.} 2018, \aj, 156,
  276, \dodoi{10.3847/1538-3881/aae651}

\bibitem[{{De Rosa} {et~al.}(2020){De Rosa}, {Dawson}, \& {Nielsen}}]{derosa20}
{De Rosa}, R.~J., {Dawson}, R., \& {Nielsen}, E.~L. 2020, \aap, 640, A73,
  \dodoi{10.1051/0004-6361/202038496}

\bibitem[{{Demangeon} {et~al.}(2021){Demangeon}, {Dalal}, {H{\'e}brard},
  {Nsamba}, {Kiefer}, {Camacho}, {Sahlmann}, {Arnold}, {Astudillo-Defru},
  {Bonfils}, {Boisse}, {Bouchy}, {Bourrier}, {Campante}, {Delfosse}, {Deleuil},
  {D{\'\i}az}, {Faria}, {Forveille}, {Hara}, {Heidari}, {Hobson}, {Lopez},
  {Moutou}, {Rey}, {Santerne}, {Sousa}, {Santos}, {Str{\o}m}, {Tsantaki}, \&
  {Udry}}]{demangeon21}
{Demangeon}, O.~D.~S., {Dalal}, S., {H{\'e}brard}, G., {et~al.} 2021, \aap,
  653, A78, \dodoi{10.1051/0004-6361/202141079}

\bibitem[{{Desidera} {et~al.}(2012){Desidera}, {Gratton}, {Carolo}, {Martinez
  Fiorenzano}, {Endl}, {Mesa}, {Cecconi}, {Claudi}, {Cosentino}, {Scuderi},
  {Sozzetti}, \& {Zurlo}}]{desidera12}
{Desidera}, S., {Gratton}, R., {Carolo}, E., {et~al.} 2012, \aap, 546, A108,
  \dodoi{10.1051/0004-6361/201220038}

\bibitem[{{D{\'\i}az} {et~al.}(2012){D{\'\i}az}, {Santerne}, {Sahlmann},
  {H{\'e}brard}, {Eggenberger}, {Santos}, {Moutou}, {Arnold}, {Boisse},
  {Bonfils}, {Bouchy}, {Delfosse}, {Desort}, {Ehrenreich}, {Forveille},
  {Lagrange}, {Lovis}, {Pepe}, {Perrier}, {Queloz}, {S{\'e}gransan}, {Udry}, \&
  {Vidal-Madjar}}]{diaz12}
{D{\'\i}az}, R.~F., {Santerne}, A., {Sahlmann}, J., {et~al.} 2012, \aap, 538,
  A113, \dodoi{10.1051/0004-6361/201117935}

\bibitem[{{D{\'\i}az} {et~al.}(2016{\natexlab{a}}){D{\'\i}az}, {Rey},
  {Demangeon}, {H{\'e}brard}, {Boisse}, {Arnold}, {Astudillo-Defru}, {Beuzit},
  {Bonfils}, {Borgniet}, {Bouchy}, {Bourrier}, {Courcol}, {Deleuil},
  {Delfosse}, {Ehrenreich}, {Forveille}, {Lagrange}, {Mayor}, {Moutou}, {Pepe},
  {Queloz}, {Santerne}, {Santos}, {Sahlmann}, {S{\'e}gransan}, {Udry}, \&
  {Wilson}}]{diaz16}
{D{\'\i}az}, R.~F., {Rey}, J., {Demangeon}, O., {et~al.} 2016{\natexlab{a}},
  \aap, 591, A146, \dodoi{10.1051/0004-6361/201628331}

\bibitem[{{D{\'\i}az} {et~al.}(2016{\natexlab{b}}){D{\'\i}az}, {Rey},
  {Demangeon}, {H{\'e}brard}, {Boisse}, {Arnold}, {Astudillo-Defru}, {Beuzit},
  {Bonfils}, {Borgniet}, {Bouchy}, {Bourrier}, {Courcol}, {Deleuil},
  {Delfosse}, {Ehrenreich}, {Forveille}, {Lagrange}, {Mayor}, {Moutou}, {Pepe},
  {Queloz}, {Santerne}, {Santos}, {Sahlmann}, {S{\'e}gransan}, {Udry}, \&
  {Wilson}}]{diaz16a}
---. 2016{\natexlab{b}}, \aap, 591, A146, \dodoi{10.1051/0004-6361/201628331}

\bibitem[{{Diego} {et~al.}(1990){Diego}, {Charalambous}, {Fish}, \&
  {Walker}}]{diego90}
{Diego}, F., {Charalambous}, A., {Fish}, A.~C., \& {Walker}, D.~D. 1990, in
  Society of Photo-Optical Instrumentation Engineers (SPIE) Conference Series,
  Vol. 1235, Instrumentation in Astronomy VII, ed. D.~L. {Crawford}, 562--576

\bibitem[{{D{\"o}llinger} {et~al.}(2009){D{\"o}llinger}, {Hatzes}, {Pasquini},
  {Guenther}, {Hartmann}, \& {Girardi}}]{dollinger09}
{D{\"o}llinger}, M.~P., {Hatzes}, A.~P., {Pasquini}, L., {et~al.} 2009, \aap,
  499, 935, \dodoi{10.1051/0004-6361/200810837}

\bibitem[{{Dumusque} {et~al.}(2011){Dumusque}, {Lovis}, {S{\'e}gransan},
  {Mayor}, {Udry}, {Benz}, {Bouchy}, {Lo Curto}, {Mordasini}, {Pepe}, {Queloz},
  {Santos}, \& {Naef}}]{dumusque11}
{Dumusque}, X., {Lovis}, C., {S{\'e}gransan}, D., {et~al.} 2011, \aap, 535,
  A55, \dodoi{10.1051/0004-6361/201117148}

\bibitem[{{Dupuy} \& {Kraus}(2013)}]{dupuy13}
{Dupuy}, T.~J., \& {Kraus}, A.~L. 2013, Science, 341, 1492,
  \dodoi{10.1126/science.1241917}

\bibitem[{{Eker} {et~al.}(2015){Eker}, {Soydugan}, {Soydugan}, {Bilir}, {Yaz
  G{\"o}k{\c c}e}, {Steer}, {T{\"u}ys{\"u}z}, {{\c S}eny{\"u}z}, \&
  {Demircan}}]{eker15}
{Eker}, Z., {Soydugan}, F., {Soydugan}, E., {et~al.} 2015, \aj, 149, 131,
  \dodoi{10.1088/0004-6256/149/4/131}

\bibitem[{{El-Badry} {et~al.}(2021){El-Badry}, {Rix}, \& {Heintz}}]{elbadry21}
{El-Badry}, K., {Rix}, H.-W., \& {Heintz}, T.~M. 2021, \mnras, 506, 2269,
  \dodoi{10.1093/mnras/stab323}

\bibitem[{Elzhov {et~al.}(2016)Elzhov, Mullen, Spiess, Bolker, Mullen, \&
  Suggests}]{elzhov16}
Elzhov, T.~V., Mullen, K.~M., Spiess, A.-N., {et~al.} 2016

\bibitem[{{Feng} {et~al.}(2019{\natexlab{a}}){Feng}, {Anglada-Escud{\'e}},
  {Tuomi}, {Jones}, {Chanam{\'e}}, {Butler}, \& {Janson}}]{feng19b}
{Feng}, F., {Anglada-Escud{\'e}}, G., {Tuomi}, M., {et~al.} 2019{\natexlab{a}},
  \mnras, 490, 5002, \dodoi{10.1093/mnras/stz2912}

\bibitem[{{Feng} {et~al.}(2017){Feng}, {Tuomi}, \& {Jones}}]{feng17b}
{Feng}, F., {Tuomi}, M., \& {Jones}, H.~R.~A. 2017, \mnras, 470, 4794,
  \dodoi{10.1093/mnras/stx1126}

\bibitem[{{Feng} {et~al.}(2016){Feng}, {Tuomi}, {Jones}, {Butler}, \&
  {Vogt}}]{feng16}
{Feng}, F., {Tuomi}, M., {Jones}, H.~R.~A., {Butler}, R.~P., \& {Vogt}, S.
  2016, \mnras, 461, 2440, \dodoi{10.1093/mnras/stw1478}

\bibitem[{{Feng} {et~al.}(2019{\natexlab{b}}){Feng}, {Crane}, {Xuesong Wang},
  {Teske}, {Shectman}, {D{\'\i}az}, {Thompson}, {Jones}, \& {Butler}}]{feng19a}
{Feng}, F., {Crane}, J.~D., {Xuesong Wang}, S., {et~al.} 2019{\natexlab{b}},
  \apjs, 242, 25, \dodoi{10.3847/1538-4365/ab1b16}

\bibitem[{{Feng} {et~al.}(2020{\natexlab{a}}){Feng}, {Butler}, {Shectman},
  {Crane}, {Vogt}, {Chambers}, {Jones}, {Xuesong Wang}, {Teske}, {Burt},
  {D{\'\i}az}, \& {Thompson}}]{feng20a}
{Feng}, F., {Butler}, R.~P., {Shectman}, S.~A., {et~al.} 2020{\natexlab{a}},
  \apjs, 246, 11, \dodoi{10.3847/1538-4365/ab5e7c}

\bibitem[{{Feng} {et~al.}(2020{\natexlab{b}}){Feng}, {Butler}, {Shectman},
  {Crane}, {Vogt}, {Chambers}, {Jones}, {Xuesong Wang}, {Teske}, {Burt},
  {D{\'\i}az}, \& {Thompson}}]{feng20}
---. 2020{\natexlab{b}}, \apjs, 246, 11, \dodoi{10.3847/1538-4365/ab5e7c}

\bibitem[{{Feng} {et~al.}(2021){Feng}, {Butler}, {Jones}, {Phillips}, {Vogt},
  {Oppenheimer}, {Holden}, {Burt}, \& {Boss}}]{feng21}
{Feng}, F., {Butler}, R.~P., {Jones}, H. R.~A., {et~al.} 2021, \mnras, 507,
  2856, \dodoi{10.1093/mnras/stab2225}

\bibitem[{{Feng} {et~al.}(2015){Feng}, {Wright}, {Nelson}, {Wang}, {Ford},
  {Marcy}, {Isaacson}, \& {Howard}}]{feng15}
{Feng}, Y.~K., {Wright}, J.~T., {Nelson}, B., {et~al.} 2015, \apj, 800, 22,
  \dodoi{10.1088/0004-637X/800/1/22}

\bibitem[{{Fischer} {et~al.}(2009){Fischer}, {Driscoll}, {Isaacson}, {Giguere},
  {Marcy}, {Valenti}, {Wright}, {Henry}, {Johnson}, {Howard}, {Peek}, \&
  {McCarthy}}]{fischer09}
{Fischer}, D., {Driscoll}, P., {Isaacson}, H., {et~al.} 2009, \apj, 703, 1545,
  \dodoi{10.1088/0004-637X/703/2/1545}

\bibitem[{{Fischer} {et~al.}(2001){Fischer}, {Marcy}, {Butler}, {Vogt},
  {Frink}, \& {Apps}}]{fischer01}
{Fischer}, D.~A., {Marcy}, G.~W., {Butler}, R.~P., {et~al.} 2001, \apj, 551,
  1107, \dodoi{10.1086/320224}

\bibitem[{{Fischer} {et~al.}(2002){Fischer}, {Marcy}, {Butler}, {Vogt}, {Walp},
  \& {Apps}}]{fischer02}
---. 2002, \pasp, 114, 529, \dodoi{10.1086/341677}

\bibitem[{{Fischer} {et~al.}(2014){Fischer}, {Marcy}, \& {Spronck}}]{fischer14}
{Fischer}, D.~A., {Marcy}, G.~W., \& {Spronck}, J. F.~P. 2014, \apjs, 210, 5,
  \dodoi{10.1088/0067-0049/210/1/5}

\bibitem[{{Fischer} {et~al.}(2003){Fischer}, {Marcy}, {Butler}, {Vogt},
  {Henry}, {Pourbaix}, {Walp}, {Misch}, \& {Wright}}]{fischer03}
{Fischer}, D.~A., {Marcy}, G.~W., {Butler}, R.~P., {et~al.} 2003, \apj, 586,
  1394, \dodoi{10.1086/367889}

\bibitem[{{Fischer} {et~al.}(2007){Fischer}, {Vogt}, {Marcy}, {Butler}, {Sato},
  {Henry}, {Robinson}, {Laughlin}, {Ida}, {Toyota}, {Omiya}, {Driscoll},
  {Takeda}, {Wright}, \& {Johnson}}]{fischer07}
{Fischer}, D.~A., {Vogt}, S.~S., {Marcy}, G.~W., {et~al.} 2007, \apj, 669,
  1336, \dodoi{10.1086/521869}

\bibitem[{{Fontanive} {et~al.}(2019){Fontanive}, {Rice}, {Bonavita}, {Lopez},
  {Mu{\v{z}}i{\'c}}, {}, \& {Biller}}]{fontanive19}
{Fontanive}, C., {Rice}, K., {Bonavita}, M., {et~al.} 2019, \mnras, 485, 4967,
  \dodoi{10.1093/mnras/stz671}

\bibitem[{{Foreman-Mackey} {et~al.}(2015){Foreman-Mackey}, {Montet}, {Hogg},
  {Morton}, {Wang}, \& {Sch{\"o}lkopf}}]{foreman-mackey15}
{Foreman-Mackey}, D., {Montet}, B.~T., {Hogg}, D.~W., {et~al.} 2015, \apj, 806,
  215, \dodoi{10.1088/0004-637X/806/2/215}

\bibitem[{{Forveille} {et~al.}(2011){Forveille}, {Bonfils}, {Lo Curto},
  {Delfosse}, {Udry}, {Bouchy}, {Lovis}, {Mayor}, {Moutou}, {Naef}, {Pepe},
  {Perrier}, {Queloz}, \& {Santos}}]{forveille11}
{Forveille}, T., {Bonfils}, X., {Lo Curto}, G., {et~al.} 2011, \aap, 526, A141,
  \dodoi{10.1051/0004-6361/201016034}

\bibitem[{{Franson} {et~al.}(2022){Franson}, {Bowler}, {Brandt}, {Dupuy},
  {Tran}, {Brandt}, {Li}, \& {Kraus}}]{franson22}
{Franson}, K., {Bowler}, B.~P., {Brandt}, T.~D., {et~al.} 2022, \aj, 163, 50,
  \dodoi{10.3847/1538-3881/ac35e8}

\bibitem[{{Fuhrmann}(2004)}]{fuhrmann04}
{Fuhrmann}, K. 2004, Astronomische Nachrichten, 325, 3,
  \dodoi{10.1002/asna.200310173}

\bibitem[{{Gaia Collaboration} {et~al.}(2020){Gaia Collaboration}, {Brown},
  {Vallenari}, {Prusti}, {de Bruijne}, {Babusiaux}, \& {Biermann}}]{gaia20}
{Gaia Collaboration}, {Brown}, A.~G.~A., {Vallenari}, A., {et~al.} 2020, arXiv
  e-prints, arXiv:2012.01533.
\newblock \doarXiv{2012.01533}

\bibitem[{{Gaia Collaboration} {et~al.}(2016){Gaia Collaboration}, {Prusti},
  {de Bruijne}, {Brown}, {Vallenari}, {Babusiaux}, {Bailer-Jones}, {Bastian},
  {Biermann}, {Evans}, {Eyer}, {Jansen}, {Jordi}, {Klioner}, {Lammers},
  {Lindegren}, {Luri}, {Mignard}, {Milligan}, {Panem}, {Poinsignon},
  {Pourbaix}, {Randich}, {Sarri}, {Sartoretti}, {Siddiqui}, {Soubiran},
  {Valette}, {van Leeuwen}, {Walton}, {Aerts}, {Arenou}, {Cropper}, {Drimmel},
  {H{\o}g}, {Katz}, {Lattanzi}, {O'Mullane}, {Grebel}, {Holland}, {Huc},
  {Passot}, {Bramante}, {Cacciari}, {Casta{\~n}eda}, {Chaoul}, {Cheek}, {De
  Angeli}, {Fabricius}, {Guerra}, {Hern{\'a}ndez}, {Jean-Antoine-Piccolo},
  {Masana}, {Messineo}, {Mowlavi}, {Nienartowicz}, {Ord{\'o}{\~n}ez-Blanco},
  {Panuzzo}, {Portell}, {Richards}, {Riello}, {Seabroke}, {Tanga},
  {Th{\'e}venin}, {Torra}, {Els}, {Gracia-Abril}, {Comoretto},
  {Garcia-Reinaldos}, {Lock}, {Mercier}, {Altmann}, {Andrae}, {Astraatmadja},
  {Bellas-Velidis}, {Benson}, {Berthier}, {Blomme}, {Busso}, {Carry},
  {Cellino}, {Clementini}, {Cowell}, {Creevey}, {Cuypers}, {Davidson}, {De
  Ridder}, {de Torres}, {Delchambre}, {Dell'Oro}, {Ducourant}, {Fr{\'e}mat},
  {Garc{\'\i}a-Torres}, {Gosset}, {Halbwachs}, {Hambly}, {Harrison}, {Hauser},
  {Hestroffer}, {Hodgkin}, {Huckle}, {Hutton}, {Jasniewicz}, {Jordan},
  {Kontizas}, {Korn}, {Lanzafame}, {Manteiga}, {Moitinho}, {Muinonen},
  {Osinde}, {Pancino}, {Pauwels}, {Petit}, {Recio-Blanco}, {Robin}, {Sarro},
  {Siopis}, {Smith}, {Smith}, {Sozzetti}, {Thuillot}, {van Reeven}, {Viala},
  {Abbas}, {Abreu Aramburu}, {Accart}, {Aguado}, {Allan}, {Allasia},
  {Altavilla}, {{\'A}lvarez}, {Alves}, {Anderson}, {Andrei}, {Anglada Varela},
  {Antiche}, {Antoja}, {Ant{\'o}n}, {Arcay}, {Atzei}, {Ayache}, {Bach},
  {Baker}, {Balaguer-N{\'u}{\~n}ez}, {Barache}, {Barata}, {Barbier}, {Barblan},
  {Baroni}, {Barrado y Navascu{\'e}s}, {Barros}, {Barstow}, {Becciani},
  {Bellazzini}, {Bellei}, {Bello Garc{\'\i}a}, {Belokurov}, {Bendjoya},
  {Berihuete}, {Bianchi}, {Bienaym{\'e}}, {Billebaud}, {Blagorodnova},
  {Blanco-Cuaresma}, {Boch}, {Bombrun}, {Borrachero}, {Bouquillon}, {Bourda},
  {Bouy}, {Bragaglia}, {Breddels}, {Brouillet}, {Br{\"u}semeister},
  {Bucciarelli}, {Budnik}, {Burgess}, {Burgon}, {Burlacu}, {Busonero}, {Buzzi},
  {Caffau}, {Cambras}, {Campbell}, {Cancelliere}, {Cantat-Gaudin}, {Carlucci},
  {Carrasco}, {Castellani}, {Charlot}, {Charnas}, {Charvet}, {Chassat},
  {Chiavassa}, {Clotet}, {Cocozza}, {Collins}, {Collins}, {Costigan}, {Crifo},
  {Cross}, {Crosta}, {Crowley}, {Dafonte}, {Damerdji}, {Dapergolas}, {David},
  {David}, {De Cat}, {de Felice}, {de Laverny}, {De Luise}, {De March}, {de
  Martino}, {de Souza}, {Debosscher}, {del Pozo}, {Delbo}, {Delgado},
  {Delgado}, {di Marco}, {Di Matteo}, {Diakite}, {Distefano}, {Dolding}, {Dos
  Anjos}, {Drazinos}, {Dur{\'a}n}, {Dzigan}, {Ecale}, {Edvardsson}, {Enke},
  {Erdmann}, {Escolar}, {Espina}, {Evans}, {Eynard Bontemps}, {Fabre},
  {Fabrizio}, {Faigler}, {Falc{\~a}o}, {Farr{\`a}s Casas}, {Faye}, {Federici},
  {Fedorets}, {Fern{\'a}ndez-Hern{\'a}ndez}, {Fernique}, {Fienga}, {Figueras},
  {Filippi}, {Findeisen}, {Fonti}, {Fouesneau}, {Fraile}, {Fraser}, {Fuchs},
  {Furnell}, {Gai}, {Galleti}, {Galluccio}, {Garabato}, {Garc{\'\i}a-Sedano},
  {Gar{\'e}}, {Garofalo}, {Garralda}, {Gavras}, {Gerssen}, {Geyer}, {Gilmore},
  {Girona}, {Giuffrida}, {Gomes}, {Gonz{\'a}lez-Marcos},
  {Gonz{\'a}lez-N{\'u}{\~n}ez}, {Gonz{\'a}lez-Vidal}, {Granvik}, {Guerrier},
  {Guillout}, {Guiraud}, {G{\'u}rpide}, {Guti{\'e}rrez-S{\'a}nchez}, {Guy},
  {Haigron}, {Hatzidimitriou}, {Haywood}, {Heiter}, {Helmi}, {Hobbs},
  {Hofmann}, {Holl}, {Holland}, {Hunt}, {Hypki}, {Icardi}, {Irwin}, {Jevardat
  de Fombelle}, {Jofr{\'e}}, {Jonker}, {Jorissen}, {Julbe}, {Karampelas},
  {Kochoska}, {Kohley}, {Kolenberg}, {Kontizas}, {Koposov}, {Kordopatis},
  {Koubsky}, {Kowalczyk}, {Krone-Martins}, {Kudryashova}, {Kull}, {Bachchan},
  {Lacoste-Seris}, {Lanza}, {Lavigne}, {Le Poncin-Lafitte}, {Lebreton},
  {Lebzelter}, {Leccia}, {Leclerc}, {Lecoeur-Taibi}, {Lemaitre}, {Lenhardt},
  {Leroux}, {Liao}, {Licata}, {Lindstr{\o}m}, {Lister}, {Livanou}, {Lobel},
  {L{\"o}ffler}, {L{\'o}pez}, {Lopez-Lozano}, {Lorenz}, {Loureiro},
  {MacDonald}, {Magalh{\~a}es Fernandes}, {Managau}, {Mann}, {Mantelet},
  {Marchal}, {Marchant}, {Marconi}, {Marie}, {Marinoni}, {Marrese},
  {Marschalk{\'o}}, {Marshall}, {Mart{\'\i}n-Fleitas}, {Martino}, {Mary},
  {Matijevi{\v{c}}}, {Mazeh}, {McMillan}, {Messina}, {Mestre}, {Michalik},
  {Millar}, {Miranda}, {Molina}, {Molinaro}, {Molinaro}, {Moln{\'a}r},
  {Moniez}, {Montegriffo}, {Monteiro}, {Mor}, {Mora}, {Morbidelli}, {Morel},
  {Morgenthaler}, {Morley}, {Morris}, {Mulone}, {Muraveva}, {Musella},
  {Narbonne}, {Nelemans}, {Nicastro}, {Noval}, {Ord{\'e}novic},
  {Ordieres-Mer{\'e}}, {Osborne}, {Pagani}, {Pagano}, {Pailler}, {Palacin},
  {Palaversa}, {Parsons}, {Paulsen}, {Pecoraro}, {Pedrosa}, {Pentik{\"a}inen},
  {Pereira}, {Pichon}, {Piersimoni}, {Pineau}, {Plachy}, {Plum}, {Poujoulet},
  {Pr{\v{s}}a}, {Pulone}, {Ragaini}, {Rago}, {Rambaux}, {Ramos-Lerate},
  {Ranalli}, {Rauw}, {Read}, {Regibo}, {Renk}, {Reyl{\'e}}, {Ribeiro},
  {Rimoldini}, {Ripepi}, {Riva}, {Rixon}, {Roelens}, {Romero-G{\'o}mez},
  {Rowell}, {Royer}, {Rudolph}, {Ruiz-Dern}, {Sadowski}, {Sagrist{\`a}
  Sell{\'e}s}, {Sahlmann}, {Salgado}, {Salguero}, {Sarasso}, {Savietto},
  {Schnorhk}, {Schultheis}, {Sciacca}, {Segol}, {Segovia}, {Segransan},
  {Serpell}, {Shih}, {Smareglia}, {Smart}, {Smith}, {Solano}, {Solitro},
  {Sordo}, {Soria Nieto}, {Souchay}, {Spagna}, {Spoto}, {Stampa}, {Steele},
  {Steidelm{\"u}ller}, {Stephenson}, {Stoev}, {Suess}, {S{\"u}veges}, {Surdej},
  {Szabados}, {Szegedi-Elek}, {Tapiador}, {Taris}, {Tauran}, {Taylor},
  {Teixeira}, {Terrett}, {Tingley}, {Trager}, {Turon}, {Ulla}, {Utrilla},
  {Valentini}, {van Elteren}, {Van Hemelryck}, {van Leeuwen}, {Varadi},
  {Vecchiato}, {Veljanoski}, {Via}, {Vicente}, {Vogt}, {Voss}, {Votruba},
  {Voutsinas}, {Walmsley}, {Weiler}, {Weingrill}, {Werner}, {Wevers},
  {Whitehead}, {Wyrzykowski}, {Yoldas}, {{\v{Z}}erjal}, {Zucker}, {Zurbach},
  {Zwitter}, {Alecu}, {Allen}, {Allende Prieto}, {Amorim},
  {Anglada-Escud{\'e}}, {Arsenijevic}, {Azaz}, {Balm}, {Beck}, {Bernstein},
  {Bigot}, {Bijaoui}, {Blasco}, {Bonfigli}, {Bono}, {Boudreault}, {Bressan},
  {Brown}, {Brunet}, {Bunclark}, {Buonanno}, {Butkevich}, {Carret}, {Carrion},
  {Chemin}, {Ch{\'e}reau}, {Corcione}, {Darmigny}, {de Boer}, {de Teodoro}, {de
  Zeeuw}, {Delle Luche}, {Domingues}, {Dubath}, {Fodor}, {Fr{\'e}zouls},
  {Fries}, {Fustes}, {Fyfe}, {Gallardo}, {Gallegos}, {Gardiol}, {Gebran},
  {Gomboc}, {G{\'o}mez}, {Grux}, {Gueguen}, {Heyrovsky}, {Hoar}, {Iannicola},
  {Isasi Parache}, {Janotto}, {Joliet}, {Jonckheere}, {Keil}, {Kim},
  {Klagyivik}, {Klar}, {Knude}, {Kochukhov}, {Kolka}, {Kos}, {Kutka}, {Lainey},
  {LeBouquin}, {Liu}, {Loreggia}, {Makarov}, {Marseille}, {Martayan},
  {Martinez-Rubi}, {Massart}, {Meynadier}, {Mignot}, {Munari}, {Nguyen},
  {Nordlander}, {Ocvirk}, {O'Flaherty}, {Olias Sanz}, {Ortiz}, {Osorio},
  {Oszkiewicz}, {Ouzounis}, {Palmer}, {Park}, {Pasquato}, {Peltzer}, {Peralta},
  {P{\'e}turaud}, {Pieniluoma}, {Pigozzi}, {Poels}, {Prat}, {Prod'homme},
  {Raison}, {Rebordao}, {Risquez}, {Rocca-Volmerange}, {Rosen}, {Ruiz-Fuertes},
  {Russo}, {Sembay}, {Serraller Vizcaino}, {Short}, {Siebert}, {Silva},
  {Sinachopoulos}, {Slezak}, {Soffel}, {Sosnowska}, {Strai{\v{z}}ys}, {ter
  Linden}, {Terrell}, {Theil}, {Tiede}, {Troisi}, {Tsalmantza}, {Tur},
  {Vaccari}, {Vachier}, {Valles}, {Van Hamme}, {Veltz}, {Virtanen}, {Wallut},
  {Wichmann}, {Wilkinson}, {Ziaeepour}, \& {Zschocke}}]{gaia16}
{Gaia Collaboration}, {Prusti}, T., {de Bruijne}, J.~H.~J., {et~al.} 2016,
  \aap, 595, A1, \dodoi{10.1051/0004-6361/201629272}

\bibitem[{{Gaia Collaboration} {et~al.}(2018){Gaia Collaboration}, {Brown},
  {Vallenari}, {Prusti}, {de Bruijne}, {Babusiaux}, {Bailer-Jones}, {Biermann},
  {Evans}, {Eyer}, {Jansen}, {Jordi}, {Klioner}, {Lammers}, {Lindegren},
  {Luri}, {Mignard}, {Panem}, {Pourbaix}, {Randich}, {Sartoretti}, {Siddiqui},
  {Soubiran}, {van Leeuwen}, {Walton}, {Arenou}, {Bastian}, {Cropper},
  {Drimmel}, {Katz}, {Lattanzi}, {Bakker}, {Cacciari}, {Casta{\~n}eda},
  {Chaoul}, {Cheek}, {De Angeli}, {Fabricius}, {Guerra}, {Holl}, {Masana},
  {Messineo}, {Mowlavi}, {Nienartowicz}, {Panuzzo}, {Portell}, {Riello},
  {Seabroke}, {Tanga}, {Th{\'e}venin}, {Gracia-Abril}, {Comoretto},
  {Garcia-Reinaldos}, {Teyssier}, {Altmann}, {Andrae}, {Audard},
  {Bellas-Velidis}, {Benson}, {Berthier}, {Blomme}, {Burgess}, {Busso},
  {Carry}, {Cellino}, {Clementini}, {Clotet}, {Creevey}, {Davidson}, {De
  Ridder}, {Delchambre}, {Dell'Oro}, {Ducourant},
  {Fern{\'a}ndez-Hern{\'a}ndez}, {Fouesneau}, {Fr{\'e}mat}, {Galluccio},
  {Garc{\'\i}a-Torres}, {Gonz{\'a}lez-N{\'u}{\~n}ez}, {Gonz{\'a}lez-Vidal},
  {Gosset}, {Guy}, {Halbwachs}, {Hambly}, {Harrison}, {Hern{\'a}ndez},
  {Hestroffer}, {Hodgkin}, {Hutton}, {Jasniewicz}, {Jean-Antoine-Piccolo},
  {Jordan}, {Korn}, {Krone-Martins}, {Lanzafame}, {Lebzelter}, {L{\"o}ffler},
  {Manteiga}, {Marrese}, {Mart{\'\i}n-Fleitas}, {Moitinho}, {Mora}, {Muinonen},
  {Osinde}, {Pancino}, {Pauwels}, {Petit}, {Recio-Blanco}, {Richards},
  {Rimoldini}, {Robin}, {Sarro}, {Siopis}, {Smith}, {Sozzetti}, {S{\"u}veges},
  {Torra}, {van Reeven}, {Abbas}, {Abreu Aramburu}, {Accart}, {Aerts},
  {Altavilla}, {{\'A}lvarez}, {Alvarez}, {Alves}, {Anderson}, {Andrei},
  {Anglada Varela}, {Antiche}, {Antoja}, {Arcay}, {Astraatmadja}, {Bach},
  {Baker}, {Balaguer-N{\'u}{\~n}ez}, {Balm}, {Barache}, {Barata}, {Barbato},
  {Barblan}, {Barklem}, {Barrado}, {Barros}, {Barstow}, {Bartholom{\'e}
  Mu{\~n}oz}, {Bassilana}, {Becciani}, {Bellazzini}, {Berihuete}, {Bertone},
  {Bianchi}, {Bienaym{\'e}}, {Blanco-Cuaresma}, {Boch}, {Boeche}, {Bombrun},
  {Borrachero}, {Bossini}, {Bouquillon}, {Bourda}, {Bragaglia}, {Bramante},
  {Breddels}, {Bressan}, {Brouillet}, {Br{\"u}semeister}, {Brugaletta},
  {Bucciarelli}, {Burlacu}, {Busonero}, {Butkevich}, {Buzzi}, {Caffau},
  {Cancelliere}, {Cannizzaro}, {Cantat-Gaudin}, {Carballo}, {Carlucci},
  {Carrasco}, {Casamiquela}, {Castellani}, {Castro-Ginard}, {Charlot},
  {Chemin}, {Chiavassa}, {Cocozza}, {Costigan}, {Cowell}, {Crifo}, {Crosta},
  {Crowley}, {Cuypers}, {Dafonte}, {Damerdji}, {Dapergolas}, {David}, {David},
  {de Laverny}, {De Luise}, {De March}, {de Martino}, {de Souza}, {de Torres},
  {Debosscher}, {del Pozo}, {Delbo}, {Delgado}, {Delgado}, {Di Matteo},
  {Diakite}, {Diener}, {Distefano}, {Dolding}, {Drazinos}, {Dur{\'a}n},
  {Edvardsson}, {Enke}, {Eriksson}, {Esquej}, {Eynard Bontemps}, {Fabre},
  {Fabrizio}, {Faigler}, {Falc{\~a}o}, {Farr{\`a}s Casas}, {Federici},
  {Fedorets}, {Fernique}, {Figueras}, {Filippi}, {Findeisen}, {Fonti},
  {Fraile}, {Fraser}, {Fr{\'e}zouls}, {Gai}, {Galleti}, {Garabato},
  {Garc{\'\i}a-Sedano}, {Garofalo}, {Garralda}, {Gavel}, {Gavras}, {Gerssen},
  {Geyer}, {Giacobbe}, {Gilmore}, {Girona}, {Giuffrida}, {Glass}, {Gomes},
  {Granvik}, {Gueguen}, {Guerrier}, {Guiraud}, {Guti{\'e}rrez-S{\'a}nchez},
  {Haigron}, {Hatzidimitriou}, {Hauser}, {Haywood}, {Heiter}, {Helmi}, {Heu},
  {Hilger}, {Hobbs}, {Hofmann}, {Holland}, {Huckle}, {Hypki}, {Icardi},
  {Jan{\ss}en}, {Jevardat de Fombelle}, {Jonker}, {Juh{\'a}sz}, {Julbe},
  {Karampelas}, {Kewley}, {Klar}, {Kochoska}, {Kohley}, {Kolenberg},
  {Kontizas}, {Kontizas}, {Koposov}, {Kordopatis}, {Kostrzewa-Rutkowska},
  {Koubsky}, {Lambert}, {Lanza}, {Lasne}, {Lavigne}, {Le Fustec}, {Le
  Poncin-Lafitte}, {Lebreton}, {Leccia}, {Leclerc}, {Lecoeur-Taibi},
  {Lenhardt}, {Leroux}, {Liao}, {Licata}, {Lindstr{\o}m}, {Lister}, {Livanou},
  {Lobel}, {L{\'o}pez}, {Managau}, {Mann}, {Mantelet}, {Marchal}, {Marchant},
  {Marconi}, {Marinoni}, {Marschalk{\'o}}, {Marshall}, {Martino}, {Marton},
  {Mary}, {Massari}, {Matijevi{\v{c}}}, {Mazeh}, {McMillan}, {Messina},
  {Michalik}, {Millar}, {Molina}, {Molinaro}, {Moln{\'a}r}, {Montegriffo},
  {Mor}, {Morbidelli}, {Morel}, {Morris}, {Mulone}, {Muraveva}, {Musella},
  {Nelemans}, {Nicastro}, {Noval}, {O'Mullane}, {Ord{\'e}novic},
  {Ord{\'o}{\~n}ez-Blanco}, {Osborne}, {Pagani}, {Pagano}, {Pailler},
  {Palacin}, {Palaversa}, {Panahi}, {Pawlak}, {Piersimoni}, {Pineau}, {Plachy},
  {Plum}, {Poggio}, {Poujoulet}, {Pr{\v{s}}a}, {Pulone}, {Racero}, {Ragaini},
  {Rambaux}, {Ramos-Lerate}, {Regibo}, {Reyl{\'e}}, {Riclet}, {Ripepi}, {Riva},
  {Rivard}, {Rixon}, {Roegiers}, {Roelens}, {Romero-G{\'o}mez}, {Rowell},
  {Royer}, {Ruiz-Dern}, {Sadowski}, {Sagrist{\`a} Sell{\'e}s}, {Sahlmann},
  {Salgado}, {Salguero}, {Sanna}, {Santana-Ros}, {Sarasso}, {Savietto},
  {Schultheis}, {Sciacca}, {Segol}, {Segovia}, {S{\'e}gransan}, {Shih},
  {Siltala}, {Silva}, {Smart}, {Smith}, {Solano}, {Solitro}, {Sordo}, {Soria
  Nieto}, {Souchay}, {Spagna}, {Spoto}, {Stampa}, {Steele},
  {Steidelm{\"u}ller}, {Stephenson}, {Stoev}, {Suess}, {Surdej}, {Szabados},
  {Szegedi-Elek}, {Tapiador}, {Taris}, {Tauran}, {Taylor}, {Teixeira},
  {Terrett}, {Teyssand ier}, {Thuillot}, {Titarenko}, {Torra Clotet}, {Turon},
  {Ulla}, {Utrilla}, {Uzzi}, {Vaillant}, {Valentini}, {Valette}, {van Elteren},
  {Van Hemelryck}, {van Leeuwen}, {Vaschetto}, {Vecchiato}, {Veljanoski},
  {Viala}, {Vicente}, {Vogt}, {von Essen}, {Voss}, {Votruba}, {Voutsinas},
  {Walmsley}, {Weiler}, {Wertz}, {Wevers}, {Wyrzykowski}, {Yoldas},
  {{\v{Z}}erjal}, {Ziaeepour}, {Zorec}, {Zschocke}, {Zucker}, {Zurbach}, \&
  {Zwitter}}]{gaia18}
{Gaia Collaboration}, {Brown}, A.~G.~A., {Vallenari}, A., {et~al.} 2018, \aap,
  616, A1, \dodoi{10.1051/0004-6361/201833051}

\bibitem[{{Giguere} {et~al.}(2015){Giguere}, {Fischer}, {Payne}, {Brewer},
  {Johnson}, {Howard}, \& {Isaacson}}]{giguere15}
{Giguere}, M.~J., {Fischer}, D.~A., {Payne}, M.~J., {et~al.} 2015, \apj, 799,
  89, \dodoi{10.1088/0004-637X/799/1/89}

\bibitem[{{Ginsburg} {et~al.}(2019){Ginsburg}, {Sip{\H o}cz}, {Brasseur},
  {Cowperthwaite}, {Craig}, {Deil}, {Guillochon}, {Guzman}, {Liedtke}, {Lian
  Lim}, {Lockhart}, {Mommert}, {Morris}, {Norman}, {Parikh}, {Persson},
  {Robitaille}, {Segovia}, {Singer}, {Tollerud}, {de Val-Borro}, {Valtchanov},
  {Woillez}, {The Astroquery collaboration}, \& {a subset of the astropy
  collaboration}}]{ginsburg19}
{Ginsburg}, A., {Sip{\H o}cz}, B.~M., {Brasseur}, C.~E., {et~al.} 2019, \aj,
  157, 98, \dodoi{10.3847/1538-3881/aafc33}

\bibitem[{{Ginski} {et~al.}(2016){Ginski}, {Mugrauer}, {Seeliger}, {Buder},
  {Errmann}, {Avenhaus}, {Mouillet}, {Maire}, \& {Raetz}}]{ginski16}
{Ginski}, C., {Mugrauer}, M., {Seeliger}, M., {et~al.} 2016, \mnras, 457, 2173,
  \dodoi{10.1093/mnras/stw049}

\bibitem[{{Goldman} {et~al.}(2010){Goldman}, {Marsat}, {Henning}, {Clemens}, \&
  {Greiner}}]{goldman10}
{Goldman}, B., {Marsat}, S., {Henning}, T., {Clemens}, C., \& {Greiner}, J.
  2010, \mnras, 405, 1140, \dodoi{10.1111/j.1365-2966.2010.16524.x}

\bibitem[{{Gong} {et~al.}(2019){Gong}, {Liu}, {Cao}, {Chen}, {Fan}, {Li}, {Li},
  {Li}, {Zhang}, \& {Zhan}}]{gong19}
{Gong}, Y., {Liu}, X., {Cao}, Y., {et~al.} 2019, \apj, 883, 203,
  \dodoi{10.3847/1538-4357/ab391e}

\bibitem[{{Grieves} {et~al.}(2017){Grieves}, {Ge}, {Thomas}, {Ma}, {Sithajan},
  {Ghezzi}, {Kimock}, {Willis}, {De Lee}, {Lee}, {Fleming}, {Agol}, {Troup},
  {Paegert}, {Schneider}, {Stassun}, {Varosi}, {Zhao}, {Jian}, {Li}, {Porto de
  Mello}, {Bizyaev}, {Pan}, {Dutra-Ferreira}, {Lorenzo-Oliveira}, {Santiago},
  {da Costa}, {Maia}, {Ogando}, \& {del Peloso}}]{grieves17}
{Grieves}, N., {Ge}, J., {Thomas}, N., {et~al.} 2017, \mnras, 467, 4264,
  \dodoi{10.1093/mnras/stx334}

\bibitem[{{Griffin}(2012)}]{griffin12}
{Griffin}, R.~F. 2012, Journal of Astrophysics and Astronomy, 33, 29,
  \dodoi{10.1007/s12036-012-9137-5}

\bibitem[{Haario {et~al.}(2001)Haario, Saksman, \& Tamminen}]{haario01}
Haario, H., Saksman, E., \& Tamminen, J. 2001, Bernoulli, 223

\bibitem[{{Halbwachs} {et~al.}(2000){Halbwachs}, {Arenou}, {Mayor}, {Udry}, \&
  {Queloz}}]{halbwachs00}
{Halbwachs}, J.~L., {Arenou}, F., {Mayor}, M., {Udry}, S., \& {Queloz}, D.
  2000, \aap, 355, 581

\bibitem[{{Han} {et~al.}(2001){Han}, {Black}, \& {Gatewood}}]{han01}
{Han}, I., {Black}, D.~C., \& {Gatewood}, G. 2001, \apjl, 548, L57,
  \dodoi{10.1086/318927}

\bibitem[{{Harris} {et~al.}(2020){Harris}, {Millman}, {van der Walt},
  {Gommers}, {Virtanen}, {Cournapeau}, {Wieser}, {Taylor}, {Berg}, {Smith},
  {Kern}, {Picus}, {Hoyer}, {van Kerkwijk}, {Brett}, {Haldane}, {del R{\'\i}o},
  {Wiebe}, {Peterson}, {G{\'e}rard-Marchant}, {Sheppard}, {Reddy}, {Weckesser},
  {Abbasi}, {Gohlke}, \& {Oliphant}}]{harris20}
{Harris}, C.~R., {Millman}, K.~J., {van der Walt}, S.~J., {et~al.} 2020, \nat,
  585, 357, \dodoi{10.1038/s41586-020-2649-2}

\bibitem[{{Hartmann} {et~al.}(2010){Hartmann}, {Guenther}, \&
  {Hatzes}}]{hartmann10}
{Hartmann}, M., {Guenther}, E.~W., \& {Hatzes}, A.~P. 2010, \apj, 717, 348,
  \dodoi{10.1088/0004-637X/717/1/348}

\bibitem[{{Hatzes} {et~al.}(2022){Hatzes}, {Gandolfi}, {Korth}, {Rodler},
  {Sabotta}, {Esposito}, {Barragan}, {Livingston}, {Serrano}, {Luque}, {Smith},
  {Redfield}, {Persson}, {Paetzold}, {Palle}, {Nowak}, {Osborne}, {Narita},
  {Mathur}, {Lam}, {Kabath}, {Johnson}, {Guenther}, {Grziwa}, {Goffo},
  {Fridlund}, {Endl}, {Deeg}, {Csizmadia}, {Cochran}, {Gonzalez Cuesta},
  {Chaturvedi}, {Carleo}, {Cabrera}, {Beck}, \& {Albrecht}}]{hatzes22}
{Hatzes}, A.~P., {Gandolfi}, D., {Korth}, J., {et~al.} 2022, arXiv e-prints,
  arXiv:2203.01018.
\newblock \doarXiv{2203.01018}

\bibitem[{{Hayashi} \& {Nakano}(1963)}]{hayashi63}
{Hayashi}, C., \& {Nakano}, T. 1963, Progress of Theoretical Physics, 30, 460,
  \dodoi{10.1143/PTP.30.460}

\bibitem[{{Heintz}(1994)}]{heintz94}
{Heintz}, W.~D. 1994, \aj, 108, 2338, \dodoi{10.1086/117247}

\bibitem[{{Hinkel} {et~al.}(2019){Hinkel}, {Unterborn}, {Kane}, {Somers}, \&
  {Galvez}}]{hinkel19}
{Hinkel}, N.~R., {Unterborn}, C., {Kane}, S.~R., {Somers}, G., \& {Galvez}, R.
  2019, \apj, 880, 49, \dodoi{10.3847/1538-4357/ab27c0}

\bibitem[{{Howard} {et~al.}(2010){Howard}, {Marcy}, {Johnson}, {Fischer},
  {Wright}, {Isaacson}, {Valenti}, {Anderson}, {Lin}, \& {Ida}}]{howard10}
{Howard}, A.~W., {Marcy}, G.~W., {Johnson}, J.~A., {et~al.} 2010, Science, 330,
  653, \dodoi{10.1126/science.1194854}

\bibitem[{{Huang} {et~al.}(2018){Huang}, {Burt}, {Vanderburg}, {G{\"u}nther},
  {Shporer}, {Dittmann}, {Winn}, {Wittenmyer}, {Sha}, {Kane}, {Ricker},
  {Vanderspek}, {Latham}, {Seager}, {Jenkins}, {Caldwell}, {Collins},
  {Guerrero}, {Smith}, {Quinn}, {Udry}, {Pepe}, {Bouchy}, {S{\'e}gransan},
  {Lovis}, {Ehrenreich}, {Marmier}, {Mayor}, {Wohler}, {Haworth}, {Morgan},
  {Fausnaugh}, {Ciardi}, {Christiansen}, {Charbonneau}, {Dragomir}, {Deming},
  {Glidden}, {Levine}, {McCullough}, {Yu}, {Narita}, {Nguyen}, {Morton},
  {Pepper}, {P{\'a}l}, {Rodriguez}, {Stassun}, {Torres}, {Sozzetti}, {Doty},
  {Christensen-Dalsgaard}, {Laughlin}, {Clampin}, {Bean}, {Buchhave}, {Bakos},
  {Sato}, {Ida}, {Kaltenegger}, {Palle}, {Sasselov}, {Butler}, {Lissauer},
  {Ge}, \& {Rinehart}}]{huang18}
{Huang}, C.~X., {Burt}, J., {Vanderburg}, A., {et~al.} 2018, \apjl, 868, L39,
  \dodoi{10.3847/2041-8213/aaef91}

\bibitem[{{Jenkins} {et~al.}(2015){Jenkins}, {D{\'\i}az}, {Jones}, {Butler},
  {Tinney}, {O'Toole}, {Carter}, {Wittenmyer}, \& {Pinfield}}]{jenkins15}
{Jenkins}, J.~S., {D{\'\i}az}, M., {Jones}, H.~R.~A., {et~al.} 2015, \mnras,
  453, 1439, \dodoi{10.1093/mnras/stv1596}

\bibitem[{{Jenkins} {et~al.}(2017){Jenkins}, {Jones}, {Tuomi}, {D{\'\i}az},
  {Cordero}, {Aguayo}, {Pantoja}, {Arriagada}, {Mahu}, {Brahm}, {Rojo}, {Soto},
  {Ivanyuk}, {Becerra Yoma}, {Day-Jones}, {Ruiz}, {Pavlenko}, {Barnes},
  {Murgas}, {Pinfield}, {Jones}, {L{\'o}pez-Morales}, {Shectman}, {Butler}, \&
  {Minniti}}]{jenkins17}
{Jenkins}, J.~S., {Jones}, H.~R.~A., {Tuomi}, M., {et~al.} 2017, \mnras, 466,
  443, \dodoi{10.1093/mnras/stw2811}

\bibitem[{{Johnson-Groh} {et~al.}(2017{\natexlab{a}}){Johnson-Groh}, {Marois},
  {De Rosa}, {Nielsen}, {Rameau}, {Blunt}, {Vargas}, {Ammons}, {Bailey},
  {Barman}, {Bulger}, {Chilcote}, {Cotten}, {Doyon}, {Duch{\^e}ne},
  {Fitzgerald}, {Follette}, {Goodsell}, {Graham}, {Greenbaum}, {Hibon}, {Hung},
  {Ingraham}, {Kalas}, {Konopacky}, {Larkin}, {Macintosh}, {Maire}, {Marchis},
  {Marley}, {Metchev}, {Millar-Blanchaer}, {Oppenheimer}, {Palmer}, {Patience},
  {Perrin}, {Poyneer}, {Pueyo}, {Rajan}, {Rantakyr{\"o}}, {Savransky},
  {Schneider}, {Sivaramakrishnan}, {Song}, {Soummer}, {Thomas}, {Vega},
  {Wallace}, {Wang}, {Ward-Duong}, {Wiktorowicz}, \& {Wolff}}]{johnsongroh17}
{Johnson-Groh}, M., {Marois}, C., {De Rosa}, R.~J., {et~al.}
  2017{\natexlab{a}}, \aj, 153, 190, \dodoi{10.3847/1538-3881/aa6480}

\bibitem[{{Johnson-Groh} {et~al.}(2017{\natexlab{b}}){Johnson-Groh}, {Marois},
  {De Rosa}, {Nielsen}, {Rameau}, {Blunt}, {Vargas}, {Ammons}, {Bailey},
  {Barman}, {Bulger}, {Chilcote}, {Cotten}, {Doyon}, {Duch{\^e}ne},
  {Fitzgerald}, {Follette}, {Goodsell}, {Graham}, {Greenbaum}, {Hibon}, {Hung},
  {Ingraham}, {Kalas}, {Konopacky}, {Larkin}, {Macintosh}, {Maire}, {Marchis},
  {Marley}, {Metchev}, {Millar-Blanchaer}, {Oppenheimer}, {Palmer}, {Patience},
  {Perrin}, {Poyneer}, {Pueyo}, {Rajan}, {Rantakyr{\"o}}, {Savransky},
  {Schneider}, {Sivaramakrishnan}, {Song}, {Soummer}, {Thomas}, {Vega},
  {Wallace}, {Wang}, {Ward-Duong}, {Wiktorowicz}, \& {Wolff}}]{johnson17}
---. 2017{\natexlab{b}}, \aj, 153, 190, \dodoi{10.3847/1538-3881/aa6480}

\bibitem[{{Jones} {et~al.}(2002){Jones}, {Paul Butler}, {Tinney}, {Marcy},
  {Penny}, {McCarthy}, {Carter}, \& {Pourbaix}}]{jones02}
{Jones}, H. R.~A., {Paul Butler}, R., {Tinney}, C.~G., {et~al.} 2002, \mnras,
  333, 871, \dodoi{10.1046/j.1365-8711.2002.05459.x}

\bibitem[{{Jones} {et~al.}(2017){Jones}, {Brahm}, {Wittenmyer}, {Drass},
  {Jenkins}, {Melo}, {Vos}, \& {Rojo}}]{jones17}
{Jones}, M.~I., {Brahm}, R., {Wittenmyer}, R.~A., {et~al.} 2017, \aap, 602,
  A58, \dodoi{10.1051/0004-6361/201630278}

\bibitem[{{Jones} {et~al.}(2015{\natexlab{a}}){Jones}, {Jenkins}, {Rojo},
  {Melo}, \& {Bluhm}}]{jones15b}
{Jones}, M.~I., {Jenkins}, J.~S., {Rojo}, P., {Melo}, C.~H.~F., \& {Bluhm}, P.
  2015{\natexlab{a}}, \aap, 573, A3, \dodoi{10.1051/0004-6361/201424771}

\bibitem[{{Jones} {et~al.}(2015{\natexlab{b}}){Jones}, {Jenkins}, {Rojo},
  {Melo}, \& {Bluhm}}]{jones14}
---. 2015{\natexlab{b}}, \aap, 573, A3, \dodoi{10.1051/0004-6361/201424771}

\bibitem[{{Jones} {et~al.}(2015{\natexlab{c}}){Jones}, {Jenkins}, {Rojo},
  {Olivares}, \& {Melo}}]{jones15}
{Jones}, M.~I., {Jenkins}, J.~S., {Rojo}, P., {Olivares}, F., \& {Melo},
  C.~H.~F. 2015{\natexlab{c}}, \aap, 580, A14,
  \dodoi{10.1051/0004-6361/201525853}

\bibitem[{{Jones} {et~al.}(2016){Jones}, {Jenkins}, {Brahm}, {Wittenmyer},
  {Olivares E.}, {Melo}, {Rojo}, {Jord{\'a}n}, {Drass}, {Butler}, \&
  {Wang}}]{jones16}
{Jones}, M.~I., {Jenkins}, J.~S., {Brahm}, R., {et~al.} 2016, \aap, 590, A38,
  \dodoi{10.1051/0004-6361/201628067}

\bibitem[{{Jones} {et~al.}(2021){Jones}, {Wittenmyer}, {Aguilera-G{\'o}mez},
  {Soto}, {Torres}, {Trifonov}, {Jenkins}, {Zapata}, {Sarkis}, {Zakhozhay},
  {Brahm}, {Ram{\'\i}rez}, {Santana}, {Vines}, {D{\'\i}az},
  {Vu{\v{c}}kovi{\'c}}, \& {Pantoja}}]{jones21}
{Jones}, M.~I., {Wittenmyer}, R., {Aguilera-G{\'o}mez}, C., {et~al.} 2021,
  \aap, 646, A131, \dodoi{10.1051/0004-6361/202038555}

\bibitem[{{Jovanovic} {et~al.}(2015){Jovanovic}, {Martinache}, {Guyon},
  {Clergeon}, {Singh}, {Kudo}, {Garrel}, {Newman}, {Doughty}, {Lozi}, {Males},
  {Minowa}, {Hayano}, {Takato}, {Morino}, {Kuhn}, {Serabyn}, {Norris},
  {Tuthill}, {Schworer}, {Stewart}, {Close}, {Huby}, {Perrin}, {Lacour},
  {Gauchet}, {Vievard}, {Murakami}, {Oshiyama}, {Baba}, {Matsuo}, {Nishikawa},
  {Tamura}, {Lai}, {Marchis}, {Duchene}, {Kotani}, \& {Woillez}}]{jovanovic15}
{Jovanovic}, N., {Martinache}, F., {Guyon}, O., {et~al.} 2015, \pasp, 127, 890,
  \dodoi{10.1086/682989}

\bibitem[{{Kane} {et~al.}(2019){Kane}, {Dalba}, {Li}, {Horch}, {Hirsch},
  {Horner}, {Wittenmyer}, {Howell}, {Everett}, {Butler}, {Tinney}, {Carter},
  {Wright}, {Jones}, {Bailey}, \& {O'Toole}}]{kane19}
{Kane}, S.~R., {Dalba}, P.~A., {Li}, Z., {et~al.} 2019, \aj, 157, 252,
  \dodoi{10.3847/1538-3881/ab1ddf}

\bibitem[{Kass \& Raftery(1995)}]{kass95}
Kass, R.~E., \& Raftery, A.~E. 1995, Journal of the american statistical
  association, 90, 773

\bibitem[{{Kervella} {et~al.}(2019){Kervella}, {Arenou}, {Mignard}, \&
  {Th{\'e}venin}}]{kervella19}
{Kervella}, P., {Arenou}, F., {Mignard}, F., \& {Th{\'e}venin}, F. 2019, \aap,
  623, A72, \dodoi{10.1051/0004-6361/201834371}

\bibitem[{{Kervella} {et~al.}(2022){Kervella}, {Arenou}, \&
  {Th{\'e}venin}}]{kervella22}
{Kervella}, P., {Arenou}, F., \& {Th{\'e}venin}, F. 2022, \aap, 657, A7,
  \dodoi{10.1051/0004-6361/202142146}

\bibitem[{{Kiefer} {et~al.}(2021){Kiefer}, {H{\'e}brard}, {Lecavelier des
  Etangs}, {Martioli}, {Dalal}, \& {Vidal-Madjar}}]{kiefer21}
{Kiefer}, F., {H{\'e}brard}, G., {Lecavelier des Etangs}, A., {et~al.} 2021,
  \aap, 645, A7, \dodoi{10.1051/0004-6361/202039168}

\bibitem[{{Kiefer} {et~al.}(2019){Kiefer}, {H{\'e}brard}, {Sahlmann}, {Sousa},
  {Forveille}, {Santos}, {Mayor}, {Deleuil}, {Wilson}, {Dalal}, {D{\'\i}az},
  {Henry}, {Hagelberg}, {Hobson}, {Demangeon}, {Bourrier}, {Delfosse},
  {Arnold}, {Astudillo-Defru}, {Beuzit}, {Boisse}, {Bonfils}, {Borgniet},
  {Bouchy}, {Courcol}, {Ehrenreich}, {Hara}, {Lagrange}, {Lovis}, {Montagnier},
  {Moutou}, {Pepe}, {Perrier}, {Rey}, {Santerne}, {S{\'e}gransan}, {Udry}, \&
  {Vidal-Madjar}}]{kiefer19}
{Kiefer}, F., {H{\'e}brard}, G., {Sahlmann}, J., {et~al.} 2019, \aap, 631,
  A125, \dodoi{10.1051/0004-6361/201935113}

\bibitem[{{Kirkpatrick} {et~al.}(2021){Kirkpatrick}, {Gelino}, {Faherty},
  {Meisner}, {Caselden}, {Schneider}, {Marocco}, {Cayago}, {Smart},
  {Eisenhardt}, {Kuchner}, {Wright}, {Cushing}, {Allers}, {Bardalez Gagliuffi},
  {Burgasser}, {Gagn{\'e}}, {Logsdon}, {Martin}, {Ingalls}, {Lowrance},
  {Abrahams}, {Aganze}, {Gerasimov}, {Gonzales}, {Hsu}, {Kamraj}, {Kiman},
  {Rees}, {Theissen}, {Ammar}, {Andersen}, {Beaulieu}, {Colin}, {Elachi},
  {Goodman}, {Gramaize}, {Hamlet}, {Hong}, {Jonkeren}, {Khalil}, {Martin},
  {Pendrill}, {Pumphrey}, {Rothermich}, {Sainio}, {Stenner}, {Tanner},
  {Th{\'e}venot}, {Voloshin}, {Walla}, {W{\k{e}}dracki}, \& {Backyard Worlds:
  Planet 9 Collaboration}}]{kirkpatrick21}
{Kirkpatrick}, J.~D., {Gelino}, C.~R., {Faherty}, J.~K., {et~al.} 2021, \apjs,
  253, 7, \dodoi{10.3847/1538-4365/abd107}

\bibitem[{{Kopparapu} {et~al.}(2013){Kopparapu}, {Ramirez}, {Kasting}, {Eymet},
  {Robinson}, {Mahadevan}, {Terrien}, {Domagal-Goldman}, {Meadows}, \&
  {Deshpande}}]{kopparapu13}
{Kopparapu}, R.~K., {Ramirez}, R., {Kasting}, J.~F., {et~al.} 2013, \apj, 765,
  131, \dodoi{10.1088/0004-637X/765/2/131}

\bibitem[{{Kumar}(1963)}]{kumar63a}
{Kumar}, S.~S. 1963, \apj, 137, 1121, \dodoi{10.1086/147589}

\bibitem[{{Kunovac Hod{\v{z}}i{\'c}} {et~al.}(2021){Kunovac Hod{\v{z}}i{\'c}},
  {Triaud}, {Cegla}, {Chaplin}, \& {Davies}}]{kunovac21}
{Kunovac Hod{\v{z}}i{\'c}}, V., {Triaud}, A. H.~M.~J., {Cegla}, H.~M.,
  {Chaplin}, W.~J., \& {Davies}, G.~R. 2021, \mnras, 502, 2893,
  \dodoi{10.1093/mnras/stab237}

\bibitem[{{K{\"u}rster} {et~al.}(2015){K{\"u}rster}, {Trifonov}, {Reffert},
  {Kostogryz}, \& {Rodler}}]{kurster15}
{K{\"u}rster}, M., {Trifonov}, T., {Reffert}, S., {Kostogryz}, N.~M., \&
  {Rodler}, F. 2015, \aap, 577, A103, \dodoi{10.1051/0004-6361/201525872}

\bibitem[{{Lacour} {et~al.}(2021){Lacour}, {Wang}, {Rodet}, {Nowak},
  {Shangguan}, {Beust}, {Lagrange}, {Abuter}, {Amorim}, {Asensio-Torres},
  {Benisty}, {Berger}, {Blunt}, {Boccaletti}, {Bohn}, {Bolzer}, {Bonnefoy},
  {Bonnet}, {Bourdarot}, {Brandner}, {Cantalloube}, {Caselli}, {Charnay},
  {Chauvin}, {Choquet}, {Christiaens}, {Cl{\'e}net}, {Coud{\'e} Du Foresto},
  {Cridland}, {Dembet}, {Dexter}, {de Zeeuw}, {Drescher}, {Duvert}, {Eckart},
  {Eisenhauer}, {Gao}, {Garcia}, {Garcia Lopez}, {Gendron}, {Genzel},
  {Gillessen}, {Girard}, {Haubois}, {Hei{\ss}el}, {Henning}, {Hinkley},
  {Hippler}, {Horrobin}, {Houll{\'e}}, {Hubert}, {Jocou}, {Kammerer},
  {Keppler}, {Kervella}, {Kreidberg}, {Lapeyr{\`e}re}, {Le Bouquin},
  {L{\'e}na}, {Lutz}, {Maire}, {M{\'e}rand}, {Molli{\`e}re}, {Monnier},
  {Mouillet}, {Nasedkin}, {Ott}, {Otten}, {Paladini}, {Paumard}, {Perraut},
  {Perrin}, {Pfuhl}, {Rickman}, {Pueyo}, {Rameau}, {Rousset}, {Rustamkulov},
  {Samland}, {Shimizu}, {Sing}, {Stadler}, {Stolker}, {Straub}, {Straubmeier},
  {Sturm}, {Tacconi}, {van Dishoeck}, {Vigan}, {Vincent}, {von Fellenberg},
  {Ward-Duong}, {Widmann}, {Wieprecht}, {Wiezorrek}, {Woillez}, {Yazici},
  {Young}, \& {Gravity Collaboration}}]{lacour21}
{Lacour}, S., {Wang}, J.~J., {Rodet}, L., {et~al.} 2021, \aap, 654, L2,
  \dodoi{10.1051/0004-6361/202141889}

\bibitem[{{Lagrange} {et~al.}(2009){Lagrange}, {Gratadour}, {Chauvin}, {Fusco},
  {Ehrenreich}, {Mouillet}, {Rousset}, {Rouan}, {Allard}, {Gendron}, {Charton},
  {Mugnier}, {Rabou}, {Montri}, \& {Lacombe}}]{lagrange09}
{Lagrange}, A.~M., {Gratadour}, D., {Chauvin}, G., {et~al.} 2009, \aap, 493,
  L21, \dodoi{10.1051/0004-6361:200811325}

\bibitem[{{Lagrange} {et~al.}(2019){Lagrange}, {Meunier}, {Rubini}, {Keppler},
  {Galland}, {Chapellier}, {Michel}, {Balona}, {Beust}, {Guillot}, {Grandjean},
  {Borgniet}, {M{\'e}karnia}, {Wilson}, {Kiefer}, {Bonnefoy}, {Lillo-Box},
  {Pantoja}, {Jones}, {Iglesias}, {Rodet}, {Diaz}, {Zapata}, {Abe}, \&
  {Schmider}}]{lagrange19}
{Lagrange}, A.~M., {Meunier}, N., {Rubini}, P., {et~al.} 2019, Nature
  Astronomy, 3, 1135, \dodoi{10.1038/s41550-019-0857-1}

\bibitem[{{Lagrange} {et~al.}(2020){Lagrange}, {Rubini}, {Nowak}, {Lacour},
  {Grandjean}, {Boccaletti}, {Langlois}, {Delorme}, {Gratton}, {Wang},
  {Flasseur}, {Galicher}, {Kral}, {Meunier}, {Beust}, {Babusiaux}, {Le
  Coroller}, {Thebault}, {Kervella}, {Zurlo}, {Maire}, {Wahhaj}, {Amorim},
  {Asensio-Torres}, {Benisty}, {Berger}, {Bonnefoy}, {Brandner}, {Cantalloube},
  {Charnay}, {Chauvin}, {Choquet}, {Cl{\'e}net}, {Christiaens}, {Coud{\'e} Du
  Foresto}, {de Zeeuw}, {Desidera}, {Duvert}, {Eckart}, {Eisenhauer},
  {Galland}, {Gao}, {Garcia}, {Garcia Lopez}, {Gendron}, {Genzel}, {Gillessen},
  {Girard}, {Hagelberg}, {Haubois}, {Henning}, {Heissel}, {Hippler},
  {Horrobin}, {Janson}, {Kammerer}, {Kenworthy}, {Keppler}, {Kreidberg},
  {Lapeyr{\`e}re}, {Le Bouquin}, {L{\'e}na}, {M{\'e}rand}, {Messina},
  {Molli{\`e}re}, {Monnier}, {Ott}, {Otten}, {Paumard}, {Paladini}, {Perraut},
  {Perrin}, {Pueyo}, {Pfuhl}, {Rodet}, {Rodriguez-Coira}, {Rousset}, {Samland},
  {Shangguan}, {Schmidt}, {Straub}, {Straubmeier}, {Stolker}, {Vigan},
  {Vincent}, {Widmann}, {Woillez}, \& {Gravity Collaboration}}]{lagrange20}
{Lagrange}, A.~M., {Rubini}, P., {Nowak}, M., {et~al.} 2020, \aap, 642, A18,
  \dodoi{10.1051/0004-6361/202038823}

\bibitem[{{Li} {et~al.}(2021){Li}, {Brandt}, {Brandt}, {Dupuy}, {Michalik},
  {Jensen-Clem}, {Zeng}, {Faherty}, \& {Mitra}}]{li21}
{Li}, Y., {Brandt}, T.~D., {Brandt}, G.~M., {et~al.} 2021, \aj, 162, 266,
  \dodoi{10.3847/1538-3881/ac27ab}

\bibitem[{{Lo Curto} {et~al.}(2010){Lo Curto}, {Mayor}, {Benz}, {Bouchy},
  {Lovis}, {Moutou}, {Naef}, {Pepe}, {Queloz}, {Santos}, {Segransan}, \&
  {Udry}}]{locurto10}
{Lo Curto}, G., {Mayor}, M., {Benz}, W., {et~al.} 2010, \aap, 512, A48,
  \dodoi{10.1051/0004-6361/200913523}

\bibitem[{{Lo Curto} {et~al.}(2015){Lo Curto}, {Pepe}, {Avila}, {Boffin},
  {Bovay}, {Chazelas}, {Coffinet}, {Fleury}, {Hughes}, {Lovis}, {Maire},
  {Manescau}, {Pasquini}, {Rihs}, {Sinclaire}, \& {Udry}}]{curto15}
{Lo Curto}, G., {Pepe}, F., {Avila}, G., {et~al.} 2015, The Messenger, 162, 9

\bibitem[{{Luhman}(2007)}]{luhman07}
{Luhman}, K.~L. 2007, \apjs, 173, 104, \dodoi{10.1086/520114}

\bibitem[{{Luhn} {et~al.}(2019){Luhn}, {Bastien}, {Wright}, {Johnson},
  {Howard}, \& {Isaacson}}]{luhn19}
{Luhn}, J.~K., {Bastien}, F.~A., {Wright}, J.~T., {et~al.} 2019, \aj, 157, 149,
  \dodoi{10.3847/1538-3881/aaf5d0}

\bibitem[{{Ma} \& {Ge}(2014)}]{ma14}
{Ma}, B., \& {Ge}, J. 2014, \mnras, 439, 2781, \dodoi{10.1093/mnras/stu134}

\bibitem[{{Mann} {et~al.}(2019){Mann}, {Dupuy}, {Kraus}, {Gaidos}, {Ansdell},
  {Ireland}, {Rizzuto}, {Hung}, {Dittmann}, {Factor}, {Feiden}, {Martinez},
  {Ru{\'\i}z-Rodr{\'\i}guez}, \& {Thao}}]{mann19}
{Mann}, A.~W., {Dupuy}, T., {Kraus}, A.~L., {et~al.} 2019, \apj, 871, 63,
  \dodoi{10.3847/1538-4357/aaf3bc}

\bibitem[{{Marcy} \& {Benitz}(1989)}]{marcy89}
{Marcy}, G.~W., \& {Benitz}, K.~J. 1989, \apj, 344, 441, \dodoi{10.1086/167812}

\bibitem[{{Marcy} \& {Butler}(1995)}]{marcy95}
{Marcy}, G.~W., \& {Butler}, R.~P. 1995, in The Bottom of the Main Sequence -
  and Beyond, ed. C.~G. {Tinney}, 98

\bibitem[{{Marcy} \& {Butler}(2000)}]{marcy00}
{Marcy}, G.~W., \& {Butler}, R.~P. 2000, \pasp, 112, 137,
  \dodoi{10.1086/316516}

\bibitem[{{Marcy} {et~al.}(1999){Marcy}, {Butler}, {Vogt}, {Fischer}, \&
  {Liu}}]{marcy99}
{Marcy}, G.~W., {Butler}, R.~P., {Vogt}, S.~S., {Fischer}, D., \& {Liu}, M.~C.
  1999, \apj, 520, 239, \dodoi{10.1086/307451}

\bibitem[{{Marcy} {et~al.}(2005){Marcy}, {Butler}, {Vogt}, {Fischer}, {Henry},
  {Laughlin}, {Wright}, \& {Johnson}}]{marcy05}
{Marcy}, G.~W., {Butler}, R.~P., {Vogt}, S.~S., {et~al.} 2005, \apj, 619, 570,
  \dodoi{10.1086/426384}

\bibitem[{{Marley} {et~al.}(2021){Marley}, {Saumon}, {Visscher}, {Lupu},
  {Freedman}, {Morley}, {Fortney}, {Seay}, {Smith}, {Teal}, \&
  {Wang}}]{marley21}
{Marley}, M.~S., {Saumon}, D., {Visscher}, C., {et~al.} 2021, \apj, 920, 85,
  \dodoi{10.3847/1538-4357/ac141d}

\bibitem[{{Marmier} {et~al.}(2013){Marmier}, {S{\'e}gransan}, {Udry}, {Mayor},
  {Pepe}, {Queloz}, {Lovis}, {Naef}, {Santos}, {Alonso}, {Alves}, {Berthet},
  {Chazelas}, {Demory}, {Dumusque}, {Eggenberger}, {Figueira}, {Gillon},
  {Hagelberg}, {Lendl}, {Mardling}, {M{\'e}gevand}, {Neveu}, {Sahlmann},
  {Sosnowska}, {Tewes}, \& {Triaud}}]{marmier13}
{Marmier}, M., {S{\'e}gransan}, D., {Udry}, S., {et~al.} 2013, \aap, 551, A90,
  \dodoi{10.1051/0004-6361/201219639}

\bibitem[{{Martioli} {et~al.}(2010){Martioli}, {McArthur}, {Benedict}, {Bean},
  {Harrison}, \& {Armstrong}}]{martioli10}
{Martioli}, E., {McArthur}, B.~E., {Benedict}, G.~F., {et~al.} 2010, \apj, 708,
  625, \dodoi{10.1088/0004-637X/708/1/625}

\bibitem[{{Mason} {et~al.}(2001){Mason}, {Wycoff}, {Hartkopf}, {Douglass}, \&
  {Worley}}]{mason01}
{Mason}, B.~D., {Wycoff}, G.~L., {Hartkopf}, W.~I., {Douglass}, G.~G., \&
  {Worley}, C.~E. 2001, \aj, 122, 3466, \dodoi{10.1086/323920}

\bibitem[{{Mayor} {et~al.}(2004){Mayor}, {Udry}, {Naef}, {Pepe}, {Queloz},
  {Santos}, \& {Burnet}}]{mayor04}
{Mayor}, M., {Udry}, S., {Naef}, D., {et~al.} 2004, \aap, 415, 391,
  \dodoi{10.1051/0004-6361:20034250}

\bibitem[{{Mayor} {et~al.}(2011){Mayor}, {Marmier}, {Lovis}, {Udry},
  {S{\'e}gransan}, {Pepe}, {Benz}, {Bertaux}, {Bouchy}, {Dumusque}, {Lo Curto},
  {Mordasini}, {Queloz}, \& {Santos}}]{mayor11}
{Mayor}, M., {Marmier}, M., {Lovis}, C., {et~al.} 2011, ArXiv e-prints.
\newblock \doarXiv{1109.2497}

\bibitem[{{Melo} {et~al.}(2007){Melo}, {Santos}, {Gieren}, {Pietrzynski},
  {Ruiz}, {Sousa}, {Bouchy}, {Lovis}, {Mayor}, {Pepe}, {Queloz}, {da Silva}, \&
  {Udry}}]{melo07}
{Melo}, C., {Santos}, N.~C., {Gieren}, W., {et~al.} 2007, \aap, 467, 721,
  \dodoi{10.1051/0004-6361:20066845}

\bibitem[{{Ment} {et~al.}(2018){Ment}, {Fischer}, {Bakos}, {Howard}, \&
  {Isaacson}}]{ment18}
{Ment}, K., {Fischer}, D.~A., {Bakos}, G., {Howard}, A.~W., \& {Isaacson}, H.
  2018, \aj, 156, 213, \dodoi{10.3847/1538-3881/aae1f5}

\bibitem[{{Metchev} {et~al.}(2009){Metchev}, {Marois}, \&
  {Zuckerman}}]{metchev09}
{Metchev}, S., {Marois}, C., \& {Zuckerman}, B. 2009, \apjl, 705, L204,
  \dodoi{10.1088/0004-637X/705/2/L204}

\bibitem[{{Michalik} {et~al.}(2015){Michalik}, {Lindegren}, \&
  {Hobbs}}]{michalik15}
{Michalik}, D., {Lindegren}, L., \& {Hobbs}, D. 2015, \aap, 574, A115,
  \dodoi{10.1051/0004-6361/201425310}

\bibitem[{{Minniti} {et~al.}(2009){Minniti}, {Butler}, {L{\'o}pez-Morales},
  {Shectman}, {Adams}, {Arriagada}, {Boss}, \& {Chambers}}]{minniti09}
{Minniti}, D., {Butler}, R.~P., {L{\'o}pez-Morales}, M., {et~al.} 2009, \apj,
  693, 1424, \dodoi{10.1088/0004-637X/693/2/1424}

\bibitem[{{Moe} \& {Kratter}(2021)}]{moe21}
{Moe}, M., \& {Kratter}, K.~M. 2021, \mnras, 507, 3593,
  \dodoi{10.1093/mnras/stab2328}

\bibitem[{{Moutou} {et~al.}(2017){Moutou}, {Vigan}, {Mesa}, {Desidera},
  {Th{\'e}bault}, {Zurlo}, \& {Salter}}]{moutou17}
{Moutou}, C., {Vigan}, A., {Mesa}, D., {et~al.} 2017, \aap, 602, A87,
  \dodoi{10.1051/0004-6361/201630173}

\bibitem[{{Moutou} {et~al.}(2005){Moutou}, {Mayor}, {Bouchy}, {Lovis}, {Pepe},
  {Queloz}, {Santos}, {Udry}, {Benz}, {Lo Curto}, {Naef}, {S{\'e}gransan}, \&
  {Sivan}}]{moutou05}
{Moutou}, C., {Mayor}, M., {Bouchy}, F., {et~al.} 2005, \aap, 439, 367,
  \dodoi{10.1051/0004-6361:20052826}

\bibitem[{{Moutou} {et~al.}(2009){Moutou}, {Mayor}, {Lo Curto}, {Udry},
  {Bouchy}, {Benz}, {Lovis}, {Naef}, {Pepe}, {Queloz}, \& {Santos}}]{moutou09}
{Moutou}, C., {Mayor}, M., {Lo Curto}, G., {et~al.} 2009, \aap, 496, 513,
  \dodoi{10.1051/0004-6361:200810941}

\bibitem[{{Moutou} {et~al.}(2011){Moutou}, {Mayor}, {Lo Curto},
  {S{\'e}gransan}, {Udry}, {Bouchy}, {Benz}, {Lovis}, {Naef}, {Pepe}, {Queloz},
  {Santos}, \& {Sousa}}]{moutou11}
---. 2011, \aap, 527, A63, \dodoi{10.1051/0004-6361/201015371}

\bibitem[{{Mugrauer} {et~al.}(2014){Mugrauer}, {Ginski}, \&
  {Seeliger}}]{mugrauer14}
{Mugrauer}, M., {Ginski}, C., \& {Seeliger}, M. 2014, \mnras, 439, 1063,
  \dodoi{10.1093/mnras/stu044}

\bibitem[{{Naef} {et~al.}(2004){Naef}, {Mayor}, {Beuzit}, {Perrier}, {Queloz},
  {Sivan}, \& {Udry}}]{naef04}
{Naef}, D., {Mayor}, M., {Beuzit}, J.~L., {et~al.} 2004, \aap, 414, 351,
  \dodoi{10.1051/0004-6361:20034091}

\bibitem[{{Naef} {et~al.}(2001){Naef}, {Mayor}, {Pepe}, {Queloz}, {Santos},
  {Udry}, \& {Burnet}}]{naef01}
{Naef}, D., {Mayor}, M., {Pepe}, F., {et~al.} 2001, \aap, 375, 205,
  \dodoi{10.1051/0004-6361:20010841}

\bibitem[{{Naef} {et~al.}(2010){Naef}, {Mayor}, {Lo Curto}, {Bouchy}, {Lovis},
  {Moutou}, {Benz}, {Pepe}, {Queloz}, {Santos}, {S{\'e}gransan}, {Udry},
  {Bonfils}, {Delfosse}, {Forveille}, {H{\'e}brard}, {Mordasini}, {Perrier},
  {Boisse}, \& {Sosnowska}}]{naef10}
{Naef}, D., {Mayor}, M., {Lo Curto}, G., {et~al.} 2010, \aap, 523, A15,
  \dodoi{10.1051/0004-6361/200913616}

\bibitem[{{Nakajima} {et~al.}(1995){Nakajima}, {Oppenheimer}, {Kulkarni},
  {Golimowski}, {Matthews}, \& {Durrance}}]{nakajima95}
{Nakajima}, T., {Oppenheimer}, B.~R., {Kulkarni}, S.~R., {et~al.} 1995, \nat,
  378, 463, \dodoi{10.1038/378463a0}

\bibitem[{{Nielsen} {et~al.}(2014){Nielsen}, {Liu}, {Wahhaj}, {Biller},
  {Hayward}, {Males}, {Close}, {Morzinski}, {Skemer}, {Kuchner}, {Rodigas},
  {Hinz}, {Chun}, {Ftaclas}, \& {Toomey}}]{nielsen14}
{Nielsen}, E.~L., {Liu}, M.~C., {Wahhaj}, Z., {et~al.} 2014, \apj, 794, 158,
  \dodoi{10.1088/0004-637X/794/2/158}

\bibitem[{{Nielsen} {et~al.}(2020){Nielsen}, {De Rosa}, {Wang}, {Sahlmann},
  {Kalas}, {Duch{\^e}ne}, {Rameau}, {Marley}, {Saumon}, {Macintosh},
  {Millar-Blanchaer}, {Nguyen}, {Ammons}, {Bailey}, {Barman}, {Bulger},
  {Chilcote}, {Cotten}, {Doyon}, {Esposito}, {Fitzgerald}, {Follette},
  {Gerard}, {Goodsell}, {Graham}, {Greenbaum}, {Hibon}, {Hung}, {Ingraham},
  {Konopacky}, {Larkin}, {Maire}, {Marchis}, {Marois}, {Metchev},
  {Oppenheimer}, {Palmer}, {Patience}, {Perrin}, {Poyneer}, {Pueyo}, {Rajan},
  {Rantakyr{\"o}}, {Ruffio}, {Savransky}, {Schneider}, {Sivaramakrishnan},
  {Song}, {Soummer}, {Thomas}, {Wallace}, {Ward-Duong}, {Wiktorowicz}, \&
  {Wolff}}]{nielsen20}
{Nielsen}, E.~L., {De Rosa}, R.~J., {Wang}, J.~J., {et~al.} 2020, \aj, 159, 71,
  \dodoi{10.3847/1538-3881/ab5b92}

\bibitem[{{Noguchi} {et~al.}(2002){Noguchi}, {Aoki}, {Kawanomoto}, {Ando},
  {Honda}, {Izumiura}, {Kambe}, {Okita}, {Sadakane}, {Sato}, {Tajitsu},
  {Takada-Hidai}, {Tanaka}, {Watanabe}, \& {Yoshida}}]{noguchi02}
{Noguchi}, K., {Aoki}, W., {Kawanomoto}, S., {et~al.} 2002, \pasj, 54, 855,
  \dodoi{10.1093/pasj/54.6.855}

\bibitem[{{Nowak} {et~al.}(2020){Nowak}, {Lacour}, {Lagrange}, {Rubini},
  {Wang}, {Stolker}, {Abuter}, {Amorim}, {Asensio-Torres}, {Baub{\"o}ck},
  {Benisty}, {Berger}, {Beust}, {Blunt}, {Boccaletti}, {Bonnefoy}, {Bonnet},
  {Brandner}, {Cantalloube}, {Charnay}, {Choquet}, {Christiaens}, {Cl{\'e}net},
  {Coud{\'e} Du Foresto}, {Cridland}, {de Zeeuw}, {Dembet}, {Dexter},
  {Drescher}, {Duvert}, {Eckart}, {Eisenhauer}, {Gao}, {Garcia}, {Garcia
  Lopez}, {Gardner}, {Gendron}, {Genzel}, {Gillessen}, {Girard}, {Grandjean},
  {Haubois}, {Hei{\ss}el}, {Henning}, {Hinkley}, {Hippler}, {Horrobin},
  {Houll{\'e}}, {Hubert}, {Jim{\'e}nez-Rosales}, {Jocou}, {Kammerer},
  {Kervella}, {Keppler}, {Kreidberg}, {Kulikauskas}, {Lapeyr{\`e}re}, {Le
  Bouquin}, {L{\'e}na}, {M{\'e}rand}, {Maire}, {Molli{\`e}re}, {Monnier},
  {Mouillet}, {M{\"u}ller}, {Nasedkin}, {Ott}, {Otten}, {Paumard}, {Paladini},
  {Perraut}, {Perrin}, {Pueyo}, {Pfuhl}, {Rameau}, {Rodet},
  {Rodr{\'\i}guez-Coira}, {Rousset}, {Scheithauer}, {Shangguan}, {Stadler},
  {Straub}, {Straubmeier}, {Sturm}, {Tacconi}, {van Dishoeck}, {Vigan},
  {Vincent}, {von Fellenberg}, {Ward-Duong}, {Widmann}, {Wieprecht},
  {Wiezorrek}, {Woillez}, \& {Gravity Collaboration}}]{nowak20}
{Nowak}, M., {Lacour}, S., {Lagrange}, A.~M., {et~al.} 2020, \aap, 642, L2,
  \dodoi{10.1051/0004-6361/202039039}

\bibitem[{Nychka {et~al.}(2018)Nychka, Furrer, Paige, Sain, \& Nychka}]{fields}
Nychka, D., Furrer, R., Paige, J., Sain, S., \& Nychka, M.~D. 2018

\bibitem[{Nystrom {et~al.}(2015)Nystrom, Levine, Roskies, \& Scott}]{xsede}
Nystrom, N.~A., Levine, M.~J., Roskies, R.~Z., \& Scott, J.~R. 2015, in
  Proceedings of the 2015 XSEDE Conference: Scientific Advancements Enabled by
  Enhanced Cyberinfrastructure, XSEDE '15 (New York, NY, USA: ACM), 30:1--30:8.
\newblock \url{http://doi.acm.org/10.1145/2792745.2792775}

\bibitem[{{Patel} {et~al.}(2007){Patel}, {Vogt}, {Marcy}, {Johnson}, {Fischer},
  {Wright}, \& {Butler}}]{patel07}
{Patel}, S.~G., {Vogt}, S.~S., {Marcy}, G.~W., {et~al.} 2007, \apj, 665, 744,
  \dodoi{10.1086/519066}

\bibitem[{{Pepe} {et~al.}(2000){Pepe}, {Mayor}, {Delabre}, {Kohler}, {Lacroix},
  {Queloz}, {Udry}, {Benz}, {Bertaux}, \& {Sivan}}]{pepe00}
{Pepe}, F., {Mayor}, M., {Delabre}, B., {et~al.} 2000, in Society of
  Photo-Optical Instrumentation Engineers (SPIE) Conference Series, Vol. 4008,
  Optical and IR Telescope Instrumentation and Detectors, ed. M.~{Iye} \& A.~F.
  {Moorwood}, 582--592

\bibitem[{{Pepe} {et~al.}(2010){Pepe}, {Cristiani}, {Rebolo Lopez}, {Santos},
  {Amorim}, {Avila}, {Benz}, {Bonifacio}, {Cabral}, {Carvas}, {Cirami},
  {Coelho}, {Comari}, {Coretti}, {De Caprio}, {Dekker}, {Delabre}, {Di
  Marcantonio}, {D'Odorico}, {Fleury}, {Garc{\'{\i}}a}, {Herreros Linares},
  {Hughes}, {Iwert}, {Lima}, {Lizon}, {Lo Curto}, {Lovis}, {Manescau},
  {Martins}, {M{\'e}gevand}, {Moitinho}, {Molaro}, {Monteiro}, {Monteiro},
  {Pasquini}, {Mordasini}, {Queloz}, {Rasilla}, {Rebord{\~a}o}, {Santana
  Tschudi}, {Santin}, {Sosnowska}, {Span{\`o}}, {Tenegi}, {Udry}, {Vanzella},
  {Viel}, {Zapatero Osorio}, \& {Zerbi}}]{pepe10}
{Pepe}, F.~A., {Cristiani}, S., {Rebolo Lopez}, R., {et~al.} 2010, in
  \procspie, Vol. 7735, Ground-based and Airborne Instrumentation for Astronomy
  III, 77350F

\bibitem[{{Perrier} {et~al.}(2003){Perrier}, {Sivan}, {Naef}, {Beuzit},
  {Mayor}, {Queloz}, \& {Udry}}]{perrier03}
{Perrier}, C., {Sivan}, J.~P., {Naef}, D., {et~al.} 2003, \aap, 410, 1039,
  \dodoi{10.1051/0004-6361:20031340}

\bibitem[{{Perruchot} {et~al.}(2008){Perruchot}, {Kohler}, {Bouchy}, {Richaud},
  {Richaud}, {Moreaux}, {Merzougui}, {Sottile}, {Hill}, {Knispel}, {Regal},
  {Meunier}, {Ilovaisky}, {Le Coroller}, {Gillet}, {Schmitt}, {Pepe}, {Fleury},
  {Sosnowska}, {Vors}, {M{\'e}gevand}, {Blanc}, {Carol}, {Point}, {Laloge}, \&
  {Brunel}}]{perruchot08}
{Perruchot}, S., {Kohler}, D., {Bouchy}, F., {et~al.} 2008, Society of
  Photo-Optical Instrumentation Engineers (SPIE) Conference Series, Vol. 7014,
  {The SOPHIE spectrograph: design and technical key-points for high throughput
  and high stability}, 70140J

\bibitem[{{Perryman} {et~al.}(1997){Perryman}, {Lindegren}, {Kovalevsky},
  {Hog}, {Bastian}, {Bernacca}, {Creze}, {Donati}, {Grenon}, {Grewing}, {van
  Leeuwen}, {van der Marel}, {Mignard}, {Murray}, {Le Poole}, {Schrijver},
  {Turon}, {Arenou}, {Froeschle}, \& {Petersen}}]{perryman97}
{Perryman}, M.~A.~C., {Lindegren}, L., {Kovalevsky}, J., {et~al.} 1997, \aap,
  500, 501

\bibitem[{{Persson} {et~al.}(2019){Persson}, {Csizmadia}, {Mustill},
  {Fridlund}, {Hatzes}, {Nowak}, {Georgieva}, {Gandolfi}, {Davies},
  {Livingston}, {Palle}, {Monta{\~n}es Rodr{\'\i}guez}, {Endl}, {Hirano},
  {Prieto-Arranz}, {Korth}, {Grziwa}, {Esposito}, {Albrecht}, {Johnson},
  {Barrag{\'a}n}, {Parviainen}, {Van Eylen}, {Alonso Sobrino}, {Beck},
  {Cabrera}, {Carleo}, {Cochran}, {Dai}, {Deeg}, {de Leon}, {Eigm{\"u}ller},
  {Erikson}, {Fukui}, {Gonz{\'a}lez-Cuesta}, {Guenther}, {Hidalgo}, {Hjorth},
  {Kabath}, {Knudstrup}, {Kusakabe}, {Lam}, {Lund}, {Luque}, {Mathur},
  {Murgas}, {Narita}, {Nespral}, {Niraula}, {Olofsson}, {P{\"a}tzold}, {Rauer},
  {Redfield}, {Ribas}, {Skarka}, {Smith}, {Subjak}, \& {Tamura}}]{persson19}
{Persson}, C.~M., {Csizmadia}, S., {Mustill}, A.~J., {et~al.} 2019, \aap, 628,
  A64, \dodoi{10.1051/0004-6361/201935505}

\bibitem[{{Phillips} {et~al.}(2020){Phillips}, {Tremblin}, {Baraffe},
  {Chabrier}, {Allard}, {Spiegelman}, {Goyal}, {Drummond}, \&
  {H{\'e}brard}}]{phillips20}
{Phillips}, M.~W., {Tremblin}, P., {Baraffe}, I., {et~al.} 2020, \aap, 637,
  A38, \dodoi{10.1051/0004-6361/201937381}

\bibitem[{{Pilyavsky} {et~al.}(2011){Pilyavsky}, {Mahadevan}, {Kane}, {Howard},
  {Ciardi}, {de Pree}, {Dragomir}, {Fischer}, {Henry}, {Jensen}, {Laughlin},
  {Marlowe}, {Rabus}, {von Braun}, {Wright}, \& {Wang}}]{pilyavsky11}
{Pilyavsky}, G., {Mahadevan}, S., {Kane}, S.~R., {et~al.} 2011, \apj, 743, 162,
  \dodoi{10.1088/0004-637X/743/2/162}

\bibitem[{{Ramsey} {et~al.}(1998){Ramsey}, {Adams}, {Barnes}, {Booth},
  {Cornell}, {Fowler}, {Gaffney}, {Glaspey}, {Good}, {Hill}, {Kelton},
  {Krabbendam}, {Long}, {MacQueen}, {Ray}, {Ricklefs}, {Sage}, {Sebring},
  {Spiesman}, \& {Steiner}}]{ramsey98}
{Ramsey}, L.~W., {Adams}, M.~T., {Barnes}, T.~G., {et~al.} 1998, in Society of
  Photo-Optical Instrumentation Engineers (SPIE) Conference Series, Vol. 3352,
  Advanced Technology Optical/IR Telescopes VI, ed. L.~M. {Stepp}, 34--42

\bibitem[{{Rebolo} {et~al.}(1995){Rebolo}, {Zapatero Osorio}, \&
  {Mart{\'\i}n}}]{rebolo95}
{Rebolo}, R., {Zapatero Osorio}, M.~R., \& {Mart{\'\i}n}, E.~L. 1995, \nat,
  377, 129, \dodoi{10.1038/377129a0}

\bibitem[{{Reffert} \& {Quirrenbach}(2011)}]{reffert11}
{Reffert}, S., \& {Quirrenbach}, A. 2011, \aap, 527, A140,
  \dodoi{10.1051/0004-6361/201015861}

\bibitem[{{Rickman} {et~al.}(2020){Rickman}, {S{\'e}gransan}, {Hagelberg},
  {Beuzit}, {Cheetham}, {Delisle}, {Forveille}, \& {Udry}}]{rickman20}
{Rickman}, E.~L., {S{\'e}gransan}, D., {Hagelberg}, J., {et~al.} 2020, \aap,
  635, A203, \dodoi{10.1051/0004-6361/202037524}

\bibitem[{{Rickman} {et~al.}(2019){Rickman}, {S{\'e}gransan}, {Marmier},
  {Udry}, {Bouchy}, {Lovis}, {Mayor}, {Pepe}, {Queloz}, {Santos}, {Allart},
  {Bonvin}, {Bratschi}, {Cersullo}, {Chazelas}, {Choplin}, {Conod}, {Deline},
  {Delisle}, {Dos Santos}, {Figueira}, {Giles}, {Girard}, {Lavie}, {Martin},
  {Motalebi}, {Nielsen}, {Osborn}, {Ottoni}, {Raimbault}, {Rey}, {Roger},
  {Seidel}, {Stalport}, {Su{\'a}rez Mascare{\~n}o}, {Triaud}, {Turner},
  {Weber}, \& {Wyttenbach}}]{rickman19}
{Rickman}, E.~L., {S{\'e}gransan}, D., {Marmier}, M., {et~al.} 2019, \aap, 625,
  A71, \dodoi{10.1051/0004-6361/201935356}

\bibitem[{Ripley {et~al.}(2013)Ripley, Venables, Bates, Hornik, Gebhardt,
  Firth, \& Ripley}]{ripley13}
Ripley, B., Venables, B., Bates, D.~M., {et~al.} 2013, Cran R, 538

\bibitem[{{Roberts} {et~al.}(2015{\natexlab{a}}){Roberts}, {Tokovinin},
  {Mason}, {Hartkopf}, \& {Riddle}}]{roberts15b}
{Roberts}, Lewis~C., J., {Tokovinin}, A., {Mason}, B.~D., {Hartkopf}, W.~I., \&
  {Riddle}, R.~L. 2015{\natexlab{a}}, \aj, 150, 130,
  \dodoi{10.1088/0004-6256/150/4/130}

\bibitem[{{Roberts} {et~al.}(2015{\natexlab{b}}){Roberts}, {Mason}, {Neyman},
  {Wu}, {Riddle}, {Shelton}, {Angione}, {Baranec}, {Bouchez}, {Bui}, {Burruss},
  {Burse}, {Chordia}, {Croner}, {Das}, {Dekany}, {Guiwits}, {Hale}, {Henning},
  {Kulkarni}, {Law}, {McKenna}, {Milburn}, {Palmer}, {Punnadi}, {Ramaprakash},
  {Roberts}, {Tendulkar}, {Trinh}, {Troy}, {Truong}, \& {Zolkower}}]{roberts15}
{Roberts}, Lewis~C., J., {Mason}, B.~D., {Neyman}, C.~R., {et~al.}
  2015{\natexlab{b}}, \aj, 149, 144, \dodoi{10.1088/0004-6256/149/4/144}

\bibitem[{{Robertson} {et~al.}(2012){Robertson}, {Horner}, {Wittenmyer},
  {Endl}, {Cochran}, {MacQueen}, {Brugamyer}, {Simon}, {Barnes}, \&
  {Caldwell}}]{robertson12}
{Robertson}, P., {Horner}, J., {Wittenmyer}, R.~A., {et~al.} 2012, \apj, 754,
  50, \dodoi{10.1088/0004-637X/754/1/50}

\bibitem[{{Robotham}(2016)}]{robotham16}
{Robotham}, A.~S.~G. 2016, {magicaxis: Pretty scientific plotting with
  minor-tick and log minor-tick support}, Astrophysics Source Code Library.
\newblock \doeprint{1604.004}

\bibitem[{{Rodigas} {et~al.}(2016){Rodigas}, {Arriagada}, {Faherty},
  {Anglada-Escud{\'e}}, {Kaib}, {Butler}, {Shectman}, {Weinberger}, {Males},
  {Morzinski}, {Close}, {Hinz}, {Crane}, {Thompson}, {Teske}, {D{\'\i}az},
  {Minniti}, {Lopez-Morales}, {Adams}, \& {Boss}}]{rodigas16}
{Rodigas}, T.~J., {Arriagada}, P., {Faherty}, J., {et~al.} 2016, \apj, 818,
  106, \dodoi{10.3847/0004-637X/818/2/106}

\bibitem[{{Rosenthal} {et~al.}(2021){Rosenthal}, {Fulton}, {Hirsch},
  {Isaacson}, {Howard}, {Dedrick}, {Sherstyuk}, {Blunt}, {Petigura}, {Knutson},
  {Behmard}, {Chontos}, {Crepp}, {Crossfield}, {Dalba}, {Fischer}, {Henry},
  {Kane}, {Kosiarek}, {Marcy}, {Rubenzahl}, {Weiss}, \& {Wright}}]{rosenthal21}
{Rosenthal}, L.~J., {Fulton}, B.~J., {Hirsch}, L.~A., {et~al.} 2021, \apjs,
  255, 8, \dodoi{10.3847/1538-4365/abe23c}

\bibitem[{{Sahlmann} {et~al.}(2011){Sahlmann}, {S{\'e}gransan}, {Queloz},
  {Udry}, {Santos}, {Marmier}, {Mayor}, {Naef}, {Pepe}, \&
  {Zucker}}]{sahlmann11}
{Sahlmann}, J., {S{\'e}gransan}, D., {Queloz}, D., {et~al.} 2011, \aap, 525,
  A95, \dodoi{10.1051/0004-6361/201015427}

\bibitem[{{Santos} {et~al.}(2001){Santos}, {Mayor}, {Naef}, {Pepe}, {Queloz},
  {Udry}, \& {Burnet}}]{santos01}
{Santos}, N.~C., {Mayor}, M., {Naef}, D., {et~al.} 2001, \aap, 379, 999,
  \dodoi{10.1051/0004-6361:20011366}

\bibitem[{{Sato} {et~al.}(2013){Sato}, {Omiya}, {Harakawa}, {Liu}, {Izumiura},
  {Kambe}, {Takeda}, {Yoshida}, {Itoh}, {Ando}, {Kokubo}, \& {Ida}}]{sato13}
{Sato}, B., {Omiya}, M., {Harakawa}, H., {et~al.} 2013, \pasj, 65, 85,
  \dodoi{10.1093/pasj/65.4.85}

\bibitem[{{S{\'e}gransan} {et~al.}(2010){S{\'e}gransan}, {Udry}, {Mayor},
  {Naef}, {Pepe}, {Queloz}, {Santos}, {Demory}, {Figueira}, {Gillon},
  {Marmier}, {M{\'e}gevand}, {Sosnowska}, {Tamuz}, \& {Triaud}}]{segransan10}
{S{\'e}gransan}, D., {Udry}, S., {Mayor}, M., {et~al.} 2010, \aap, 511, A45,
  \dodoi{10.1051/0004-6361/200912136}

\bibitem[{{Shahaf} \& {Mazeh}(2019)}]{shahaf19}
{Shahaf}, S., \& {Mazeh}, T. 2019, \mnras, 487, 3356,
  \dodoi{10.1093/mnras/stz1517}

\bibitem[{{Simpson} {et~al.}(2010){Simpson}, {Baliunas}, {Henry}, \&
  {Watson}}]{simpson10}
{Simpson}, E.~K., {Baliunas}, S.~L., {Henry}, G.~W., \& {Watson}, C.~A. 2010,
  \mnras, 408, 1666, \dodoi{10.1111/j.1365-2966.2010.17230.x}

\bibitem[{{Snellen} {et~al.}(2014){Snellen}, {Brandl}, {de Kok}, {Brogi},
  {Birkby}, \& {Schwarz}}]{snellen14}
{Snellen}, I. A.~G., {Brandl}, B.~R., {de Kok}, R.~J., {et~al.} 2014, \nat,
  509, 63, \dodoi{10.1038/nature13253}

\bibitem[{{Snellen} \& {Brown}(2018)}]{snellen18}
{Snellen}, I.~A.~G., \& {Brown}, A.~G.~A. 2018, Nature Astronomy, 2, 883,
  \dodoi{10.1038/s41550-018-0561-6}

\bibitem[{{Sozzetti} \& {Desidera}(2010)}]{sozzetti10}
{Sozzetti}, A., \& {Desidera}, S. 2010, \aap, 509, A103,
  \dodoi{10.1051/0004-6361/200912717}

\bibitem[{{Sozzetti} {et~al.}(2006){Sozzetti}, {Udry}, {Zucker}, {Torres},
  {Beuzit}, {Latham}, {Mayor}, {Mazeh}, {Naef}, {Perrier}, {Queloz}, \&
  {Sivan}}]{sozzetti06}
{Sozzetti}, A., {Udry}, S., {Zucker}, S., {et~al.} 2006, \aap, 449, 417,
  \dodoi{10.1051/0004-6361:20054303}

\bibitem[{{Spiegel} \& {Burrows}(2012)}]{spiegel12}
{Spiegel}, D.~S., \& {Burrows}, A. 2012, \apj, 745, 174,
  \dodoi{10.1088/0004-637X/745/2/174}

\bibitem[{{Spiegel} {et~al.}(2011){Spiegel}, {Burrows}, \&
  {Milsom}}]{spiegel11}
{Spiegel}, D.~S., {Burrows}, A., \& {Milsom}, J.~A. 2011, \apj, 727, 57,
  \dodoi{10.1088/0004-637X/727/1/57}

\bibitem[{{Stassun} {et~al.}(2017){Stassun}, {Collins}, \& {Gaudi}}]{stassun17}
{Stassun}, K.~G., {Collins}, K.~A., \& {Gaudi}, B.~S. 2017, \aj, 153, 136,
  \dodoi{10.3847/1538-3881/aa5df3}

\bibitem[{{Stassun} {et~al.}(2019){Stassun}, {Oelkers}, {Paegert}, {Torres},
  {Pepper}, {De Lee}, {Collins}, {Latham}, {Muirhead}, {Chittidi},
  {Rojas-Ayala}, {Fleming}, {Rose}, {Tenenbaum}, {Ting}, {Kane}, {Barclay},
  {Bean}, {Brassuer}, {Charbonneau}, {Ge}, {Lissauer}, {Mann}, {McLean},
  {Mullally}, {Narita}, {Plavchan}, {Ricker}, {Sasselov}, {Seager}, {Sharma},
  {Shiao}, {Sozzetti}, {Stello}, {Vanderspek}, {Wallace}, \&
  {Winn}}]{stassun19}
{Stassun}, K.~G., {Oelkers}, R.~J., {Paegert}, M., {et~al.} 2019, \aj, 158,
  138, \dodoi{10.3847/1538-3881/ab3467}

\bibitem[{{Tamuz} {et~al.}(2008){Tamuz}, {S{\'e}gransan}, {Udry}, {Mayor},
  {Eggenberger}, {Naef}, {Pepe}, {Queloz}, {Santos}, {Demory}, {Figuera},
  {Marmier}, \& {Montagnier}}]{tamuz08}
{Tamuz}, O., {S{\'e}gransan}, D., {Udry}, S., {et~al.} 2008, \aap, 480, L33,
  \dodoi{10.1051/0004-6361:20078737}

\bibitem[{{Thalmann} {et~al.}(2009){Thalmann}, {Carson}, {Janson}, {Goto},
  {McElwain}, {Egner}, {Feldt}, {Hashimoto}, {Hayano}, {Henning}, {Hodapp},
  {Kandori}, {Klahr}, {Kudo}, {Kusakabe}, {Mordasini}, {Morino}, {Suto},
  {Suzuki}, \& {Tamura}}]{thalmann09}
{Thalmann}, C., {Carson}, J., {Janson}, M., {et~al.} 2009, \apjl, 707, L123,
  \dodoi{10.1088/0004-637X/707/2/L123}

\bibitem[{{Tinney} {et~al.}(2002){Tinney}, {Butler}, {Marcy}, {Jones}, {Penny},
  {McCarthy}, \& {Carter}}]{tinney02}
{Tinney}, C.~G., {Butler}, R.~P., {Marcy}, G.~W., {et~al.} 2002, \apj, 571,
  528, \dodoi{10.1086/339916}

\bibitem[{{Tokovinin}(2014{\natexlab{a}})}]{tokovinin14a}
{Tokovinin}, A. 2014{\natexlab{a}}, \aj, 147, 86,
  \dodoi{10.1088/0004-6256/147/4/86}

\bibitem[{{Tokovinin}(2014{\natexlab{b}})}]{tokovinin14b}
---. 2014{\natexlab{b}}, \aj, 147, 87, \dodoi{10.1088/0004-6256/147/4/87}

\bibitem[{{Tokovinin} \& {Kiyaeva}(2016)}]{tokovinin16}
{Tokovinin}, A., \& {Kiyaeva}, O. 2016, \mnras, 456, 2070,
  \dodoi{10.1093/mnras/stv2825}

\bibitem[{{Trifonov} {et~al.}(2020){Trifonov}, {Tal-Or}, {Zechmeister},
  {Kaminski}, {Zucker}, \& {Mazeh}}]{trifonov20}
{Trifonov}, T., {Tal-Or}, L., {Zechmeister}, M., {et~al.} 2020, \aap, 636, A74,
  \dodoi{10.1051/0004-6361/201936686ARXIV: 2001.05942OPEN}

\bibitem[{{Trifonov} {et~al.}(2017){Trifonov}, {K{\"u}rster}, {Zechmeister},
  {Zakhozhay}, {Reffert}, {Lee}, {Rodler}, {Vogt}, \& {Brems}}]{trifonov17}
{Trifonov}, T., {K{\"u}rster}, M., {Zechmeister}, M., {et~al.} 2017, \aap, 602,
  L8, \dodoi{10.1051/0004-6361/201731044}

\bibitem[{{Tull}(1998)}]{tull98}
{Tull}, R.~G. 1998, in Society of Photo-Optical Instrumentation Engineers
  (SPIE) Conference Series, Vol. 3355, Optical Astronomical Instrumentation,
  ed. S.~{D'Odorico}, 387--398

\bibitem[{{Tuomi} {et~al.}(2013){Tuomi}, {Jones}, {Jenkins}, {Tinney},
  {Butler}, {Vogt}, {Barnes}, {Wittenmyer}, {O'Toole}, {Horner}, {Bailey},
  {Carter}, {Wright}, {Salter}, \& {Pinfield}}]{tuomi12}
{Tuomi}, M., {Jones}, H.~R.~A., {Jenkins}, J.~S., {et~al.} 2013, \aap, 551,
  A79, \dodoi{10.1051/0004-6361/201220509}

\bibitem[{{Udry} {et~al.}(2000){Udry}, {Mayor}, {Queloz}, {Naef}, \&
  {Santos}}]{udry00}
{Udry}, S., {Mayor}, M., {Queloz}, D., {Naef}, D., \& {Santos}, N. 2000, in
  From Extrasolar Planets to Cosmology: The VLT Opening Symposium, ed.
  J.~{Bergeron} \& A.~{Renzini}, 571

\bibitem[{{van der Walt} {et~al.}(2011){van der Walt}, {Colbert}, \&
  {Varoquaux}}]{walt11}
{van der Walt}, S., {Colbert}, S.~C., \& {Varoquaux}, G. 2011, Computing in
  Science and Engineering, 13, 22, \dodoi{10.1109/MCSE.2011.37}

\bibitem[{{van Leeuwen}(2007)}]{leeuwen07}
{van Leeuwen}, F. 2007, \aap, 474, 653, \dodoi{10.1051/0004-6361:20078357}

\bibitem[{{Vandal} {et~al.}(2020){Vandal}, {Rameau}, \& {Doyon}}]{vandal20}
{Vandal}, T., {Rameau}, J., \& {Doyon}, R. 2020, \aj, 160, 243,
  \dodoi{10.3847/1538-3881/abba30}

\bibitem[{{Venner} {et~al.}(2021){Venner}, {Vanderburg}, \&
  {Pearce}}]{venner21}
{Venner}, A., {Vanderburg}, A., \& {Pearce}, L.~A. 2021, \aj, 162, 12,
  \dodoi{10.3847/1538-3881/abf932}

\bibitem[{{Vigan} {et~al.}(2016){Vigan}, {Bonnefoy}, {Ginski}, {Beust},
  {Galicher}, {Janson}, {Baudino}, {Buenzli}, {Hagelberg}, {D'Orazi},
  {Desidera}, {Maire}, {Gratton}, {Sauvage}, {Chauvin}, {Thalmann}, {Malo},
  {Salter}, {Zurlo}, {Antichi}, {Baruffolo}, {Baudoz}, {Blanchard},
  {Boccaletti}, {Beuzit}, {Carle}, {Claudi}, {Costille}, {Delboulb{\'e}},
  {Dohlen}, {Dominik}, {Feldt}, {Fusco}, {Gluck}, {Girard}, {Giro}, {Gry},
  {Henning}, {Hubin}, {Hugot}, {Jaquet}, {Kasper}, {Lagrange}, {Langlois}, {Le
  Mignant}, {Llored}, {Madec}, {Martinez}, {Mawet}, {Mesa}, {Milli},
  {Mouillet}, {Moulin}, {Moutou}, {Orign{\'e}}, {Pavlov}, {Perret}, {Petit},
  {Pragt}, {Puget}, {Rabou}, {Rochat}, {Roelfsema}, {Salasnich}, {Schmid},
  {Sevin}, {Siebenmorgen}, {Smette}, {Stadler}, {Suarez}, {Turatto}, {Udry},
  {Vakili}, {Wahhaj}, {Weber}, \& {Wildi}}]{vigan16}
{Vigan}, A., {Bonnefoy}, M., {Ginski}, C., {et~al.} 2016, \aap, 587, A55,
  \dodoi{10.1051/0004-6361/201526465}

\bibitem[{{Vogt}(1987)}]{vogt87}
{Vogt}, S.~S. 1987, \pasp, 99, 1214, \dodoi{10.1086/132107}

\bibitem[{{Vogt} {et~al.}(2002){Vogt}, {Butler}, {Marcy}, {Fischer},
  {Pourbaix}, {Apps}, \& {Laughlin}}]{vogt02}
{Vogt}, S.~S., {Butler}, R.~P., {Marcy}, G.~W., {et~al.} 2002, \apj, 568, 352,
  \dodoi{10.1086/338768}

\bibitem[{{Vogt} {et~al.}(2000){Vogt}, {Marcy}, {Butler}, \& {Apps}}]{vogt00}
{Vogt}, S.~S., {Marcy}, G.~W., {Butler}, R.~P., \& {Apps}, K. 2000, \apj, 536,
  902, \dodoi{10.1086/308981}

\bibitem[{Vogt {et~al.}(1994)Vogt, Allen, Bigelow, Bresee, Brown, Cantrall,
  Conrad, Couture, Delaney, Epps, Hilyard, Hilyard, Horn, Jern, Kanto, Keane,
  Kibrick, Lewis, Osborne, Pardeilhan, Pfister, Ricketts, Robinson, Stover,
  Tucker, Ward, \& Wei}]{vogt94}
Vogt, S.~S., Allen, S.~L., Bigelow, B.~C., {et~al.} 1994, in Instrumentation in
  Astronomy VIII, ed. D.~L. Crawford \& E.~R. Craine, Vol. 2198, International
  Society for Optics and Photonics (SPIE), 362 -- 375.
\newblock \url{https://doi.org/10.1117/12.176725}

\bibitem[{{Vogt} {et~al.}(2014){Vogt}, {Radovan}, {Kibrick}, {Butler},
  {Alcott}, {Allen}, {Arriagada}, {Bolte}, {Burt}, {Cabak}, {Chloros},
  {Cowley}, {Deich}, {Dupraw}, {Earthman}, {Epps}, {Faber}, {Fischer}, {Gates},
  {Hilyard}, {Holden}, {Johnston}, {Keiser}, {Kanto}, {Katsuki}, {Laiterman},
  {Lanclos}, {Laughlin}, {Lewis}, {Lockwood}, {Lynam}, {Marcy}, {McLean},
  {Miller}, {Misch}, {Peck}, {Pfister}, {Phillips}, {Rivera}, {Sandford},
  {Saylor}, {Stover}, {Thompson}, {Walp}, {Ward}, {Wareham}, {Wei}, \&
  {Wright}}]{vogt14}
{Vogt}, S.~S., {Radovan}, M., {Kibrick}, R., {et~al.} 2014, Publications of the
  Astronomical Society of the Pacific, 126, 359, \dodoi{10.1086/676120}

\bibitem[{{Wagner} {et~al.}(2019){Wagner}, {Apai}, \& {Kratter}}]{wagner19}
{Wagner}, K., {Apai}, D., \& {Kratter}, K.~M. 2019, \apj, 877, 46,
  \dodoi{10.3847/1538-4357/ab1904}

\bibitem[{{Wang} {et~al.}(2016){Wang}, {Graham}, {Pueyo}, {Kalas},
  {Millar-Blanchaer}, {Ruffio}, {De Rosa}, {Ammons}, {Arriaga}, {Bailey},
  {Barman}, {Bulger}, {Burrows}, {Cardwell}, {Chen}, {Chilcote}, {Cotten},
  {Fitzgerald}, {Follette}, {Doyon}, {Duch{\^e}ne}, {Greenbaum}, {Hibon},
  {Hung}, {Ingraham}, {Konopacky}, {Larkin}, {Macintosh}, {Maire}, {Marchis},
  {Marley}, {Marois}, {Metchev}, {Nielsen}, {Oppenheimer}, {Palmer}, {Patel},
  {Patience}, {Perrin}, {Poyneer}, {Rajan}, {Rameau}, {Rantakyr{\"o}},
  {Savransky}, {Sivaramakrishnan}, {Song}, {Soummer}, {Thomas}, {Vasisht},
  {Vega}, {Wallace}, {Ward-Duong}, {Wiktorowicz}, \& {Wolff}}]{wang16}
{Wang}, J.~J., {Graham}, J.~R., {Pueyo}, L., {et~al.} 2016, \aj, 152, 97,
  \dodoi{10.3847/0004-6256/152/4/97}

\bibitem[{{Wang} {et~al.}(2014){Wang}, {Sato}, {Omiya}, {Harakawa}, {Liu},
  {Song}, {He}, {Wu}, {Izumiura}, {Kambe}, {Takeda}, {Yoshida}, {Itoh}, {Ando},
  {Kokubo}, {Ida}, \& {Zhao}}]{wang14}
{Wang}, L., {Sato}, B., {Omiya}, M., {et~al.} 2014, \pasj, 66, 118,
  \dodoi{10.1093/pasj/psu113}

\bibitem[{{Wenger} {et~al.}(2000){Wenger}, {Ochsenbein}, {Egret}, {Dubois},
  {Bonnarel}, {Borde}, {Genova}, {Jasniewicz}, {Lalo{\"e}}, {Lesteven}, \&
  {Monier}}]{wenger00}
{Wenger}, M., {Ochsenbein}, F., {Egret}, D., {et~al.} 2000, \aaps, 143, 9,
  \dodoi{10.1051/aas:2000332}

\bibitem[{{Wilson} {et~al.}(2016){Wilson}, {H{\'e}brard}, {Santos}, {Sahlmann},
  {Montagnier}, {Astudillo-Defru}, {Boisse}, {Bouchy}, {Rey}, {Arnold},
  {Bonfils}, {Bourrier}, {Courcol}, {Deleuil}, {Delfosse}, {D{\'\i}az},
  {Ehrenreich}, {Forveille}, {Moutou}, {Pepe}, {Santerne}, {S{\'e}gransan}, \&
  {Udry}}]{wilson16}
{Wilson}, P.~A., {H{\'e}brard}, G., {Santos}, N.~C., {et~al.} 2016, \aap, 588,
  A144, \dodoi{10.1051/0004-6361/201527581}

\bibitem[{{Wittenmyer} {et~al.}(2007){Wittenmyer}, {Endl}, \&
  {Cochran}}]{wittenmyer07}
{Wittenmyer}, R.~A., {Endl}, M., \& {Cochran}, W.~D. 2007, \apj, 654, 625,
  \dodoi{10.1086/509110}

\bibitem[{{Wittenmyer} {et~al.}(2009){Wittenmyer}, {Endl}, {Cochran},
  {Levison}, \& {Henry}}]{wittenmyer09}
{Wittenmyer}, R.~A., {Endl}, M., {Cochran}, W.~D., {Levison}, H.~F., \&
  {Henry}, G.~W. 2009, \apjs, 182, 97, \dodoi{10.1088/0067-0049/182/1/97}

\bibitem[{{Wittenmyer} {et~al.}(2015){Wittenmyer}, {Wang}, {Liu}, {Horner},
  {Endl}, {Johnson}, {Tinney}, \& {Carter}}]{wittenmyer15}
{Wittenmyer}, R.~A., {Wang}, L., {Liu}, F., {et~al.} 2015, \apj, 800, 74,
  \dodoi{10.1088/0004-637X/800/1/74}

\bibitem[{{Wittenmyer} {et~al.}(2012){Wittenmyer}, {Horner}, {Tuomi}, {Salter},
  {Tinney}, {Butler}, {Jones}, {O'Toole}, {Bailey}, {Carter}, {Jenkins},
  {Zhang}, {Vogt}, \& {Rivera}}]{wittenmyer12}
{Wittenmyer}, R.~A., {Horner}, J., {Tuomi}, M., {et~al.} 2012, \apj, 753, 169,
  \dodoi{10.1088/0004-637X/753/2/169}

\bibitem[{{Wittenmyer} {et~al.}(2017){Wittenmyer}, {Horner}, {Mengel},
  {Butler}, {Wright}, {Tinney}, {Carter}, {Jones}, {Anglada-Escud{\'e}},
  {Bailey}, \& {O'Toole}}]{wittenmyer17}
{Wittenmyer}, R.~A., {Horner}, J., {Mengel}, M.~W., {et~al.} 2017, \aj, 153,
  167, \dodoi{10.3847/1538-3881/aa5f17}

\bibitem[{Wittenmyer {et~al.}(2019)Wittenmyer, Wang, Horner, Butler, Tinney,
  Carter, Wright, Jones, Bailey, O’Toole, \& Johns}]{wittenmyer19}
Wittenmyer, R.~A., Wang, S., Horner, J., {et~al.} 2019, Monthly Notices of the
  Royal Astronomical Society, 492, 377

\bibitem[{{Wittenmyer} {et~al.}(2020){Wittenmyer}, {Wang}, {Horner}, {Butler},
  {Tinney}, {Carter}, {Wright}, {Jones}, {Bailey}, {O'Toole}, \&
  {Johns}}]{wittenmyer20}
{Wittenmyer}, R.~A., {Wang}, S., {Horner}, J., {et~al.} 2020, \mnras, 492, 377,
  \dodoi{10.1093/mnras/stz3436}

\bibitem[{{Xuan} {et~al.}(2020){Xuan}, {Kennedy}, {Wyatt}, \&
  {Yelverton}}]{xuan20}
{Xuan}, J.~W., {Kennedy}, G.~M., {Wyatt}, M.~C., \& {Yelverton}, B. 2020,
  \mnras, 499, 5059, \dodoi{10.1093/mnras/staa3155}

\bibitem[{{Xuan} {et~al.}(2019){Xuan}, {Knutson}, {Bryan}, {Benneke}, \&
  {Bowler}}]{xuan19}
{Xuan}, W.~J., {Knutson}, H., {Bryan}, M., {Benneke}, B., \& {Bowler}, B. 2019,
  in American Astronomical Society Meeting Abstracts, Vol. 233, American
  Astronomical Society Meeting Abstracts \#233, 247.11

\bibitem[{{Zechmeister} {et~al.}(2018){Zechmeister}, {Reiners}, {Amado},
  {Azzaro}, {Bauer}, {B{\'e}jar}, {Caballero}, {Guenther}, {Hagen}, {Jeffers},
  {Kaminski}, {K{\"u}rster}, {Launhardt}, {Montes}, {Morales}, {Quirrenbach},
  {Reffert}, {Ribas}, {Seifert}, {Tal-Or}, \& {Wolthoff}}]{zechmeister18}
{Zechmeister}, M., {Reiners}, A., {Amado}, P.~J., {et~al.} 2018, \aap, 609,
  A12, \dodoi{10.1051/0004-6361/201731483}

\bibitem[{{Zhu} {et~al.}(2012){Zhu}, {Hartmann}, {Nelson}, \& {Gammie}}]{zhu12}
{Zhu}, Z., {Hartmann}, L., {Nelson}, R.~P., \& {Gammie}, C.~F. 2012, \apj, 746,
  110, \dodoi{10.1088/0004-637X/746/1/110}

\bibitem[{{Ziegler} {et~al.}(2021){Ziegler}, {Tokovinin}, {Latiolais},
  {Brice{\~n}o}, {Law}, \& {Mann}}]{ziegler21}
{Ziegler}, C., {Tokovinin}, A., {Latiolais}, M., {et~al.} 2021, \aj, 162, 192,
  \dodoi{10.3847/1538-3881/ac17f6}

\bibitem[{{Zucker} \& {Mazeh}(2000)}]{zucker00}
{Zucker}, S., \& {Mazeh}, T. 2000, \apjl, 531, L67, \dodoi{10.1086/312523}

\bibitem[{{Zucker} \& {Mazeh}(2001)}]{zucker01}
---. 2001, \apj, 562, 549, \dodoi{10.1086/322959}

\bibitem[{{Zurlo} {et~al.}(2018){Zurlo}, {Mesa}, {Desidera}, {Messina},
  {Gratton}, {Moutou}, {Beuzit}, {Biller}, {Boccaletti}, {Bonavita},
  {Bonnefoy}, {Bhowmik}, {Brandner}, {Buenzli}, {Chauvin}, {Cudel}, {D'Orazi},
  {Feldt}, {Hagelberg}, {Janson}, {Lagrange}, {Langlois}, {Lannier}, {Lavie},
  {Lazzoni}, {Maire}, {Meyer}, {Mouillet}, {Peretti}, {Perrot}, {Potiron},
  {Salter}, {Schmidt}, {Sissa}, {Vigan}, {Delboulb{\'e}}, {Petit}, {Ramos},
  {Rigal}, \& {Rochat}}]{zurlo18}
{Zurlo}, A., {Mesa}, D., {Desidera}, S., {et~al.} 2018, \mnras, 480, 35,
  \dodoi{10.1093/mnras/sty1809}

\end{thebibliography}

\appendix
\figsetstart
\figsetnum{1}
\figsettitle{Combined fit for the 144 systems for online access. }

\figsetgrpstart
\figsetgrpnum{1.1}
\figsetgrptitle{HD105618}
\figsetplot{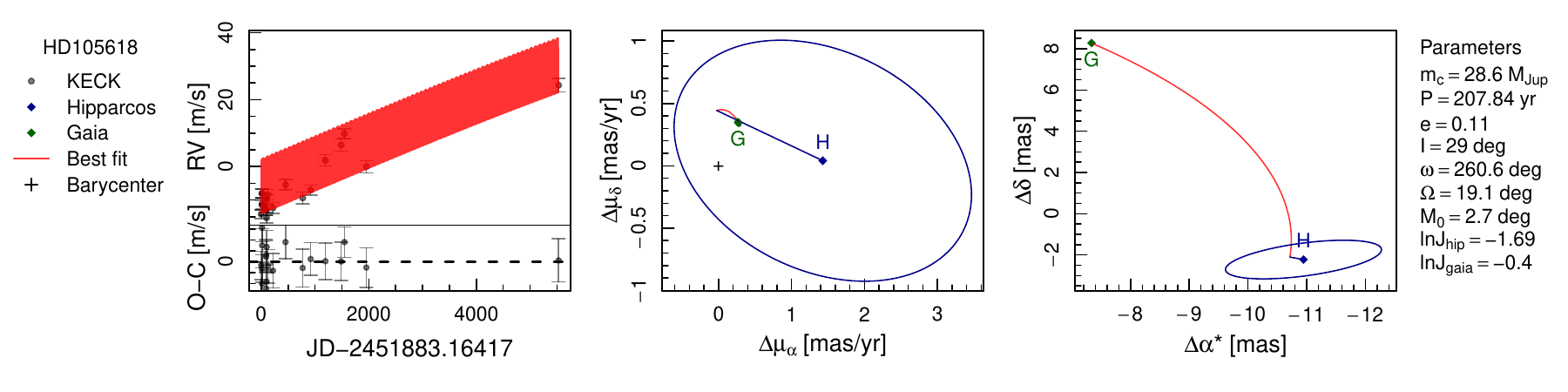}
\figsetgrpnote{Combined RV and astrometry fit. }
\figsetgrpend

\figsetgrpstart
\figsetgrpnum{1.2}
\figsetgrptitle{HD105811}
\figsetplot{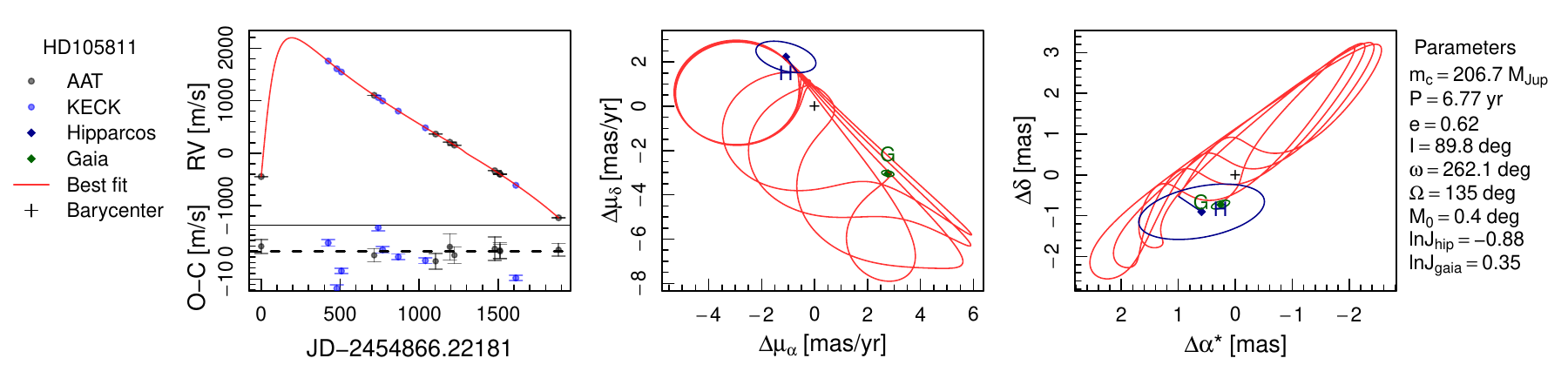}
\figsetgrpnote{Combined RV and astrometry fit. }
\figsetgrpend

\figsetgrpstart
\figsetgrpnum{1.3}
\figsetgrptitle{HD106515A}
\figsetplot{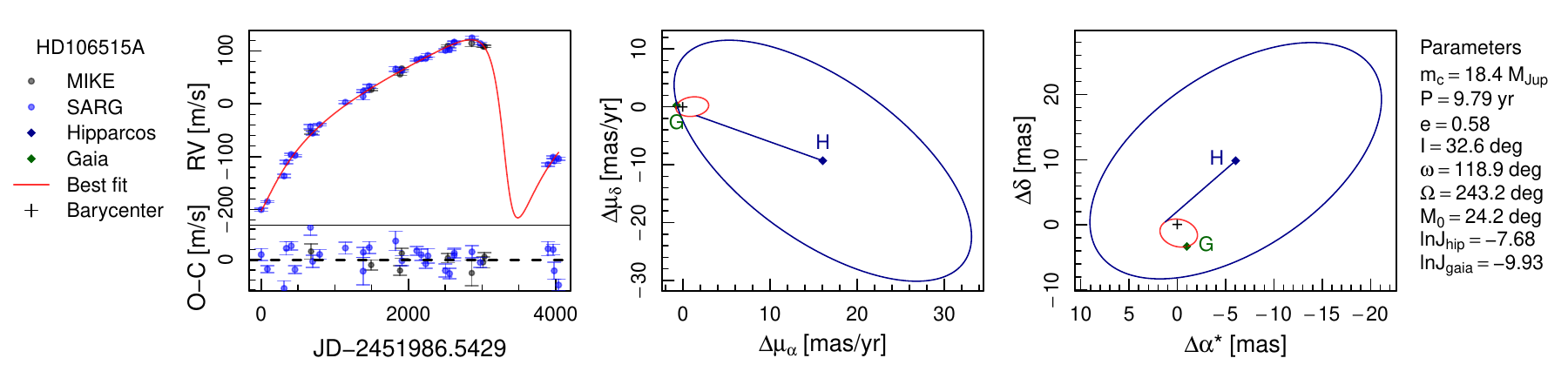}
\figsetgrpnote{Combined RV and astrometry fit. }
\figsetgrpend

\figsetgrpstart
\figsetgrpnum{1.4}
\figsetgrptitle{HD106574}
\figsetplot{HD106574_N0_fit.pdf}
\figsetgrpnote{Combined RV and astrometry fit. }
\figsetgrpend

\figsetgrpstart
\figsetgrpnum{1.5}
\figsetgrptitle{HD10697}
\figsetplot{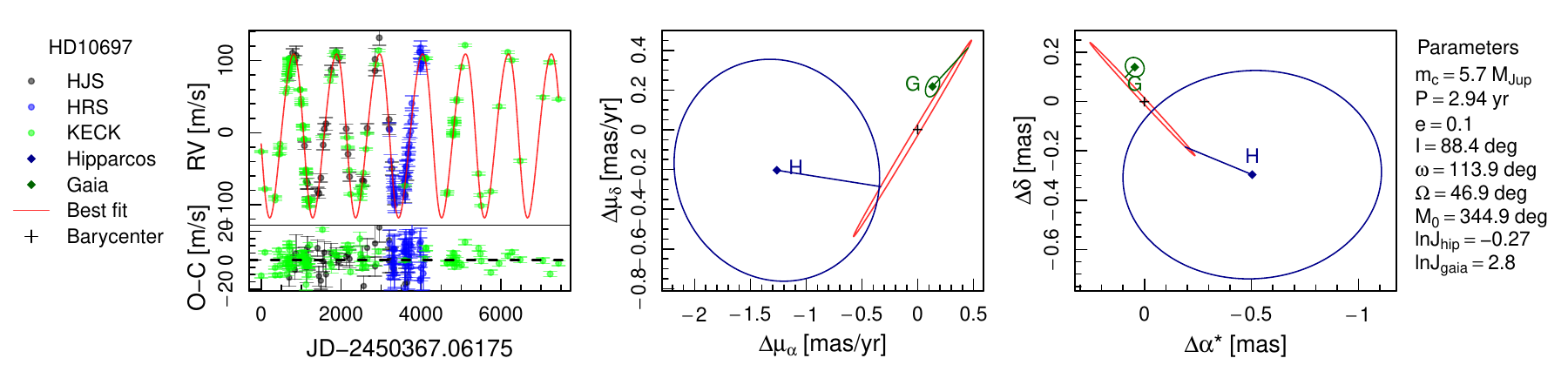}
\figsetgrpnote{Combined RV and astrometry fit. }
\figsetgrpend

\figsetgrpstart
\figsetgrpnum{1.34}
\figsetgrptitle{HD13724}
\figsetplot{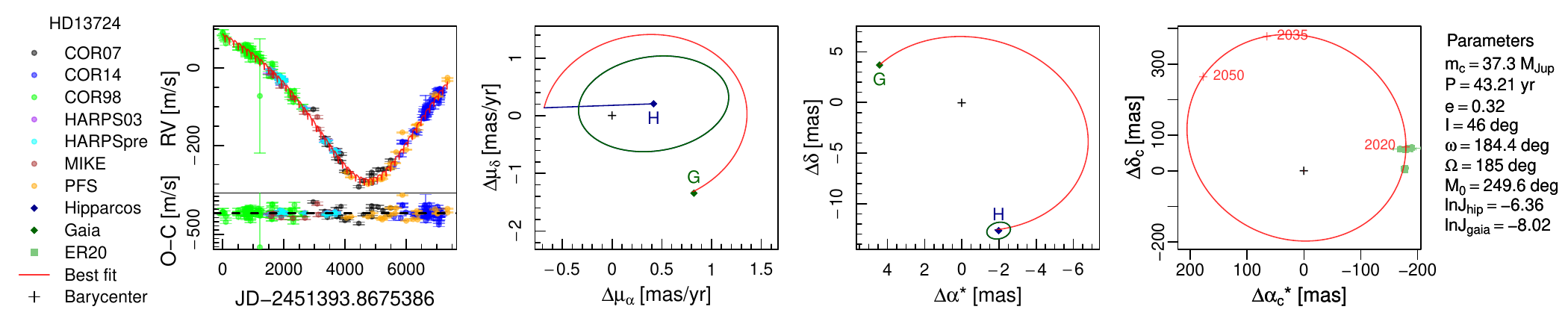}
\figsetgrpnote{Combined RV and astrometry fit. }
\figsetgrpend

\figsetgrpstart
\figsetgrpnum{1.71}
\figsetgrptitle{HD182488}
\figsetplot{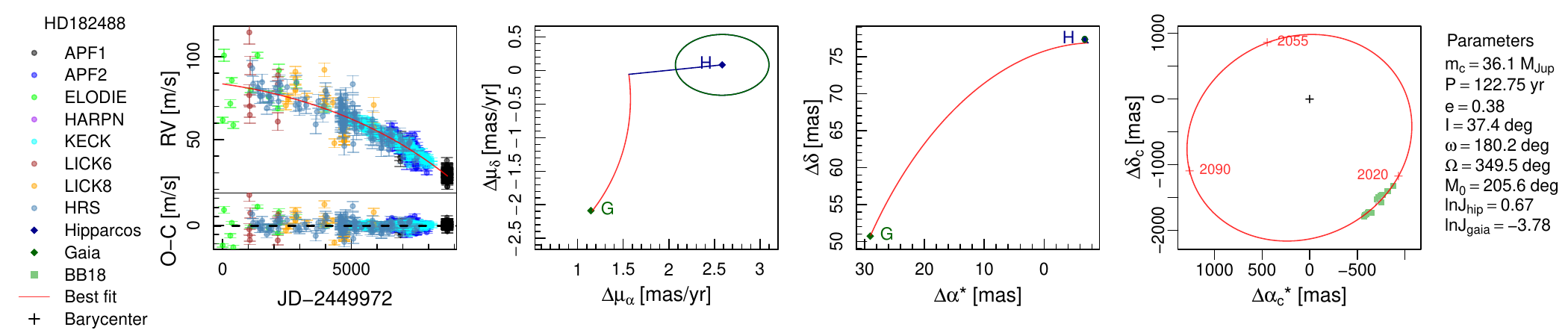}
\figsetgrpnote{Combined RV and astrometry fit. }
\figsetgrpend

\figsetgrpstart
\figsetgrpnum{1.122}
\figsetgrptitle{HD39060}
\figsetplot{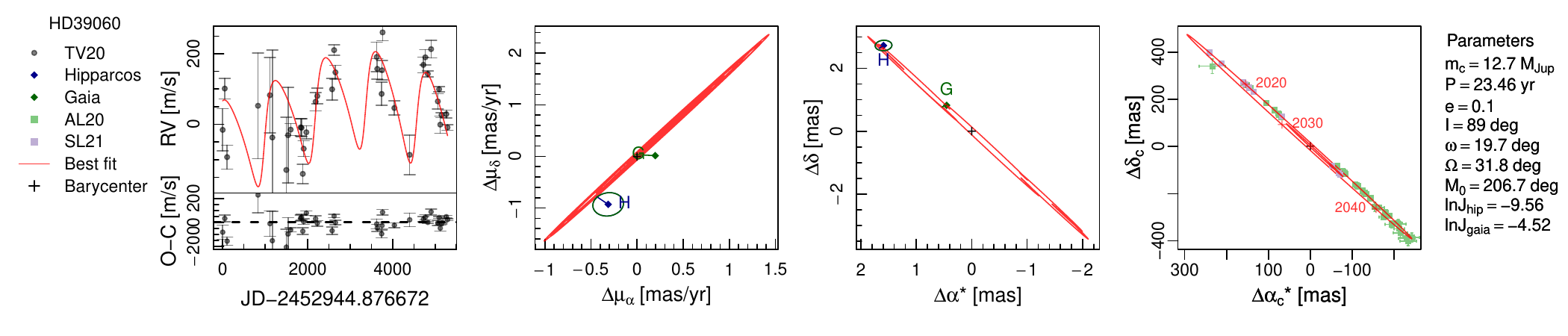}
\figsetgrpnote{Combined RV and astrometry fit. }
\figsetgrpend

\figsetgrpstart
\figsetgrpnum{1.123}
\figsetgrptitle{HD39091}
\figsetplot{HD39091_N0_fit.pdf}
\figsetgrpnote{Combined RV and astrometry fit. }
\figsetgrpend

\figsetgrpstart
\figsetgrpnum{1.124}
\figsetgrptitle{HD39213}
\figsetplot{HD39213_N0_fit.pdf}
\figsetgrpnote{Combined RV and astrometry fit. }
\figsetgrpend

\figsetgrpstart
\figsetgrpnum{1.125}
\figsetgrptitle{HD4113}
\figsetplot{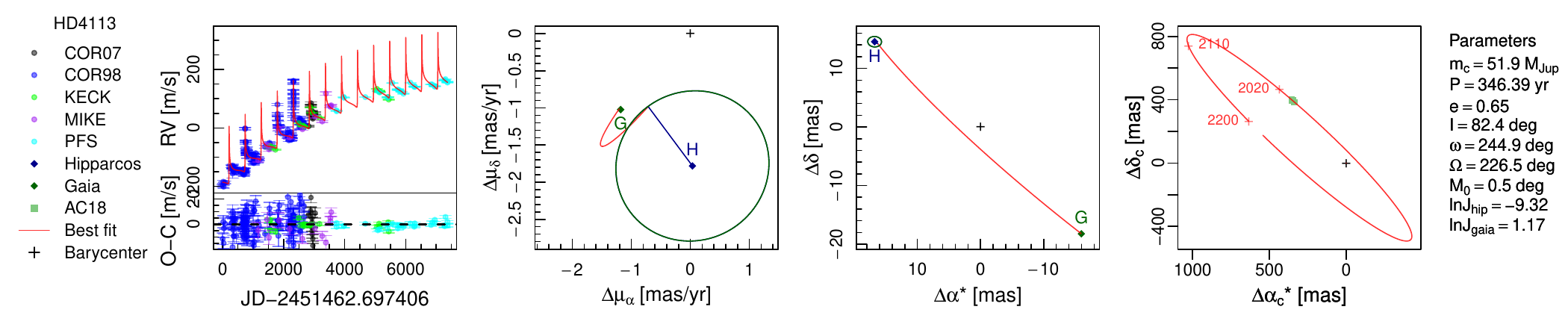}
\figsetgrpnote{Combined RV and astrometry fit. }
\figsetgrpend

\figsetgrpstart
\figsetgrpnum{1.126}
\figsetgrptitle{HD42581}
\figsetplot{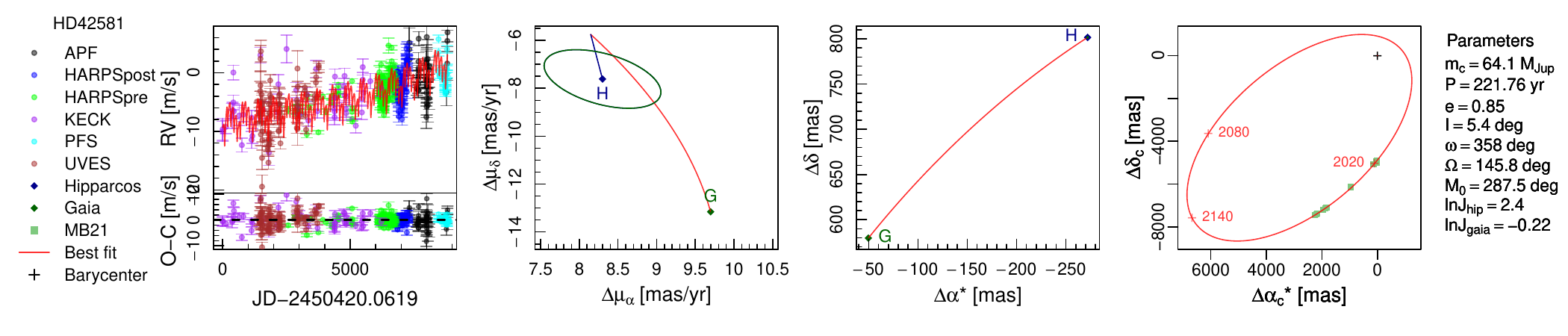}
\figsetgrpnote{Combined RV and astrometry fit. }
\figsetgrpend

\figsetgrpstart
\figsetgrpnum{1.144}
\figsetgrptitle{HD7449}
\figsetplot{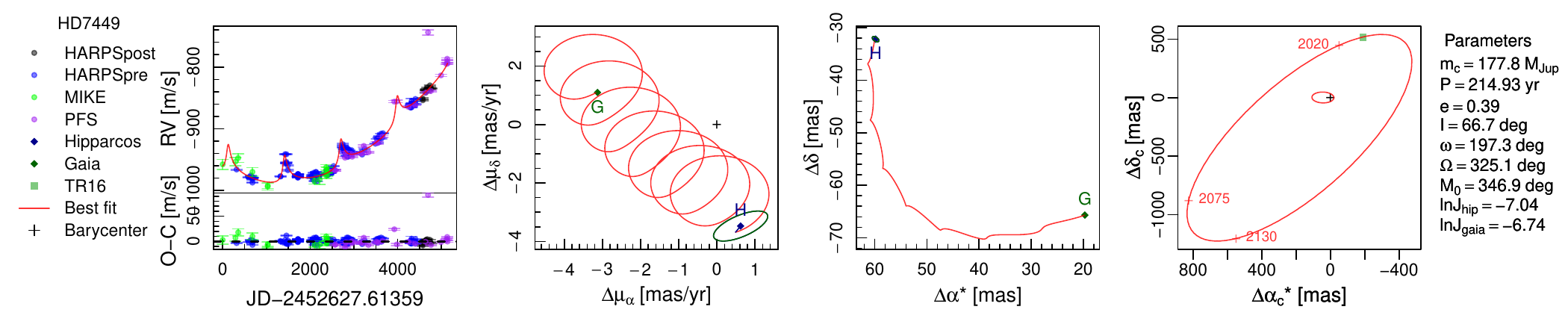}
\figsetgrpnote{Combined RV and astrometry fit. }
\figsetgrpend

\figsetend

\end{document}